\documentclass[11pt,a4paper]{article}
\pdfoutput=1
\usepackage{jheppub}
\usepackage{amsthm,amsbsy,amsfonts,mathrsfs,enumerate,float,wrapfig,amsmath}
\newcommand{\nn}{\nonumber}
\newcommand{\be}{\begin{equation}}
\newcommand{\ee}{\end{equation}}
\newcommand{\bea}{\begin{eqnarray}}
\newcommand{\eea}{\end{eqnarray}}

\usepackage{tikz}
\usetikzlibrary{calc}
\usetikzlibrary{decorations}
\pgfdeclaredecoration{ignore}{final}
{
\state{final}{}
}

 \pgfdeclaremetadecoration{middle}{initial}{
    \state{initial}[
        width={(\pgfmetadecoratedpathlength - \the\pgfdecorationsegmentlength)/2},
        next state=middle
    ]
    {\decoration{moveto}}

    \state{middle}[
        width={\the\pgfdecorationsegmentlength},
        next state=final
    ]
    {\decoration{curveto}}

    \state{final}
    {\decoration{ignore}}
}
\tikzset{middle segment/.style={decoration={middle},decorate, segment length=#1}}

\newcommand{\U}{\mathrm{U}}
\newcommand{\SU}{\mathrm{SU}}
\newcommand{\SO}{\mathrm{SO}}
\newcommand{\Sp}{\mathrm{Sp}}

\newcommand{\E}{E}
\newcommand{\D}{D}
\newcommand{\A}{A}
\graphicspath{ {./tikzfigures/} }

\renewcommand{\emptyset}{\varnothing}
\renewcommand{\hat}{\widehat}
\renewcommand{\tilde}{\widetilde}

\usepackage[T1]{fontenc}

\title{5d/6d DE instantons from trivalent gluing of web diagrams}

\author[a]{Hirotaka Hayashi,}
\author[b, c]{Kantaro Ohmori}

\affiliation[a]{Department of Physics, School of Science, Tokai University, 4-1-1 Kitakaname, Hiratsuka-shi, Kanagawa 259-1292, Japan}
\affiliation[b]{School of Natural Sciences, Institute for Advanced Study, Princeton, NJ 08540, USA}
\affiliation[c]{Department of Physics, Faculty of Science, The University of Tokyo, 7-3-1 Hongo, Bunkyo-ku, Tokyo 113-0033, Japan }

\emailAdd{h.hayashi@tokai.ac.jp}
\emailAdd{kantaro@ias.edu}

\abstract{
	We propose a new prescription for computing the Nekrasov partition functions of five-dimensional theories with eight supercharges realized by gauging non-perturbative flavor symmetries of three five-dimensional superconformal field theories. The topological vertex formalism gives a way to compute the partition functions of the matter theories with flavor instanton backgrounds, and the gauging is achieved by summing over Young diagrams. We apply the prescription to calculate the Nekrasov partition functions of various five-dimensional gauge theories such as $\SO(2N)$ gauge theories with or without hypermultiplets in the vector representation and also pure $E_6, E_7, E_8$ gauge theories. Furthermore, the technique can be applied to computations of the Nekrasov partition functions of five-dimensional theories which arise from circle compactifications of six-dimensional minimal superconformal field theories characterized by the gauge groups $\SU(3), \SO(8), E_6, E_7, E_8$. We exemplify our method by comparing some of the obtained partition functions with known results and find perfect agreement. We also present a prescription of extending the gluing rule to the refined topological vertex. 
}

\begin{document}
\preprint{
\begin{flushright}
\tt 
UT-17-06
\end{flushright}
}

\maketitle


\section{Introduction}
\label{sec:intro}
The (refined) topological vertex is a powerful tool to compute the all genus topological string amplitudes for toric Calabi--Yau threefolds \cite{Iqbal:2002we, Aganagic:2003db, Awata:2005fa, Iqbal:2007ii}.
One can compute the full topological string partition function like a Feynman diagram-like method and it can yield the full list of the Gromov--Witten invariants and the Gopakumar--Vafa invariants of a toric Calabi--Yau threefold in principle.
The topological string partition function also has a physical interpretation through string theory or M-theory.
When we consider M-theory on a non--compact Calabi--Yau threefold with a compact base that is contractible, the low energy effective field theory gives rise to a five-dimensional (5d) theory with eight supercharges which has a ultraviolet (UV) completion \cite{Witten:1996qb, Morrison:1996xf, Douglas:1996xp, Intriligator:1997pq}.
Then M2-branes wrapping various holomorphic curves in the Calabi--Yau threefold yield BPS particles in the 5d theory.
Therefore, the curve counting for a non--compact Calabi--Yau threefold is equivalent to the counting of BPS particles of the 5d theory and this implies that the topological string partition function is equal to the Nekrasov partition function up to some extra factors.
Indeed several checks of the equality have been done for example in \cite{Iqbal:2003ix, Iqbal:2003zz, Eguchi:2003sj, Hollowood:2003cv, Taki:2007dh} for 5d $\SU(N)$ gauge theories with flavors by utilizing the method of the topological vertex.

Recently, the topological vertex formalism has been extended for computing the topological string partition functions of certain non--toric Calabi--Yau threefolds \cite{Hayashi:2013qwa, Hayashi:2014wfa, Hayashi:2015xla}\footnote{There is also another vertex--like approach to compute the unrefined topological string amplitudes for some non--toric Calabi--Yau threefolds \cite{Diaconescu:2005ik, Diaconescu:2005mv}.}. The new method makes use of a Higgs prescription of the superconformal index in \cite{Gaiotto:2012uq, Gaiotto:2012xa}\footnote{In terms of geometry, the Higgsing corresponds to a topology changing transition and a similar technique has been also used in \cite{Dimofte:2010tz, Taki:2010bj, Aganagic:2011sg, Aganagic:2012hs} in the context of the refined version of the geometric transition.}. In fact, some non--toric Calabi--Yau threefold can be obtained from a topology changing transition or a Higgsing from a toric Calabi--Yau threefold. Then applying the Higgsing prescription for the topological string partition function of the ``UV'' Calabi--Yau threefold gives rise to the topological string partition function of the ``infrared'' (IR) non--toric Calabi--Yau threefold. This new technique enables us to compute the Nekrasov partition functions of the 5d rank one $E_7, E_8$ theories \cite{Hayashi:2013qwa, Hayashi:2014wfa}, the 5d $\SU(N)$ gauge theory with a hypermultiplet in the antisymmetric representation and also the 5d $\Sp(N)$ gauge theory \cite{Hayashi:2016jak}. Furthermore, it has been also applied to the calculation of the Nekrasov partition functions of 5d theories which has a six--dimensional (6d) UV completion, and non-trivial checks with the elliptic genus of the 6d self--dual strings have been done in \cite{Kim:2015jba, Hayashi:2016abm}. 

Although the new method enlarges the space of non--compact Calabi--Yau threefolds to which we can apply the topological vertex, there is still a large class of non--compact Calabi--Yau threefolds to which we have not yet known how to apply the topological vertex. An interesting class of such Calabi--Yau threefolds is the ones which yield 5d gauge theories with a gauge group $\SO(2N)$ or $\E_6, \E_7, \E_8$. In this paper, we propose a new technique which enables us to compute the Nekrasov partition functions of the 5d pure gauge theories with a gauge group $\SO(2N)$ or $\E_6, \E_7, \E_8$ from the topological vertex. The new method utilizes a dual description of the 5d pure gauge theory with a gauge group of $DE$--type. In fact, it turns out that the dual description is given by gauging the diagonal part of flavor symmetries of {\it three} 5d theories. We call such a gauging {\it trivalent gauging}. We have often encountered the case of gauging the diagonal part of flavor symmetries of two 5d theories from toric Calabi--Yau threefolds or equivalently 5-brane webs \cite{Aharony:1997ju, Aharony:1997bh, Leung:1997tw}. Gauging the flavor symmetries of three 5d theories is a natural generalization but goes beyond the standard picture of 5-brane webs.
The main aim of this paper is to formulate a novel method to compute the Nekrasov partition functions of 5d theories constructed by the trivalent gauging. The 5d theories coupled by the trivalent gauging may be considered as ``matter'' parts for the gauging. We indeed develop a way to compute the partition functions of the 5d theories as a ``matter'' contribution for the gauging from the topological vertex. Then, the trivalent gauging can be implemented by inserting the Nekrasov partition function of vector multiplets for the gauging and summing over Young diagrams. The prescription may be interpreted as a generalization of the gauging for the superconformal index in four-dimension \cite{Romelsberger:2005eg, Kinney:2005ej,Gadde:2011uv}. However, the extension to the gauging for five-dimensional partition functions is quite non-trivial compared with the four-dimensional case since we need to add instanton contributions which appear by the gauging.  

The new prescription of the trivalent gauging not only apply to the partition functions of 5d theories which have a 5d UV completion but also apply the partition functions of 5d theories which have a 6d UV completion. An interesting class of 6d superconformal field theories (SCFTs) are non-Higgsable cluster theories \cite{Morrison:2012np, Heckman:2013pva}. These 6d $\mathcal{N}=(1, 0) $SCFTs are an important ingredient for the atomic classification of general 6d SCFTs \cite{Heckman:2013pva, DelZotto:2014hpa, Heckman:2015bfa}. When the non-Higgsable cluster has only one tensor multiplet then they are called 6d minimal SCFTs and labeled by an integer $n=3, 4, 5, 6, 7, 8$ and $12$ \cite{Morrison:2012np, Heckman:2013pva}. 5d descriptions for some 6d minimal SCFTs with eight supercharges have been proposed in \cite{DelZotto:2015rca}. In fact, it turns out that the 5d descriptions for the cases of $n=4, 6, 8, 12$ can be described by gauging flavor symmetries of three or four 5d theories. Therefore, we can use the trivalent gauging method and it is possible to compute the Nekrasov partition functions of the 5d descriptions of some 6d minimal SCFTs on a circle. The Nekrasov partition function for a 5d theory with a 6d UV completion can be also interpreted as the sum of the elliptic genera of the self--dual strings in the 6d SCFT. We will give a non--trivial check between the result from the trivalent gauging and the elliptic genus for the case of $n=4$ by using the elliptic genus computed in \cite{Haghighat:2014vxa}. We will further propose a 5d description of the 6d minimal SCFT of the case $n=3$ and calculated its 5d Nekrasov partition function. Again we will see a non-trivial matching with the elliptic genus computation recently done in \cite{Kim:2016foj}. 

The organization of this paper is as follows.
In section \ref{sec:newweb}, we first determine a dual description for the 5d $\SO(2N+4)$ gauge theory with or without hypermultiplets in the vector representation and also for  the 5d pure gauge theories with a gauge group of $E$--type.
In section \ref{sec:partSO2Np4}, we present a new technique to compute the topological string partition function from the trivalent gauging of 5d theories.
We then apply the method to compute the Nekrasov partition function of the 5d $\SO(2N+4)$ gauge theory with or without flavors and perform non--trivial checks with known results.
We then apply the trivalent gauging prescription for the partition functions of the pure $\E_{6}, \E_7, \E_8$ gauge theories in section \ref{sec:5dE}.
In section \ref{sec:6dminimal}, the trivalent gauging method is applied to 5d descriptions for some minimal 6d SCFTs.
We also propose a 5d description for the 6d minimal SCFT in the case of $n=3$, and give non--trivial support for it.
We further comment on a 5d description of a non-Higgsable cluster theory with multiplet tensor multiplets in section \ref{sec:6dNHC}.
In section \ref{sec:refined}, we present a way to extend the prescription of the trivalent gauging to the refined topological vertex formalism.
We then conclude our work in section \ref{sec:concl}.
In appendix \ref{sec:SOodd}, we describe a relation between the $SO(2N+4)$ gauge theory and the $\SO(2N+3)$ gauge theory from a Higgsing, which provides  a way to compute the Nekrasov partition function of the $\SO(2N+3)$ gauge theory from the $\SO(2N+4)$ gauge theory.
We finally summarize technical tools used in this paper in appendix \ref{sec:formulae}. 

This paper accompanies a \texttt{Mathematica} notebook which is available from the arXiv web site. The notebook performs some of the computations of topological vertices exhibited in section \ref{sec:partSO2Np4}, section \ref{sec:5dE} and section \ref{sec:6dminimal}, and the computation of the Hilbert series explained in appendix \ref{sec:formulae}. The notebook utilizes the \texttt{Mathematica} application LieART \cite{Feger:2012bs}. The notebook only provides calculations related to the unrefined limit.

\bigskip

\section{A dual description of 5d gauge theory with $D, E$-type gauge group}
\label{sec:newweb}
Five-dimensional gauge theories with eight supercharges can be realized by compactifying M-theory on a singular Calabi--Yau threefolds $X_3$ \cite{Witten:1996qb, Morrison:1996xf, Douglas:1996xp, Intriligator:1997pq}. When the Calabi-Yau threefold $X_3$ has a $G$--type surface singularity over a sphere $C_B$, then the low energy effective field theory from the M-theory compactification yields a 5d pure gauge theory with a gauge group $G$. Here $G$ is either $\A_N = \SU(N+1), (N=1, 2, \cdots)$, $\D_{N+2} = \SO(2N+4), (N=2, 3, \cdots)$\footnote{An uncommon convention for $N$ is due to the construction of its dual theory in this section.} or $\E_6, \E_7, \E_8$. 

The resolution of the singularity means that the 5d gauge theory is on the Coulomb branch. The Calabi-Yau manifold $\tilde{X}_3$ after the resolution contains a collection of spheres fibered over the base sphere $C_B$. The intersections among collection of the fibered spheres form a shape of the Dykin diagram of the Lie algebra $\mathfrak{g}$ (the Lie algebra of a Lie group $G$) corresponding to the resolution of the $G$-type singularity. We denote the fiber which consists of spheres alighted along the Dynkin-diagram  of type $\mathfrak{g}$ by $F_{\mathfrak{g}}$. Each sphere in $F_{\mathfrak{g}}$ corresponds to a simple root of $\mathfrak{g}$ and let a collection of spheres corresponding to a root $\alpha$ be $C_{\alpha}$. Then an M2-brane wrapping a curve $C_{\alpha}$ in $F_{\mathfrak{g}}$ yields a massive W-boson for the root $\alpha$ of $\mathfrak{g}$ in the 5d gauge theory. Therefore, the size of $C_{\alpha}$ is a Coulomb branch modulus. On the other hand, an M2-brane wrapping the base $C_B$ yields an instanton particle of the 5d gauge theory. The size of the base $C_B$ is then related to $\frac{1}{g_{YM}^2}$ where $g_{YM}$ is the 5d gauge coupling. We also denote a complex surface which is $C_{\alpha}$ fibration over $C_B$ by $S_{\alpha}$. 

From this construction it is clear that the gauge theory information is encoded in the complex two-dimensional space $S_{\mathfrak{g}}$ which is given by the $F_{\mathfrak{g}}$ fibration over the base $C_B$.
The effect of gravity may be neglected by taking a limit where the transverse direction to $S_{\mathfrak{g}}$ is infinitely large.
We will always take the field theory limit and hence the background $\tilde{X}_3$ is a non-compact Calabi--Yau threefold whose compact base is given by the complex surface $S_{\mathfrak{g}}$.
More generally, M--theory on a non--compact Calabi--Yau manifold which is a line bundle over a compact surface $S$ will yield a 5d $\mathcal{N}=1$ supersymmetric theory.
When the complex surface $S$ is contractible then the 5d theory has a UV completion \cite{Witten:1996qb, Morrison:1996xf, Douglas:1996xp, Intriligator:1997pq} and the theory becomes a SCFT when the volume of $S$ vanishes .
We will restrict our attention to such a case in this paper. 

The case of $G=\A_N$ is special since the Calabi--Yau manifold $\tilde{X}_3$ is a toric variety. In this case, we can use the powerful technique of toric geometry or a dual picture of 5-brane webs in type IIB string theory \cite{Aharony:1997ju, Aharony:1997bh, Leung:1997tw}. In this section, we will argue that the cases of $G = \D_{N+2}, \E_6, \E_7, \E_8$ in fact have a web--like description by making use of the geometric picture, although we are not sure whether there exists any kind of brane construction which physically realizes that web-like picture. 

\subsection{5d $\SO(2N+4)$ gauge theory}
\label{sec:5dSO2Np4}

Let us first consider the case of $G=\D_{N+2}, N=2, 3, \cdots$. The Calabi--Yau geometry $\tilde{X}_3$ has the compact surface $S_{\mathfrak{so}(2N+4)}$ which is a $F_{\mathfrak{so}(2N+4)}$ fibration over the base $C_B$. The non-Abelian $SO(2N+4)$ gauge symmetry is recovered at the origin of the Coulomb branch moduli space which corresponds to the limit where the spheres forming the $F_{\mathfrak{so}(2N+4)}$ fiber shrink simultaneously over the base $C_B$, recovering the $\D_{N+2}$ surface singularity over the base $C_B$. It is possible to further shrink the base $C_B$. Then the whole complex surface $S_{\mathfrak{so}(2N+4)}$ shrinks to zero size and the gauge coupling become infinitely strong.
This limit corresponds to the conformal limit where nonperturbative particles as well as perturbative particles become simultaneously massless, and therefore the 5d theory becomes a superconformal field theory. 

In order to obtain a dual gauge theory description we consider a different order of shrinking of the surface $S_{\mathfrak{so}(2N+4)}$. The fiber $F_{\mathfrak{g}}$ consists of $N+2$ spheres whose shape is the Dynkin diagram of type $\D_{N+2}$. Among the $N+2$ spheres, there is one special sphere $C_g$ which intersect with adjacent three spheres. We then consider $C_g$ as a base and shrink the other spheres including $C_B$.
Since $C_B$ is fibered over $C_g$, the geometry develops an $\A_1$ singularity wrapping $C_g$ after shrinking $C_B$.
Hence the theory has an $\SU(2)$ gauge symmetry. Furthermore, we have three singular points on $C_g$. Two of them originate from contracting a surface $S_{\mathfrak{su}(2)}$ which has a $F_{\mathfrak{su}(2)}$ fiber over $C_B$. The other singular point originates from contracting a surface $S_{\mathfrak{su}(N)}$ which has a $F_{\mathfrak{su}(N)}$ fiber over $C_B$. Since the singularities arise from shrinking the complex surfaces, each singular point yields a 5d SCFT and they are coupled by the $\SU(2)$ gauge symmetry associated to the $\A_1$ singularity over $C_g$. Hence each of the SCFTs should have an $\SU(2)$ flavor symmetry and the diagonal part of the three $\SU(2)$ flavor symmetries is gauged. Therefore, the dual description is realized by the $\SU(2)$ gauging of the {\it three} 5d SCFTs. We call the gauging {\it trivalent gauging}. 

Let us then see the three superconformal field theories in detail. Two of them come from shrinking the complex surface $S_{\mathfrak{su}(2)}$. Hence, the 5d theory is a pure $\SU(2)$ gauge theory with its mass parameter turned on. 
The pure $\SU(2)$ gauge theory should have an $\SU(2)$ flavor symmetry in UV which can be used for the $\SU(2)$ trivalent gauging. Hence, the discrete theta angle for the pure $\SU(2)$ gauge theory should be zero. The other SCFT comes from shrinking the complex surface $S_{\mathfrak{su}(N)}$. Therefore the 5d theory is a pure $\SU(N)$ gauge theory. 
Since the pure $\SU(N)$ gauge theory should have an $\SU(2)$ flavor symmetry in UV again for the $\SU(2)$ trivalent gauging, the Chern-Simons (CS) level should be $\pm N$ \cite{Hayashi:2015fsa}.
For each case of the pure $\SU(2)$ gauge theory and the pure $\SU(N)$ gauge theory,  the $\SU(2)$ flavor symmetry arises non--perturbatively in UV.
To deal with the flavor symmetry we should directly consider the UV superconformal field theory of the pure $\SU(2)$ gauge theory and the pure $\SU(N)_{\pm N}$ gauge theory\footnotemark,
\footnotetext{$\SU(N)_{\kappa}$ implies that an $\SU(N)$ gauge theory with the CS level $\kappa$. }
which we denote by $\hat{D}_2(\SU(2))$ and $\hat{D}_N(\SU(N))$ respectively.
Here the notation $\hat{D}_p(\SU(2))$ \footnotemark\footnotetext{The notation of $\hat{D}_p(\SU(2))$ has been introduced in \cite{DelZotto:2015rca} as a 5d uplift of the 4d $D_p(\SU(2))$ theory \cite{Cecotti:2012jx, Cecotti:2013lda} which is equivalent to the 4d $(A_1, D_p)$ Argyres--Douglas theory.} implies a SCFT which arises from M-theory on an orbifold $\mathbb{C}^3/\Gamma$ where the orbifold action of $\Gamma$ is given by
\begin{equation}
g = (\omega^2, \omega^{-1}, \omega^{-1}) \label{5dorbifold}
\end{equation}
with $\omega^{2p} = 1$ and $p=2, 3, \cdots$. The three components act on the three complex coordinates of $\mathbb{C}^3$. Note that the orbifold action 
\begin{equation}
g^p =  (\omega^{2p}, \omega^{-p}, \omega^{-p}) = (1, -1, -1)
\end{equation}
yields an $A_1$ singularity, leading to an $\SU(2)$ flavor symmetry. The $\hat{D}_p(\SU(2))$ theory is then a rank $(p-1)$ 
SCFT with an $\SU(2)$  flavor symmetry. Therefore, it has $p-1$ Coulomb branch moduli and one mass parameter. In particular, $\hat{D}_2(\SU(2))$ theory is the yields the same SCFT as $E_1$ theory in \cite{Seiberg:1996bd}.

\begin{figure}[t]
\centering
\includegraphics[width=8cm]{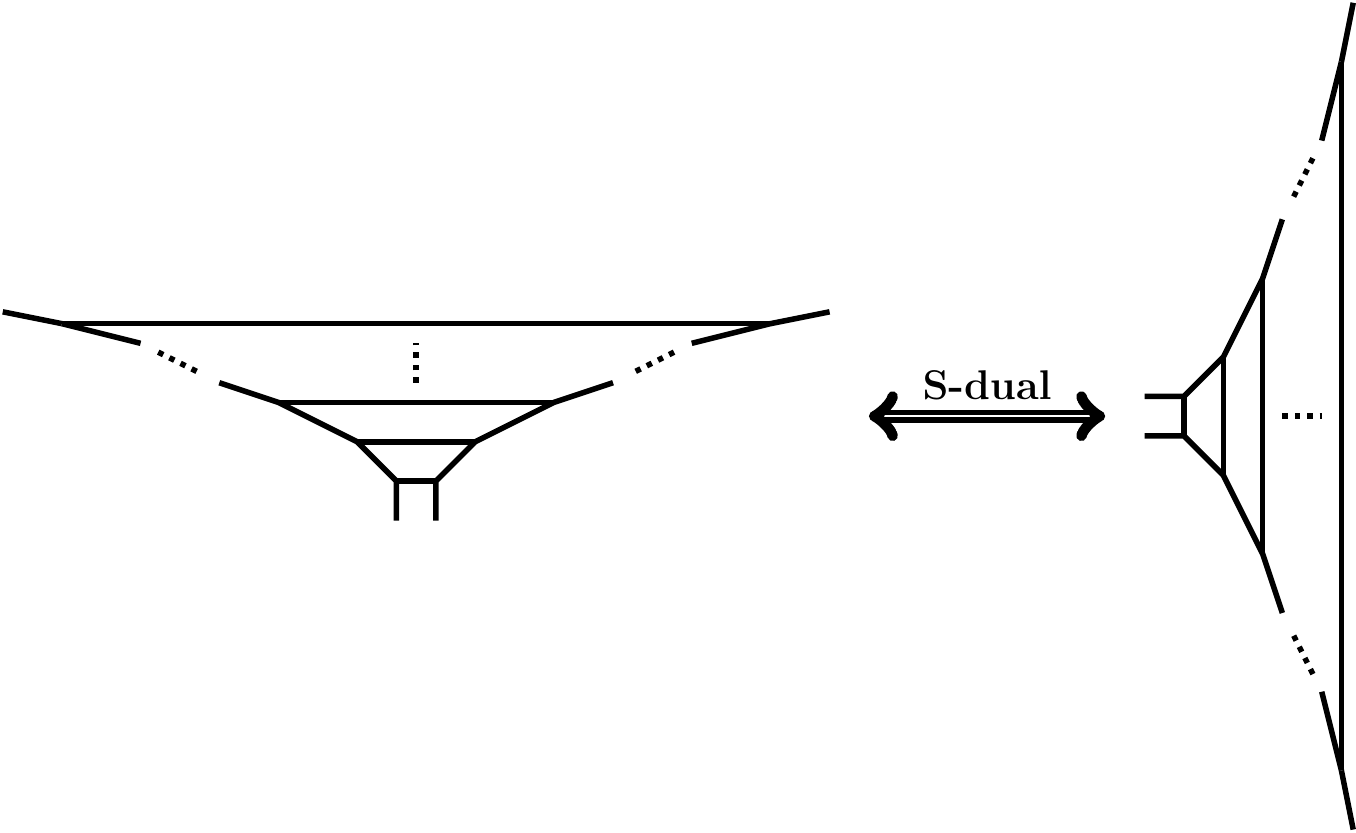}
\caption{Left: The 5-brane web for the pure $\SU(p)$ gauge theory with the $\pm p$ CS level. We have $p$ D5-branes which lie in the horizontal direction. The parallel two external NS5-branes imply the non-perturbative $\SU(2)$ flavor symmetry. Right: The S-dual configuration to the 5-brane web on the left. Namely the 5-brane web for the $\hat{D}_{p}(\SU(2))$ theory. }
\label{fig:DpSU2}
\end{figure}

It is illustrative to describe the $\hat{D}_p(\SU(2))$ theory by a 5-brane web. A 5-brane web is a dual configuration of a certain Calabi--Yau threefold $\tilde{X}_3$ \cite{Leung:1997tw}. The directions which the 5-brane extend are summarized in table \ref{tb:5brane}.
\begin{table}[t]
\begin{center}
\begin{tabular}{c|c c c c c | c c | c c c}
 & 0 & 1 & 2 & 3 & 4 & 5 & 6 & 7 & 8 & 9\\
 \hline
 D5-brane& $\times$ & $\times$ & $\times$ & $\times$ & $\times$ & $\times$ & &&& \\
 NS5-brane & $\times$ & $\times$ & $\times$ & $\times$ & $\times$ &  & $\times$ &&& \\
 $(p, q)$ 5-brane & $\times$ & $\times$ & $\times$ & $\times$ & $\times$ & \multicolumn{2}{|c|}{\text{angle}}&&& \\
7-brane & $\times$ & $\times$ & $\times$ & $\times$ & $\times$ &  &  & $\times$ &$\times$  & $\times$
 \end{tabular}
 \caption{The configuration of 5-branes in the ten-dimensional spacetime of type IIB string theory. The slope of the $(p, q)$ 5-brane is $\frac{q}{p}$ in the two-dimensional $(x_5, x_6)$-space. In particular, a horizontal line represents a horizontal line and a vertical line represents an NS5-brane. Since the structure of the 5-brane only appears in the $(x_5, x_6)$-plane, we only write the two-dimensional plane for depicting a 5-brane.}
\label{tb:5brane}
\end{center}
\end{table}
It is also useful to introduce 7-branes attached to the ends of external 5-branes in a 5-brane web configuration to read off the flavor symmetry of a 5d theory realized on a 5-brane web \cite{DeWolfe:1999hj}. The 5-brane web for the pure $\SU(p)_{\pm p}$ gauge theory is given in figure \ref{fig:DpSU2}.
To understand the $\SU(2)$ flavor symmetry "perturbatively", it might help to take a S-dual of the web, which is also depicted in figure \ref{fig:DpSU2}.
The S-duality of the 5d theory is simply given by the $\frac{\pi}{2}$ rotation of the web in the $(x_5, x_6)$-plane.
Note that in the S-dual picture, the flavor symmetry of the $\hat{D}_p(\SU(2))$ theory is realized perturbatively as background gauge field on two D7-branes attached to the ends of the external 5-branes extending in the right direction.
However, we do not have internal D5-branes and the theory does not admit a Lagrangian description.
On the other hand the $\SU(2)$ flavor symmetry appears non-perturbatively in the pure $\SU(p)$ gauge theory since it is associated to a symmetry on the two $(0, 1)$ 7-branes or the two NS5-branes.

In summary, when we regard $C_g$ as the base manifold, the geometry gives rise to the following 5d theory
\begin{align}
\hat{D}_N(\SU(2)) - {\overset{\overset{\text{\large$\hat{D}_2(\SU(2))$}}{\textstyle\vert}}{\SU(2)}} - \hat{D}_2(\SU(2)) \label{SO2Np4}
\end{align}
The $\SU(2)$ in the center of \eqref{SO2Np4} implies the $\SU(2)$ trivalent gauging which couple the two $\hat{D}_2(\SU(2))$ theories and the $\hat{D}_N(\SU(2))$ theory by the diagonal gauging of their $\SU(2)$ flavor symmetries.
We argue that this is a dual description of the pure $\SO(2N+4)$ gauge theory.
One can check that the number of the moduli and the parameters of one theory match with those of the other theory.
The pure $\SO(2N+4)$ gauge theory has $N+2$ Coulomb branch moduli and one mass parameter corresponding to the gauge coupling.
The dual theory \eqref{SO2Np4} has $(N-1) + 1 + 1 + 1 = N+2$ Coulomb branch moduli and one mass parameter from the gauging coupling of the $\SU(2)$ trivalent gauging in \eqref{SO2Np4}.
This duality between $\SO(2N+2)$ gauge theory and the $\SU(2)$ gauge theory \eqref{SO2Np4} with non-Lagrangian matter is a generalization of base-fiber dualities between 5d $\SU(N)$ linear quiver gauge theories \cite{Aharony:1997bh, Bao:2011rc} as well as 4d theories \cite{Katz:1996fh}.

\begin{figure}[t]
\centering
\includegraphics[width=5cm]{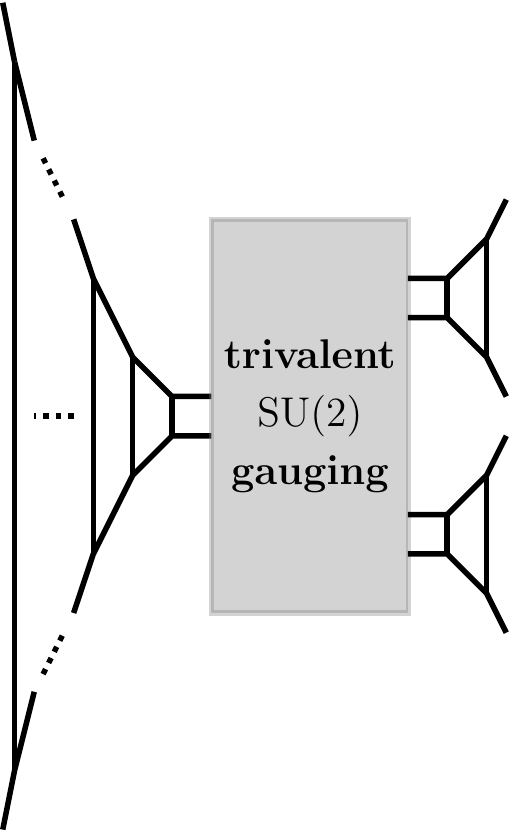}
\caption{A 5-brane web--like description of the theory \eqref{SO2Np4} which is dual to the pure $\SO(2N+4)$ gauge theory. The prescription for the ``trivalent $\SU(2)$ gauging" is going to be given in the next section. Three webs actually does not live in the same plane, and thus do not cross each other in the cases we will deal with in this paper.}
\label{fig:webforSO2Np4}
\end{figure}

It is also possible to write a web-like picture for the dual theory \eqref{SO2Np4}. Noting that the $\hat{D}_p(\SU(2))$ theory is given by the web in the right figure of figure \ref{fig:DpSU2}, we can write a web--like picture for the theory \eqref{SO2Np4} as in figure \ref{fig:webforSO2Np4}.
Due to the trivalent gauging, it is not possible to write the diagram in figure \ref{fig:webforSO2Np4} as a proper 5-brane web on a plane.
Verifying that this picture somewhat makes sense is the main purpose of this paper.
In particular, what to do with the ``trivalent $\SU(2)$ gauging" in the picture is going to be given in the next section.
Note that the lengths between the parallel horizontal legs for the three 5-brane webs are the size of $C_B$ and hence they should be equal to each other. We need to impose this condition for the partition function computation in the later sections. In the dual picture, the size of $C_B$ becomes the Coulomb branch modulus of the $\SU(2)$ trivalent gauging. In terms of the web diagram, the trivalent gauging may be thought of as {\it trivalent gluing} of the three webs which give rise to the $\hat{D}_2(\SU(2)), \hat{D}_2(\SU(2))$ and $\hat{D}_N(\SU(2))$ theories. We will use the terminology of trivalent gauging and trivalent gluing interchangeably in this paper. 

\begin{figure}[t]
\centering
\includegraphics[width=12cm]{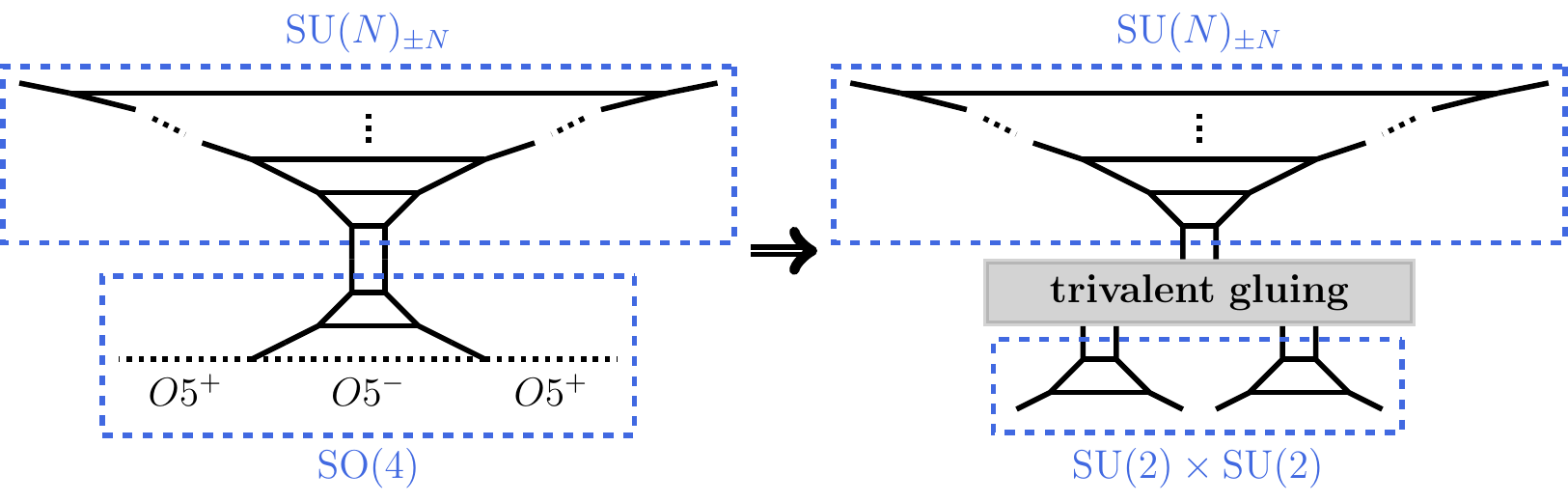}
\caption{A transition from a 5-brane web with an $O5$--plane to a web--like diagram with trivalent gluing. The left figure represents a 5-brane web of the pure $\SO(2N+4)$ gauge theory using an $O5$--plane. The right figure is a web--like description by replacing the 5-brane for the $\SO(4)$ gauge theory part in the left figure with the two 5-branes webs of the pure $\SU(2)$ gauge theory with no discrete theta angle. Now the three 5-brane webs are connected by the trivalent gluing. }
\label{fig:SO4vsSU2sq}
\end{figure}

We can further support the dual description \eqref{SO2Np4} in another manner. The pure $\SO(2N+4)$ gauge theory can be also realized by a 5-brane web with an $O5$--plane as in the left figure in figure \ref{fig:SO4vsSU2sq}. 
The 5-brane web configuration can be thought of as connecting a pure $\SU(N)$ gauge theory with the CS level $\pm N$ with a pure $\SO(4)$ gauge theory by the two NS5-branes in the middle of the diagram. Since $\mathfrak{so}(4) \cong \mathfrak{su}(2) \times \mathfrak{su}(2)$, we may replace the 5-brane web for the $\SO(4)$ gauge theory with the two 5-brane webs for the pure $\SU(2)$ gauge theory as in figure \ref{fig:SO4vsSU2sq}.
Then the web--like figure on the right in figure \ref{fig:SO4vsSU2sq} may be considered as an S--dual configuration of the web in figure \ref{fig:webforSO2Np4}. 

\begin{figure}[t]
\centering
\includegraphics[width=10cm]{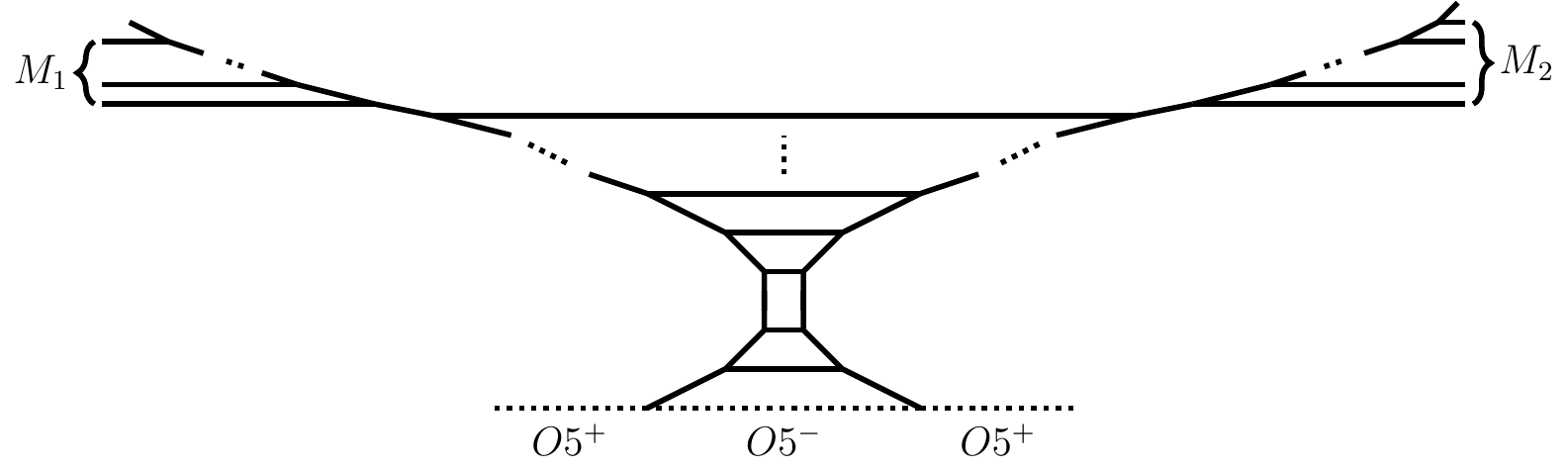}
\caption{A 5-brane web for the $\SO(2N+4)$ gauge theory with $M_1 +M_2$ hypermultiplets in the vector representation.}
\label{fig:SO2Np4wflvrs}
\end{figure}

\begin{figure}[t]
\centering
\includegraphics[width=10cm]{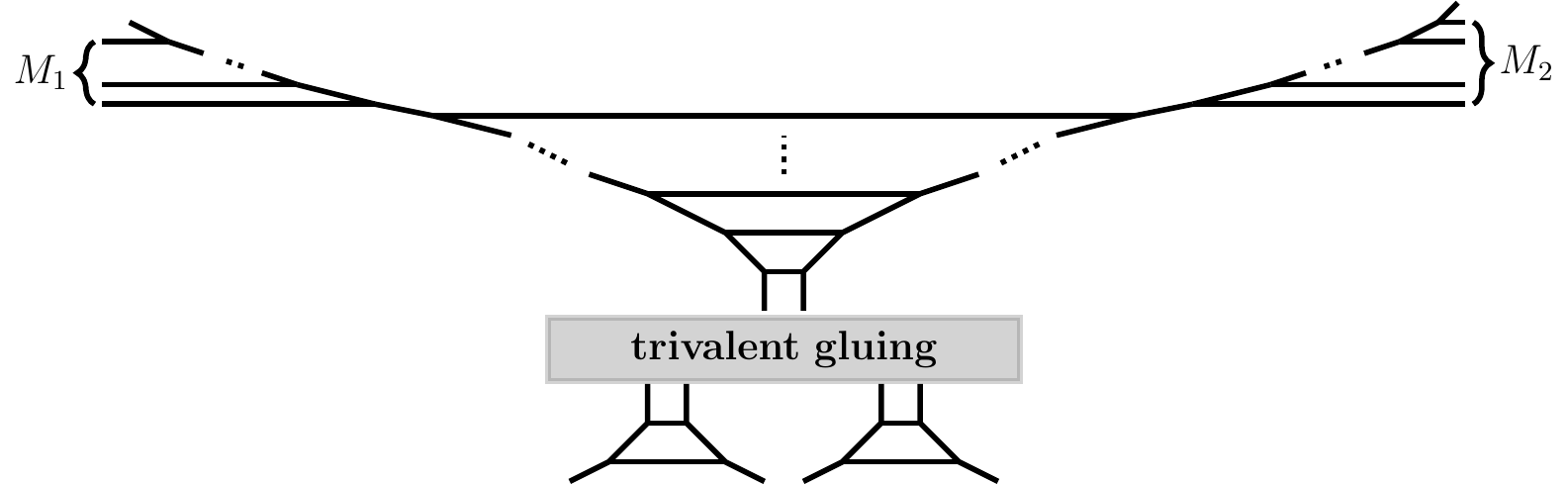}
\caption{A 5-brane web--like description for the $\SO(2N+4)$ gauge theory with $M_1 +M_2$ hypermultiplets in the vector representation by replacing the web for the $SO(4)$ part with the two webs for the pure $\SU(2)$. The three 5-brane webs are connected by the trivalent gluing. }
\label{fig:SO4vsSU2sqwflvrs}
\end{figure}

\begin{figure}[t]
\centering
\includegraphics[width=5cm]{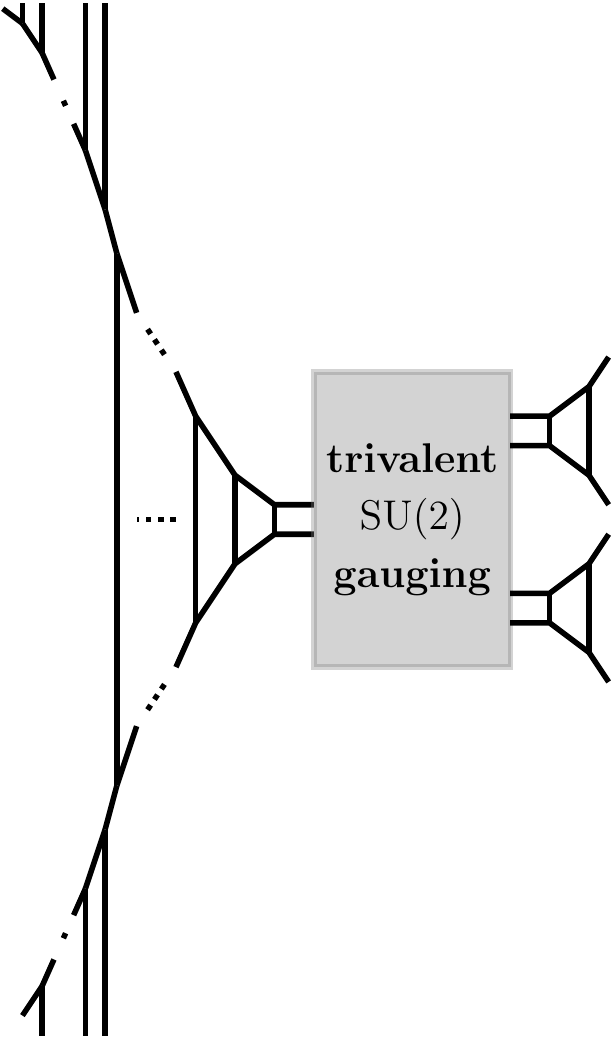}
\caption{A web--like diagram which is obtained by rotating web in figure \ref{fig:SO4vsSU2sqwflvrs} by $\frac{\pi}{2}$. }
\label{fig:webforSO2Np4wflvrs}
\end{figure}

This understanding also provides us with a way to introduce hypermultiplets in the vector representation of $\SO(2N+4)$. Starting from the 5-brane web of the pure $\SO(2N+4)$ gauge theory, $M_1 + M_2$ hypermultiplets in the vector representation can be added by introducing $M_1$ flavor 5-branes on the left and $M_2$ flavor 5-branes on the right as in figure \ref{fig:SO2Np4wflvrs}.
We here assume $M_1 \leq N+1$ and $M_2 \leq N+1$ and also $M_1 + M_2 \leq 2N+1$. In fact the $\SO(2N+4)$ gauge theory with $N_f$ hypermultiplets in the vector representation has a 5d UV completion when $N_f \leq 2N+1$ \cite{Bergman:2015dpa}\footnote{When $N_f = 2N+2$, the 5d $\SO(2N+4)$ theory has a 6d UV completion \cite{Hayashi:2015vhy}.}. In the case when the number of flavors saturates the bound $N_f = 2N+1$, the 5-brane web configuration is more involved than that in figure \ref{fig:SO2Np4wflvrs} but it is still possible to write down a 5-brane web by introducing a configuration of 5-branes jumping over other 5-branes  \cite{Bergman:2015dpa}. With the 5-brane web picture in figure \ref{fig:SO2Np4wflvrs}, one can again apply the replacement of the web of the $\SO(4)$ gauge theory with the two webs of the pure $\SU(2)$ gauge theory as in figure \ref{fig:SO4vsSU2sqwflvrs}. 
A dual picture may be obtained by simply rotating the web in figure \ref{fig:SO4vsSU2sqwflvrs} by $\frac{\pi}{2}$ as in figure \ref{fig:webforSO2Np4wflvrs}. 
By denoting the web on the left part in figure \ref{fig:webforSO2Np4wflvrs} by $\hat{D}_N^{M_1,M_2}(\SU(2))$, a 5d theory which is dual to the 5d $\SO(2N+4)$ gauge theory with $M_1+M_2$ hypermultiplets in the vector representation is given by
\begin{align}
\hat{D}_N^{M_1,M_2}(\SU(2)) - {\overset{\overset{\text{\large$\hat{D}_2(\SU(2))$}}{\textstyle\vert}}{\SU(2)}} - \hat{D}_2(\SU(2)) \label{SO2Np4wflvrs}
\end{align}
Here $\hat{D}_N^{M_1,M_2}(\SU(2))$ is the 5d rank $(N-1)$ SCFT with an $\SU(2) \times \SU(M_1+M_2) \times \U(1)$ flavor symmetry. When $M_1 = N$ and $M_2 = N$, the flavor symmetry is further enhanced to $\SU(2) \times \SU(M_1+M_2) \times \SU(2)$. 

\subsection{5d pure $\E_6, \E_7, \E_8$ gauge theories}

It is straightforward to apply the idea in the previous subsection to the cases of $G = \E_6, \E_7$ and $\E_8$. For each case, there is again one sphere $C_g$ in the fiber $F_{\mathfrak{g}}$ which intersects with three adjacent spheres. We may consider $C_g$ as a base and shrink the other spheres including $C_B$. Then the shrinking of $C_B$ yields again an $A_1$ singularity over $C_g$, leading to an $SU(2)$ gauge symmetry. $C_g$ has three singular points and each point gives rise to a certain 5d SCFT, depending on $G=\E_6, \E_7$ or $\E_8$.

When $G=\E_6$, one singular point arises by contracting a surface $S_{\mathfrak{su}(2)}$ whereas the other two singularities originate from shrinking a surface $S_{\mathfrak{su}(3)}$. Repeating the same argument in section \ref{sec:5dSO2Np4}, the former yields the $\hat{D}_{2}(\SU(2))$ theory and the latter gives rise to the $\hat{D}_3(\SU(2))$ theory. Therefore, a dual description of the pure $\E_6$ gauge theory is given by the trivalent gauging of the $\hat{D}_{2}(\SU(2))$ theory and the two $\hat{D}_3(\SU(2))$ theories, namely
\begin{align}
\hat{D}_3(\SU(2)) - {\overset{\overset{\text{\large$\hat{D}_2(\SU(2))$}}{\textstyle\vert}}{\SU(2)}} - \hat{D}_3(\SU(2))\,\,\,. \label{E6}
\end{align}
The theory \eqref{E6} has $2 + 1 + 1 + 2 = 6$ Coulomb branch moduli and one mass parameter from the gauge coupling of the $\SU(2)$ trivalent gauging. These numbers agrees with the numbers of the Coulomb branch moduli and the mass parameter of the pure $\E_6$ gauge theory.

When $G = \E_7$, the three singular points yield different 5d SCFTs, and they are the $\hat{D}_2(\SU(2))$ theory, the $\hat{D_3}(\SU(2))$ theory and the $\hat{D}_4(\SU(2))$ theory. Hence a dual description of the pure $\E_7$ gauge theory is 
\begin{align}
\hat{D}_4(\SU(2)) - {\overset{\overset{\text{\large$\hat{D}_2(\SU(2))$}}{\textstyle\vert}}{\SU(2)}} - \hat{D}_3(\SU(2)) \label{E7}
\end{align}
The dual theory \eqref{E7} has $3 + 1 + 1 + 2 = 7$ Coulomb branch moduli and one mass parameter from the gauge coupling of the $\SU(2)$ gauging. The numbers again agree with the numbers of the Coulomb branch moduli and the mass parameter of the pure $E_7$ theory. 

Finally when $G = \E_8$, three singular points give rise to the $\hat{D}_2(\SU(2))$ theory, $\hat{D}_3(\SU(2))$ and the $\hat{D}_5(\SU(2))$ theory. Then a dual picture of the pure $\E_8$ gauge theory is 
\begin{align}
\hat{D}_5(\SU(2)) - {\overset{\overset{\text{\large$\hat{D}_2(\SU(2))$}}{\textstyle\vert}}{\SU(2)}} - \hat{D}_3(\SU(2)) \label{E8}
\end{align}
The number of the Coulomb branch moduli is $4 + 1 + 1 + 2 = 8$ and it has one mass parameter. The numbers completely agrees with the eight Coulomb branch moduli and the one mass parameter of the pure $\E_8$ gauge theory.

\bigskip

\section{Gluing rule and 5d $\SO(2N+4)$ gauge theory}
\label{sec:partSO2Np4}
Having identified the dual gauge theory descriptions \eqref{SO2Np4}--\eqref{E8} for the gauge theories with a gauge group $G = \SO(2N+4), \E_6, \E_7, \E_8$ in section \ref{sec:newweb}, we will make use of the picture to compute their Nekrasov partition functions.
The main tool is the topological vertex formalism \cite{Iqbal:2002we, Aganagic:2003db, Awata:2005fa, Iqbal:2007ii}, whose basic formulae are summarized in appendix \ref{sec:topvertex}.
When a 5d theory is realized on a 5-brane web, the application of the topological vertex to the 5-brane web gives rise to its Nekrasov partition function 
\cite{Iqbal:2003ix, Iqbal:2003zz, Eguchi:2003sj, Hollowood:2003cv, Taki:2007dh}.
However, it is not possible to simply apply the topological vertex to the web-like descriptions of the theories \eqref{SO2Np4}--
\eqref{E8} due to the existence of the trivalent gauging of three 5d theories.
In this section we propose a new technique which enables us to apply the topological vertex formalism to the trivalent gauging of three 5d theories. 
The result will come in the form of double expansion of instanton fugacity and the Coulomb branch parameter corresponding to the trivalent node of the Dynkin diagram of the gauge group, and that is compared with result from the localization computations up to some orders of those two expanding parameters.

In this section, we focus on unrefined partition functions, and postpone the refined cases to section \ref{sec:refined}. 

\subsection{Trivalent gluing}
\label{sec:gluing}
In the previous section, the web-like descriptions for the gauge theory with $G=\SO(2N+4),\E_6, \E_7, \E_8$ came from the duality frame which involves the $\SU(2)$ gauging of the diagonal part of $\SU(2)$ flavor symmetries of three SCFTs. Although each SCFT is a UV SCFT of a gauge theory, the gauged $\SU(2)$ symmetry emerges non-perturbatively at UV, so we cannot have a Lagrangian description of the duality frame, and thus we need to develop a new way to compute the partition function of such a theory.

The central idea is regarding those SCFTs as ``$\SU(2)$ matter", although they do not have a Lagrangian description where the $\SU(2)$ symmetry is manifest.
Recall that the Nekrasov partition function \cite{Nekrasov:2002qd,Nekrasov:2003rj} for an $\SU(2)$ gauge theory with hypermultiplets looks like
\begin{equation}
\sum_{\lambda,\mu} Q_g^{|\lambda|+|\mu|}\, Z^{\text{hyper}}_{\lambda,\mu}(Q_B,Q_m) Z^{\text{$\SU(2)$ vector}}_{\lambda,\mu}(Q_B),
\label{eq:LagNek}
\end{equation}
where $\lambda,\mu$ are Young diagrams, $Q_g$, $Q_m$, $Q_B$ are associated to the instanton fugacity, a mass parameter and Coulomb branch parameter, respectively\footnote{More precisely, $Q$ implies $Q=e^{-k}$ where $k$ is a modulus or a parameter of a 5d theory. We will call $Q$ also for moduli and parameters of a 5d theory. }. $Z^{\text{hyper}}_{\lambda,\mu}(Q_B,Q_m)$ is the contribution from the hypermultiplets, and $Z^{\text{$\SU(2)$ vector}}_{\lambda,\mu}(Q_B)$ is that from the $\SU(2)$ vector multiplets.
What we need now is a generalization of $Z^{\text{hyper}}_{\lambda,\mu}$ to the partition function of a general SCFT with an $\SU(2)$ flavor symmetry.

The pair of Young diagrams $(\lambda,\mu)$ labels the fixed points of the $\U(1)$ action in the $\U(2)$ instanton moduli space. 
Then, $Z^\text{hyper}_{\lambda,\mu}$ is the partition function of hypermultiplets with $\SU(2)$ background with the nontrivial instanton configuration labeled by $(\lambda,\mu)$.
Therefore, this concept is manifestly generalized into a general SCFT $\mathcal{T}$, and we denote the partition function with the flavor instanton background and flavor fugacity $Q_B$ by $Z^\mathcal{T}_{\lambda,\mu}(Q_B)$.
Then, the partition function of the trivalent $\SU(2)$ gauging of $\mathcal{T}_1$, $ \mathcal{T}_2$ and $ \mathcal{T}_3$ can be obtained by
\begin{equation}
\sum_{\lambda,\mu} Q_{g}^{|\lambda|+|\mu|}\, Z^{\mathcal{T}_1}_{\lambda,\mu}(Q_B) Z^{\mathcal{T}_2}_{\lambda,\mu}(Q_B) Z^{\mathcal{T}_3}_{\lambda,\mu}(Q_B) Z^{\text{$\SU(2)$ vector}}_{\lambda,\mu}(Q_B).
\label{eq:trivgauge}
\end{equation}
This is similar to the gauging formula for 4d index \cite{Romelsberger:2005eg,Kinney:2005ej,Gadde:2011uv},
One might worry about the validity of this formula, since the formula \eqref{eq:LagNek} comes from the $\U(N)$ instanton, 
and therefore it is not clear that the formula can be generalized into gauging of SCFTs with only $\SU(2)$ flavor.
Here we just go ahead, and it will turn out this prescription almost works.
However, we occasionally need to subtract "extra factors" similar to what is discussed in subsection \ref{sec:topvertex} when the theory have flavor symmetries as we will see in subsection \ref{sec:SO2Np4wflvrs}.

The next task is understanding how to compute such a partition function $Z^\text{SCFT}_{\lambda,\mu}$ with a nontrivial flavor background.
Note again that in our case the flavor emerges non-perturbatively, and therefore methods relying on Lagrangian descriptions cannot be utilized.
This is where the topological vertex helps.
To be inspired, let us rewrite \eqref{eq:LagNek} using the topological vertex.
The web diagram representing an $\SU(2)$ gauge theory with one fundamental hypermultiplet can be depicted as 
\begin{equation}
	\begin{tikzpicture}[scale=.6,thick,baseline = -23]
		\draw (0,0)--++(1,0) coordinate (a) --++(0,1);
		\draw (a)--++(.7,-.7) coordinate (b) --++ (0,-1) coordinate (c) --++(-.7,-.7);
		\draw (b) --++(1,0) coordinate (m1) --++(1,0) coordinate (d) --++(.7,.7);
		\draw (d) --++(0,-1) coordinate (e) --++(-1,0) coordinate (m2) --(c);
		\draw (e) --++(.7,-.7);
		\draw[<->] (b) ++(-0.5,0) -- node[midway,anchor=east] {$Q_B$}  ++(0,-1);
	\draw[<->] (b) ++(0,0.5) -- node[midway,anchor=south] {$Q_g$}  ++(2,0);
	\draw[<->] (b) ++(-2,0) -- node[midway,anchor=east]{$Q_m$} ++(0,.7);
	\end{tikzpicture}
	= \sum_{\lambda,\mu} Q_g^{|\lambda|+|\mu|} f_{\lambda,\mu}\times 
	\begin{tikzpicture}[scale=.6,thick, baseline = -23]
		\draw (0,0)--++(1,0) coordinate (a) --++(0,1);
		\draw (a)--++(.7,-.7) coordinate (b) --++ (0,-1) coordinate (c) --++(-.7,-.7);
	\draw (b) --++(1,0) coordinate (m1);
	\draw (c)--++(1,0) coordinate (m2) ;
	\draw[<->] (b) ++(-0.5,0) -- node[midway,anchor=east] {$Q_B$}  ++(0,-1);
	\draw[<->] (a) ++(-.8,0) -- node[midway,anchor=east]{$Q_m$} ++(0,-.7);
	\draw[->,middle segment=.2cm] ($(m1)+(0,.2)$) -- ($(b)+(0,.2)$) node[midway,anchor=south]{$\mu$};
	\draw[->,middle segment=.2cm] ($(m2)+(0,-.2)$) -- ($(c)+(0,-.2)$) node[midway,anchor=north]{$\lambda$};
	\end{tikzpicture}
	\,\,
	\times
	\,\,
	\begin{tikzpicture}[scale=.6,thick,baseline=-23,xscale=-1]
		\draw (a)--++(.7,-.7) coordinate (b) --++ (0,-1) coordinate (c) --++(-.7,-.7);
	\draw (b) --++(1,0) coordinate (m1);
	\draw (c)--++(1,0) coordinate (m2) ;
	\draw[->,middle segment=.2cm] ($(b)+(0,.2)$) -- ($(m1)+(0,.2)$) node[midway,anchor=south]{$\mu$};
	\draw[->,middle segment=.2cm] ($(c)+(0,-.2)$) -- ($(m2)+(0,-.2)$) node[midway,anchor=north]{$\lambda$};
	\draw[<->] (b) ++(-0.5,0) -- node[midway,anchor=west] {$Q_B$}  ++(0,-1);
	\end{tikzpicture},
	\label{eq:hypervertex}
\end{equation}
where $f_{\lambda,\mu}$ is the framing factor:
\begin{equation}
	f_{\lambda,\mu}(q)= f_{\lambda^t}^{-1}(q)f_{\mu^t}(q)
\end{equation}
where $f_{\nu}(q)$ is that of \eqref{eq:framing} with unrefined limit $t=q$.
The right hand side of the equation means the summation over a pair of Young diagrams $(\lambda,\mu)$ assigned to the indicated internal edges.
This summation over $\lambda,\mu$ can be directly identified with that in \eqref{eq:LagNek} \cite{Iqbal:2003ix, Iqbal:2003zz, Eguchi:2003sj, Hollowood:2003cv, Taki:2007dh}\footnotemark.
\footnotetext{This is also true for refined case, if one is careful about the preferred direction. See Section \ref{sec:refined}.}
Decoupling the hypermultiplet, the partition function reduces to the that of the pure $\SU(2)$ gauge theory and it is given by
\begin{equation}
\sum_{\lambda,\mu} Q_g^{|\lambda|+|\mu|}\, Z^{\text{$\SU(2)$ vector}}_{\lambda,\mu}(Q_B)=
	\begin{tikzpicture}[scale=.6,thick,baseline = -23]
		\draw (a)--++(.7,-.7) coordinate (b) --++ (0,-1) coordinate (c) --++(-.7,-.7);
		\draw (b) --++(1,0) coordinate (m1) --++(1,0) coordinate (d) --++(.7,.7);
		\draw (d) --++(0,-1) coordinate (e) --++(-1,0) coordinate (m2) --(c);
		\draw (e) --++(.7,-.7);
		\draw[<->] (b) ++(-0.5,0) -- node[midway,anchor=east] {$Q_B$}  ++(0,-1);
	\draw[<->] (b) ++(0,0.5) -- node[midway,anchor=south] {$Q_g$}  ++(2,0);
	\end{tikzpicture}
	=\displaystyle \sum_{\lambda,\mu} Q_g^{|\lambda|+|\mu|} f_{\lambda,\mu}\times\,
	\begin{tikzpicture}[scale=.6,thick, baseline = -23]
		\draw (a)--++(.7,-.7) coordinate (b) --++ (0,-1) coordinate (c) --++(-.7,-.7);
	\draw (b) --++(1,0) coordinate (m1);
	\draw (c)--++(1,0) coordinate (m2) ;
	\draw[<->] (b) ++(-0.5,0) -- node[midway,anchor=east] {$Q_B$}  ++(0,-1);
	\draw[->,middle segment=.2cm] ($(m1)+(0,.2)$) -- ($(b)+(0,.2)$) node[midway,anchor=south]{$\mu$};
	\draw[->,middle segment=.2cm] ($(m2)+(0,-.2)$) -- ($(c)+(0,-.2)$) node[midway,anchor=north]{$\lambda$};
	\end{tikzpicture}
	\,\,
	\times
	\,\,
	\begin{tikzpicture}[scale=.6,thick,baseline=-23,xscale=-1]
		\draw (a)--++(.7,-.7) coordinate (b) --++ (0,-1) coordinate (c) --++(-.7,-.7);
	\draw (b) --++(1,0) coordinate (m1);
	\draw (c)--++(1,0) coordinate (m2) ;
	\draw[->,middle segment=.2cm] ($(b)+(0,.2)$) -- ($(m1)+(0,.2)$) node[midway,anchor=south]{$\mu$};
	\draw[->,middle segment=.2cm] ($(c)+(0,-.2)$) -- ($(m2)+(0,-.2)$) node[midway,anchor=north]{$\lambda$};
	\draw[<->] (b) ++(-0.5,0) -- node[midway,anchor=west] {$Q_B$}  ++(0,-1);
	\end{tikzpicture},
\end{equation}
obtaining  the equation
\begin{equation}
Z^{\text{$\SU(2)$ vector}}_{\lambda,\mu}(Q_B)=f_{\lambda,\mu} \times\,
	\begin{tikzpicture}[scale=.6,thick, baseline = -23]
		\draw (a)--++(.7,-.7) coordinate (b) --++ (0,-1) coordinate (c) --++(-.7,-.7);
	\draw (b) --++(1,0) coordinate (m1);
	\draw (c)--++(1,0) coordinate (m2) ;
	\draw[<->] (b) ++(-0.5,0) -- node[midway,anchor=east] {$Q_B$}  ++(0,-1);
	\draw[->,middle segment=.2cm] ($(m1)+(0,.2)$) -- ($(b)+(0,.2)$) node[midway,anchor=south]{$\mu$};
	\draw[->,middle segment=.2cm] ($(m2)+(0,-.2)$) -- ($(c)+(0,-.2)$) node[midway,anchor=north]{$\lambda$};
	\end{tikzpicture}
	\,\,
	\times
	\,\,
	\begin{tikzpicture}[scale=.6,thick,baseline=-23,xscale=-1]
		\draw (a)--++(.7,-.7) coordinate (b) --++ (0,-1) coordinate (c) --++(-.7,-.7);
	\draw (b) --++(1,0) coordinate (m1);
	\draw (c)--++(1,0) coordinate (m2) ;
	\draw[->,middle segment=.2cm] ($(b)+(0,.2)$) -- ($(m1)+(0,.2)$) node[midway,anchor=south]{$\mu$};
	\draw[->,middle segment=.2cm] ($(c)+(0,-.2)$) -- ($(m2)+(0,-.2)$) node[midway,anchor=north]{$\lambda$};
	\draw[<->] (b) ++(-0.5,0) -- node[midway,anchor=west] {$Q_B$}  ++(0,-1);
	\end{tikzpicture}
	.
	\label{eq:vectorvertex}
\end{equation}
Then, equating \eqref{eq:LagNek} and \eqref{eq:hypervertex} gives aa
\begin{equation}
	Z^\text{hyper}_{\lambda,\mu}(Q_B,Q_m)=
\left.
	\begin{tikzpicture}[scale=.6,thick, baseline = -23]
		\draw (0,0)--++(1,0) coordinate (a) --++(0,1);
		\draw (a)--++(.7,-.7) coordinate (b) --++ (0,-1) coordinate (c) --++(-.7,-.7);
	\draw (b) --++(1,0) coordinate (m1);
	\draw (c)--++(1,0) coordinate (m2) ;
	\draw[<->] (b) ++(-0.5,0) -- node[midway,anchor=east] {$Q_B$}  ++(0,-1);
	\draw[->,middle segment=.2cm] ($(m1)+(0,.2)$) -- ($(b)+(0,.2)$) node[midway,anchor=south]{$\mu$};
	\draw[->,middle segment=.2cm] ($(m2)+(0,-.2)$) -- ($(c)+(0,-.2)$) node[midway,anchor=north]{$\lambda$};
	\draw[<->] (a) ++(-.8,0) -- node[midway,anchor=east]{$Q_m$} ++(0,-.7);
	\end{tikzpicture}
\middle/ \,\,
	\begin{tikzpicture}[scale=.6,thick, baseline = -23]
		\draw (a)--++(.7,-.7) coordinate (b) --++ (0,-1) coordinate (c) --++(-.7,-.7);
	\draw (b) --++(1,0) coordinate (m1);
	\draw (c)--++(1,0) coordinate (m2) ;
	\draw[<->] (b) ++(-0.5,0) -- node[midway,anchor=east] {$Q_B$}  ++(0,-1);
	\draw[->,middle segment=.2cm] ($(m1)+(0,.2)$) -- ($(b)+(0,.2)$) node[midway,anchor=south]{$\mu$};
	\draw[->,middle segment=.2cm] ($(m2)+(0,-.2)$) -- ($(c)+(0,-.2)$) node[midway,anchor=north]{$\lambda$};
	\end{tikzpicture}
\right.
.
\label{eq:hypervertex2}
\end{equation}
This equation tells us that assigning nontrivial Young diagrams to parallel external edges representing the $\SU(2)$ flavor symmetry almost realizes the flavor background labeled by those Young diagrams,
but the division by the factor
\begin{equation}
\begin{split}
Z_{\lambda,\mu}^\text{Half}(Q_B)=
	\begin{tikzpicture}[scale=.6,thick, baseline = -23]
		\draw (a)--++(.7,-.7) coordinate (b) --++ (0,-1) coordinate (c) --++(-.7,-.7);
	\draw (b) --++(1,0) coordinate (m1);
	\draw (c)--++(1,0) coordinate (m2) ;
	\draw[<->] (b) ++(-0.5,0) -- node[midway,anchor=east] {$Q_B$}  ++(0,-1);
	\draw[->,middle segment=.2cm] ($(m1)+(0,.2)$) -- ($(b)+(0,.2)$) node[midway,anchor=south]{$\mu$};
	\draw[->,middle segment=.2cm] ($(m2)+(0,-.2)$) -- ($(c)+(0,-.2)$) node[midway,anchor=north]{$\lambda$};
	\end{tikzpicture}
	\,\,
	=
	\,\,
	\begin{tikzpicture}[scale=.6,thick,baseline=-23,xscale=-1]
		\draw (a)--++(.7,-.7) coordinate (b) --++ (0,-1) coordinate (c) --++(-.7,-.7);
	\draw (b) --++(1,0) coordinate (m1);
	\draw (c)--++(1,0) coordinate (m2) ;
	\draw[->,middle segment=.2cm] ($(b)+(0,.2)$) -- ($(m1)+(0,.2)$) node[midway,anchor=south]{$\mu$};
	\draw[->,middle segment=.2cm] ($(c)+(0,-.2)$) -- ($(m2)+(0,-.2)$) node[midway,anchor=north]{$\lambda$};
	\draw[<->] (b) ++(-0.5,0) -- node[midway,anchor=west] {$Q_B$}  ++(0,-1);
	\end{tikzpicture},
=\left(\frac{Z^\text{$\SU(2)$ vector}_{\lambda,\mu}(Q_B)}{f_{\lambda,\mu}}\right)^{1/2}
	\label{eq:halfvector}
\end{split}
\end{equation}
is needed. This factor is the square root of  of $Z^\text{$\SU(2)$ vector}_{\lambda,\mu}$, and thus we call this factor a contribution from a "half" vector.

Noe let us apply this division for determining the partition function of the $\hat{D}_2(\SU(2))$ matter. The web diagram is given in figure \ref{fig:DpSU2} with $p=2$. Since the theory couples to the $\SU(2)$ flavor instanton background, we assign Young diagrams to the parallel external legs. Then the consideration \eqref{eq:hypervertex2} motivate us to declare that the partition function for the $\hat{D}_2(\SU(2))$ is given by
\begin{equation}
	\begin{split}
		Z^{\hat{D}_2(\SU(2))}_{\lambda,\mu}(Q_B,Q)&=
		\hat{Z}^{\hat{D}_2(\SU(2))}_{\lambda,\mu}(Q_B,Q)/Z^\text{Half}_{\lambda,\mu}(Q_B)\\
		&=
	\left.
	\begin{tikzpicture}[scale=.6,thick, baseline = -23]
		\draw (a)--++(.7,-.7) coordinate (b) --++ (0,-1) coordinate (c) --++(-.7,-.7) coordinate (x);
		\draw (a) -- (x)--++(-.5,-1);
		\draw (a) -- ++(-.5,1);
		\draw[<->] (b) ++(1.2,0) -- node[midway,anchor=west] {$Q_B$}  ++(0,-1);
	\draw[<->] (a) ++(0,.3) -- node[midway,anchor = south] {$Q$} ++(.7,0);
	\draw[->,middle segment=.2cm] ($(m1)+(0,.2)$) -- ($(b)+(0,.2)$) node[midway,anchor=south]{$\mu$};
	\draw[->,middle segment=.2cm] ($(m2)+(0,-.2)$) -- ($(c)+(0,-.2)$) node[midway,anchor=north]{$\lambda$};
	\draw (b)-- ++(1,0);
	\draw (c)-- ++(1,0);
	\end{tikzpicture}
\middle/ \,\,
	\begin{tikzpicture}[scale=.6,thick, baseline = -23]
		\draw (a)--++(.7,-.7) coordinate (b) --++ (0,-1) coordinate (c) --++(-.7,-.7);
	\draw (b) --++(1,0) coordinate (m1);
	\draw (c)--++(1,0) coordinate (m2) ;
	\draw[<->] (b) ++(-0.5,0) -- node[midway,anchor=east] {$Q_B$}  ++(0,-1);
	\draw[->,middle segment=.2cm] ($(m1)+(0,.2)$) -- ($(b)+(0,.2)$) node[midway,anchor=south]{$\mu$};
	\draw[->,middle segment=.2cm] ($(m2)+(0,-.2)$) -- ($(c)+(0,-.2)$) node[midway,anchor=north]{$\lambda$};
	\end{tikzpicture}
\right. ,
\end{split}
\label{eq:ZD2vertex}
\end{equation}
where $\hat{Z}_{\lambda,\mu}^{\hat{D}_2(\SU(2))}(Q)$ is the quantity 
computed by the topological vertex with nontrivial Young diagrams $\lambda,\mu$ on the external edges, with Coulomb branch parameter $Q$.
When $\lambda=\mu=\emptyset$, the factor $Z_{\emptyset,\emptyset}^\text{Half}$, 
called extra factor appearing in the literature \cite{Bergman:2013ala, Hayashi:2013qwa, Bao:2013pwa, Bergman:2013aca}, which removes the constitutions coming from decoupled strings bridging the parallel 5-branes. \eqref{eq:ZD2vertex} is a natural generalization of that.
In general, if a SCFT $\mathcal{T}$ with an $\SU(2)$ flavor symmetry can be engineered by a web diagram which make the flavor symmetry manifest, we claim that then the partition function $Z^{\mathcal{T}}_{\lambda,\mu}$ with instanton flavor background can be computed by the topological vertex in the same matter, namely the ratio of the naive topological vertex computation $\hat{Z}^{\mathcal{T}}_{\lambda,\mu}$ and $Z^\text{Half}_{\lambda,\mu}$ \footnotemark.
\footnotetext{If the web of the SCFT $\mathcal{T}$ contains other manifest flavor symmetries, then the partition function should be further divided by extra factors corresponding to those symmetries.}
In particular, a generalization to the partition function of the $\hat{D}_p(\SU(2))$ matter is obvious.

Let us check that \eqref{eq:ZD2vertex} actually works.
For that, we consider a limit of Coulomb branch parameters of the pure $\SO(8)$ gauge theory which gives an $\SU(3)$ gauge theory.
In the dual frame \eqref{SO2Np4}, two of $\hat{D}_2(\SU(2))$ decouples in this limit, and thus we get a dual description
\begin{equation}
\hat{D}_2(\SU(2)) - \SU(2). \label{eq:dualSU3}
\end{equation}
of the $\SU(3)$ gauge theory.
From \eqref{eq:ZD2vertex}, the partition function of this dual description is 
\begin{equation}
	\begin{split}
&\sum_{\lambda,\mu} Q_g^{|\lambda|+|\mu|}  Z^{\hat{D}_2(\SU(2))}_{\lambda,\mu}(Q_B,Q) Z^{\text{$\SU(2)$ vector}}(Q_B)\\
&=\sum_{\lambda,\mu} Q_g^{|\lambda|+|\mu|} f_{\lambda,\mu}\times
\left(
	\begin{tikzpicture}[scale=.6,thick, baseline = -23]
		\draw (a)--++(.7,-.7) coordinate (b) --++ (0,-1) coordinate (c) --++(-.7,-.7) coordinate (x);
		\draw (a) -- (x)--++(-.5,-1);
		\draw (a) -- ++(-.5,1);
		\draw[<->] (b) ++(1.2,0) -- node[midway,anchor=west] {$Q_B$}  ++(0,-1);
	\draw[<->] (a) ++(0,.3) -- node[midway,anchor = south] {$Q$} ++(.7,0);
	\draw (b)-- ++(1,0) coordinate(m1);
	\draw (c)-- ++(1,0) coordinate(m2);
	\draw[->,middle segment=.2cm] ($(m1)+(0,.2)$) -- ($(b)+(0,.2)$) node[midway,anchor=south]{$\mu$};
	\draw[->,middle segment=.2cm] ($(m2)+(0,-.2)$) -- ($(c)+(0,-.2)$) node[midway,anchor=north]{$\lambda$};
	\end{tikzpicture}
\middle/ \,\,
	\begin{tikzpicture}[scale=.6,thick, baseline = -23]
		\draw (a)--++(.7,-.7) coordinate (b) --++ (0,-1) coordinate (c) --++(-.7,-.7);
	\draw (b) --++(1,0) coordinate (m1);
	\draw (c)--++(1,0) coordinate (m2) ;
	\draw[<->] (b) ++(-0.5,0) -- node[midway,anchor=east] {$Q_B$}  ++(0,-1);
	\draw[->,middle segment=.2cm] ($(m1)+(0,.2)$) -- ($(b)+(0,.2)$) node[midway,anchor=south]{$\mu$};
	\draw[->,middle segment=.2cm] ($(m2)+(0,-.2)$) -- ($(c)+(0,-.2)$) node[midway,anchor=north]{$\lambda$};
	\end{tikzpicture}
\right)
\left(
	\begin{tikzpicture}[scale=.6,thick, baseline = -23]
		\draw (a)--++(.7,-.7) coordinate (b) --++ (0,-1) coordinate (c) --++(-.7,-.7);
	\draw (b) --++(1,0) coordinate (m1);
	\draw (c)--++(1,0) coordinate (m2) ;
	\draw[<->] (b) ++(-0.5,0) -- node[midway,anchor=east] {$Q_B$}  ++(0,-1);
	\draw[->,middle segment=.2cm] ($(m1)+(0,.2)$) -- ($(b)+(0,.2)$) node[midway,anchor=south]{$\mu$};
	\draw[->,middle segment=.2cm] ($(m2)+(0,-.2)$) -- ($(c)+(0,-.2)$) node[midway,anchor=north]{$\lambda$};
	\end{tikzpicture}
	\,\,
	\times
	\,\,
	\begin{tikzpicture}[scale=.6,thick,baseline=-23,xscale=-1]
		\draw (a)--++(.7,-.7) coordinate (b) --++ (0,-1) coordinate (c) --++(-.7,-.7);
	\draw (b) --++(1,0) coordinate (m1);
	\draw (c)--++(1,0) coordinate (m2) ;
	\draw[->,middle segment=.2cm] ($(b)+(0,.2)$) -- ($(m1)+(0,.2)$) node[midway,anchor=south]{$\mu$};
	\draw[->,middle segment=.2cm] ($(c)+(0,-.2)$) -- ($(m2)+(0,-.2)$) node[midway,anchor=north]{$\lambda$};
	\draw[<->] (b) ++(-0.5,0) -- node[midway,anchor=west] {$Q_B$}  ++(0,-1);
	\end{tikzpicture}
\right)
\\
&=
	\begin{tikzpicture}[scale=.6,thick, baseline = -23]
		\draw (a)--++(.7,-.7) coordinate (b) --++ (0,-1) coordinate (c) --++(-.7,-.7) coordinate (x);
		\draw (a) -- (x)--++(-.5,-1);
		\draw (a) -- ++(-.5,1);
		\draw (b) -- ++(2,0) coordinate (d);
		\draw (d) --++(0,-1) coordinate (e) --(c);
		\draw (e) --++(.7,-.7);
		\draw (d) --++(.7,.7);
		\draw[<->] (d) ++(0.3,0) -- node[midway,anchor=west] {$Q_B$}  ++(0,-1);
		\draw[<->] (b) ++(0,0.5) -- node[midway,anchor=south] {$Q_g$}  ++(2,0);
	\draw[<->] (a) ++(0,.3) -- node[midway,anchor = south] {$Q$} ++(.7,0);
	\end{tikzpicture}
	\end{split}.
\end{equation}
The resulting web diagram is in fact nothing but the S--dual web for the pure  $\SU(3)_{\pm 1}$ gauge theory.
Note that $Q,Q_g$ corresponds to the two Coulomb branch parameters of $\SU(3)$, and $Q_B$ is the related to the gauge coupling of $\SU(3)$.
Therefore the parameters $Q_g, Q_B$ exchanges their roles under the duality between the $\SU(3)_1$ description and \eqref{eq:dualSU3}.

Now we can write down a prescription for partition functions for gauge theories dealt with in the previous section.
For simplicity, here we explicitly state the pure $\SO(8)$ case.
Let us denote the Coulomb branch parameters corresponding to edge nodes by $Q_1,Q_{-1},Q_{-2}$, that corresponding to the center node by $Q_g$, and the 
parameter associated to the instanton counting by $Q_B$.
From \eqref{SO2Np4} and \eqref{eq:ZD2vertex}, the partition function is
\begin{equation}
	\begin{split}
	Z_{\SO(8)}
	&=
\sum_{\lambda,\mu}  Q_g^{|\lambda|+|\mu|}Z^{\hat{D}_2(\SU(2))}_{\lambda,\mu}(Q_B,Q_1) Z^{\hat{D}_2(\SU(2))}_{\lambda,\mu}(Q_B,Q_{-1}) Z^{\hat{D}_2(\SU(2))}_{\lambda,\mu}(Q_B,Q_{-2}) Z^\text{$\SU(2)$ vector}_{\lambda,\mu}(Q_B)\\
	&=
\sum_{\lambda,\mu} Q_g^{|\lambda|+|\mu|} \hat{Z}^{\hat{D}_2(\SU(2))}_{\lambda,\mu}(Q_B,Q_1) \hat{Z}^{\hat{D}_2(\SU(2))}_{\lambda,\mu}(Q_B,Q_{-1}) \hat{Z}^{\hat{D}_2(\SU(2))}_{\lambda,\mu}(Q_B,Q_{-2}) \frac{f_{\lambda,\mu}Z^\text{Half}_{\lambda,\mu}(Q_B)^2}{Z^\text{Half}_{\lambda,\mu}(Q_B)^3}\\
	&=
\sum_{\lambda,\mu}  Q_g^{|\lambda|+|\mu|}\hat{Z}^{\hat{D}_2(\SU(2))}_{\lambda,\mu}(Q_B,Q_1) \hat{Z}^{\hat{D}_2(\SU(2))}_{\lambda,\mu}(Q_B,Q_{-1}) \hat{Z}^{\hat{D}_2(\SU(2))}_{\lambda,\mu}(Q_B,Q_{-2}) \frac{f_{\lambda,\mu}}{Z^\text{Half}_{\lambda,\mu}(Q_B)}\\
	&=\sum_{\lambda,\mu} Q_g^{|\lambda|+|\mu|}f_{\lambda,\mu}\times
	\begin{tikzpicture}[scale=.6,thick, baseline = -23]
		\draw (a)--++(.7,-.7) coordinate (b) --++ (0,-1) coordinate (c) --++(-.7,-.7) coordinate (x);
		\draw (a) -- (x)--++(-.5,-1);
		\draw (a) -- ++(-.5,1);
		\draw[<->] (b) ++(1.2,0) -- node[midway,anchor=west] {$Q_B$}  ++(0,-1);
	\draw[<->] (a) ++(0,.3) -- node[midway,anchor = south] {$Q_1$} ++(.7,0);
	\draw[->,middle segment=.2cm] ($(m1)+(0,.2)$) -- ($(b)+(0,.2)$) node[midway,anchor=south]{$\mu$};
	\draw[->,middle segment=.2cm] ($(m2)+(0,-.2)$) -- ($(c)+(0,-.2)$) node[midway,anchor=north]{$\lambda$};
	\draw (b)-- ++(1,0);
	\draw (c)-- ++(1,0);
	\end{tikzpicture}
	\times
	\begin{tikzpicture}[scale=.6,thick, baseline = -23]
		\draw (a)--++(.7,-.7) coordinate (b) --++ (0,-1) coordinate (c) --++(-.7,-.7) coordinate (x);
		\draw (a) -- (x)--++(-.5,-1);
		\draw (a) -- ++(-.5,1);
		\draw[<->] (b) ++(1.2,0) -- node[midway,anchor=west] {$Q_B$}  ++(0,-1);
	\draw[<->] (a) ++(0,.3) -- node[midway,anchor = south] {$Q_{-1}$} ++(.7,0);
	\draw[->,middle segment=.2cm] ($(m1)+(0,.2)$) -- ($(b)+(0,.2)$) node[midway,anchor=south]{$\mu$};
	\draw[->,middle segment=.2cm] ($(m2)+(0,-.2)$) -- ($(c)+(0,-.2)$) node[midway,anchor=north]{$\lambda$};
	\draw (b)-- ++(1,0);
	\draw (c)-- ++(1,0);
	\end{tikzpicture}
	\times
	\left.
	\begin{tikzpicture}[scale=.6,thick, baseline = -23]
		\draw (a)--++(.7,-.7) coordinate (b) --++ (0,-1) coordinate (c) --++(-.7,-.7) coordinate (x);
		\draw (a) -- (x)--++(-.5,-1);
		\draw (a) -- ++(-.5,1);
		\draw[<->] (b) ++(1.2,0) -- node[midway,anchor=west] {$Q_B$}  ++(0,-1);
	\draw[<->] (a) ++(0,.3) -- node[midway,anchor = south] {$Q_{-2}$} ++(.7,0);
	\draw[->,middle segment=.2cm] ($(m1)+(0,.2)$) -- ($(b)+(0,.2)$) node[midway,anchor=south]{$\mu$};
	\draw[->,middle segment=.2cm] ($(m2)+(0,-.2)$) -- ($(c)+(0,-.2)$) node[midway,anchor=north]{$\lambda$};
	\draw (b)-- ++(1,0);
	\draw (c)-- ++(1,0);
	\end{tikzpicture}
	\middle/\,\,
	\begin{tikzpicture}[scale=.6,thick, baseline = -23]
		\draw (a)--++(.7,-.7) coordinate (b) --++ (0,-1) coordinate (c) --++(-.7,-.7);
	\draw (b) --++(1,0) coordinate (m1);
	\draw (c)--++(1,0) coordinate (m2) ;
	\draw[<->] (b) ++(-0.5,0) -- node[midway,anchor=east] {$Q_B$}  ++(0,-1);
	\draw[->,middle segment=.2cm] ($(m1)+(0,.2)$) -- ($(b)+(0,.2)$) node[midway,anchor=south]{$\mu$};
	\draw[->,middle segment=.2cm] ($(m2)+(0,-.2)$) -- ($(c)+(0,-.2)$) node[midway,anchor=north]{$\lambda$};
	\end{tikzpicture}
	\right. .
	\end{split}
	\label{eq:SO8vertex}
\end{equation}
Note that we have the factor $1/Z_{\lambda,\mu}^\text{Half}(Q_B)$
in addition to the naive expectation from the dual description \eqref{SO2Np4}.
In the latter part of this paper we are going to make non--trivial checks of \eqref{eq:SO8vertex} and its generalizations by explicitly calculating the righthand side and  comparing the result with field theory computations.

\subsection{5d pure $\SO(2N+4)$ gauge theory}
\label{sec:pureSO2Np4}
We then move onto the explicit computation of the Nekrasov partition function of the pure $\SO(2N+4)$ gauge theory, making use of the trivalent gluing rule obtained in section \ref{sec:gluing}. Its dual theory is described by the trivalent gauging as in \eqref{SO2Np4}. Namely, it is realized by the trivalent $\SU(2)$ gauging of the diagonal part of the three $\SU(2)$ flavor symmetries of the $\hat{D}_N(\SU(2))$ and the two $\hat{D}_2(\SU(2))$ theories. The web-like description of the 5d theory which is dual to the pure $\SO(2N+4)$ gauge theory was given in figure \ref{fig:webforSO2Np4}. We then apply the gluing rule as well as the topological vertex to the web diagram. For that we first compute the partition function of the ``$\hat{D}_N(\SU(2))$ matter'' part with non-trivial Young diagrams on the parallel external legs representing the $\SU(2)$  instanton background. 

\begin{figure}[t]
\centering
\includegraphics[width=5cm]{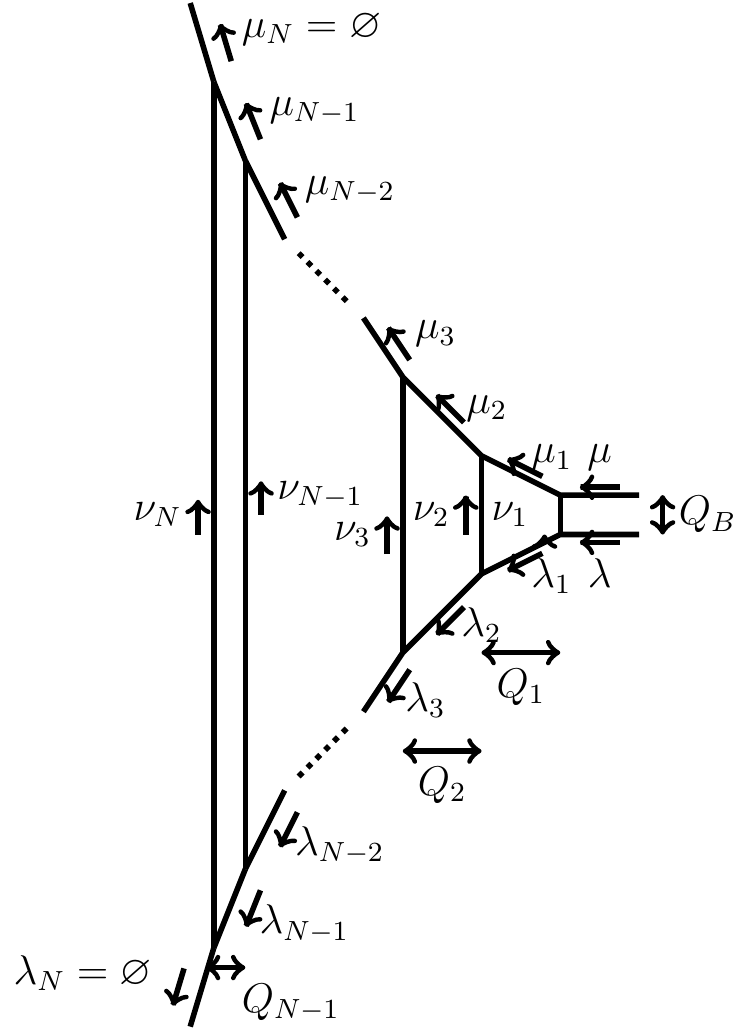}
\caption{The assignment of the Young diagrams, $\lambda, \mu, \lambda_i, \mu_i$ as well as the K$\ddot{\text{a}}$hler parameters $Q_B, Q_i$ for $i=1, \cdots, N-1$.}
\label{fig:DNSU2top}
\end{figure}

To compute the partition function of the $\hat{D}_N(\SU(2))$ matter system, we assign Young diagrams $\{\nu_a\} = \{\nu_1, \cdots, \nu_N\}$. $\{\lambda_a\} = \{\lambda_1, \cdots, \lambda_{N-1}\}$, $\{\mu_a\} = \{\mu_1, \cdots, \mu_N\}$ and also K$\ddot{\text{a}}$hler parameters $Q_B, \{Q_a\}=\{Q_1, \cdots, Q_{N-1}\}$ to the lines in the web for the $\hat{D}_N(\SU(2))$ as in figure \ref{fig:DNSU2top}.
By using the techniques in appendix \ref{sec:topvertex}, the application of the (unrefined) topological vertex to the web in figure \eqref{fig:DNSU2top} yields
\bea
\hat{Z}_{\lambda, \mu}^{\hat{D}_N(\SU(2))}(Q_B, \{Q_a\}) &=&  \sum_{\{\nu_a\}}\sum_{\{\lambda_a\}}\sum_{\{\mu_a\}}\prod_{a=1}^N\Big[(-Q_BQ_1^2Q_2^4\cdots Q_{a-1}^{2(a-1)})^{|\nu_a|}f_{\nu_a}^{-2a+1}(q)\nn\\
&&C_{\lambda_{a-1}^t\lambda_a\nu_a}(q)f_{\lambda_a^t}(q)(-Q_a)^{|\lambda_a|}C_{\mu_a\mu_{a-1}^t\nu_a^t}(q)f_{\mu_a^t}^{-1}(q)(-Q_a)^{|\mu_a|}\Big]\nn\\\label{DNSU2prepre}
\eea
where $\lambda_0 = \lambda, \mu_0 = \mu$ and $\lambda_N = \mu_N = \emptyset$. Note that we chose the last suffixes of the topological vertices as the Young diagrams assigned to the vertical lines in the web in figure \ref{fig:DNSU2top}. The choice is useful for the comparison with the Nekrasov partition function from the localization method since then \eqref{DNSU2prepre} is expanded by $Q_B$ which is eventually related to the instanton fugacity of the pure $\SO(2N+4)$ gauge theory. A straightforward computation of \eqref{DNSU2prepre} gives
\bea
\hat{Z}_{\lambda, \mu}^{\hat{D}_N(\SU(2))}(Q_B, \{Q_a\}) &=&  q^{-\frac{1}{2}||\mu||^2 + \frac{1}{2}||\mu^t||^2}\nn\\
&&\sum_{\{\nu_a\}}\prod_{a=1}^N\Big[(Q_BQ_1^2Q_2^4\cdots Q_{a-1}^{2(a-1)})^{|\nu_a|}q^{a||\nu_a^t||^2 - (a-1)||\nu_a||^2}\tilde{Z}_{\nu_a}(q)\tilde{Z}_{\nu_a^t}(q)\Big]\nn\\
&&\prod_{1 \leq a \leq b \leq N-1}\mathcal{H}\left(Q_a Q_{a+1} \cdots Q_b q^{i+j-1}\right)^{2}\nn\\
&&\prod_{1 \leq a \leq b \leq N-1}\mathcal{I}^+_{\nu_a, \nu_{b+1}}\left(Q_aQ_{a+1} \cdots Q_b\right)^{2}\prod_{1 \leq b < a \leq N}\mathcal{I}_{\nu_a, \nu_b}^-\left(Q_bQ_{b+1} \cdots Q_{a-1}\right)^{2}\nn\\
&&s_{\lambda}(q^{-\rho -\nu_1}, Q_1q^{-\rho - \nu_2}, Q_1Q_2q^{-\rho - \nu_3}, \cdots, \mathcal{Q}q^{-\rho - \nu_N})\nn\\
&&s_{\mu^t}(q^{-\rho -\nu_1}, Q_1q^{-\rho - \nu_2}, Q_1Q_2q^{-\rho - \nu_3}, \cdots, \mathcal{Q}q^{-\rho - \nu_N}),\nn\label{DNSU2pre}
\eea
where $s_{\lambda}$ is the Schur function and $q^{-\rho}$ is the specialization of its arguments, both of which are briefly reviewed in appendix \ref{sec:Schur}.
We introduced the notations
\bea
\mathcal{H}(Q) &=& \prod_{i, j=1}^{\infty}\left(1 - Qq^{i+j-1}\right)^{-1},\\
\mathcal{I}_{\nu_1, \nu_2}^{\pm}(Q) &=& \prod_{s\in \nu_1}\left(1 - Qq^{\pm\ell_{\nu_1}(s) \pm a_{\nu_2}(s) \pm 1}\right), \label{instanton.vector}\\
\mathcal{Q} &=& \prod_{a=1}^{N-1} Q_a,
\eea 
where $\pm$ signs in \eqref{instanton.vector} are taken in the same order. 

As discussed in section \ref{sec:gluing}, the partition function of \eqref{DNSU2pre} is not the one for the $\hat{D}_N(\SU(2))$ matter but one needs to divide it by the contribution of a ``half'' of the vector multiplets of \eqref{eq:halfvector}, 
and its explicit partition function is 
\bea
Z_{\lambda, \mu}^{\text{Half}}(Q_B)&=&\sum_{\nu}(-Q_B)^{|\nu|}f_{\nu^t}(q)C_{\lambda^t\emptyset\nu}(q)C_{\emptyset\mu^t\nu^t}(q)\nn\\
&=&q^{-\frac{1}{2}||\mu||^2 + \frac{1}{2}||\mu^t||^2}\sum_{\nu}Q_B^{|\nu|}q^{||\nu^t||^2}\tilde{Z}_{\nu}(q)\tilde{Z}_{\nu^t}(q)s_{\lambda}(q^{-\rho-\nu})s_{\mu^t}(q^{-\rho-\nu}).\label{half}
\eea
Therefore, the partition function of the $\hat{D}_N(\SU(2))$ matter is finally given by
\be
Z_{\lambda, \mu}^{\hat{D}_N(\SU(2))}(Q_B, \{Q_a\}) = \frac{\hat{Z}_{\lambda, \mu}^{\hat{D}_N(\SU(2))}(Q_B, \{Q_a\})}{Z_{\lambda, \mu}^{\text{Half}}(Q_B)}. \label{DNSU2}
\ee

One might worry that the contribution of the $\hat{D}_N(\SU(2))$ matter may be different when one rotates the diagram in figure \ref{fig:DNSU2top} by $\pi$ and puts Young diagrams on the parallel external legs with an orientation outward. When we consider the usual quadrivalent $\SU(2)$ gauging, we glue such a web with the web in figure \ref{fig:DNSU2top}. 
However, it turns out that the partition function \eqref{DNSU2} does not change after the $\pi$ rotation with the opposite orientation of the arrows for $\lambda, \mu$. Therefore we may use the partition function \eqref{DNSU2} both for the gluing from the left and the right. Due to this symmetric property, it is possible to use \eqref{DNSU2} even for the trivalent gauging.

Then as described in section \ref{sec:gluing}, our proposal is that the partition function of the pure $\SO(2N+4)$ gauge theory can be computed by 
treating the partition function \eqref{DNSU2} as a matter contribution for the $\SU(2)$ gauging. After inserting also the Nekrasov partition function of the $\SU(2)$ vector multiplets, we obtain
\bea
Z_{\SO(2N+4)}(Q_B, Q_g, \{Q_a\}, Q_{-1}, Q_{-2}) &=& \sum_{\lambda, \mu} Q_g^{|\lambda| + |\mu|}Z_{\lambda, \mu}^{\SU(2)\;\text{vector}}(Q_B)Z_{\lambda, \mu}^{\hat{D}_2(\SU(2))}(Q_{-1})\nn\\
&&Z_{\lambda, \mu}^{\hat{D}_2(\SU(2))}(Q_{-2})Z_{\lambda, \mu}^{\hat{D}_N(\SU(2))}(Q_B, \{Q_a\}),\label{NekSO2Np4}
\eea
where $Z_{\lambda, \mu}^{\SU(2)\;\text{vector}}(Q_B)$ is the contribution from the $\SU(2)$ vector multiplets
\bea
Z_{\lambda, \mu}^{\SU(2)\;\text{vector}}(Q_B) &=& \sum_{\nu}Q_B^{|\nu|}q^{||\nu^t||^2}\tilde{Z}_{\nu}(q)\tilde{Z}_{\nu^t}(q)s_{\lambda}(q^{-\rho-\nu})s_{\mu^t}(q^{-\rho-\nu})\nn\\
&&\sum_{\nu'}Q_B^{|\nu'|}q^{||\nu^{\prime t}||^2}\tilde{Z}_{\nu'}(q)\tilde{Z}_{\nu^{\prime t}}(q)s_{\lambda}(q^{-\rho-\nu'})s_{\mu^t}(q^{-\rho-\nu'}). \label{SU2Nek}
\eea

In the dual picture $Q_B$ corresponds to the Coulomb branch modulus of the $\SU(2)$ gauging and $Q_g/Q_B$ corresponds to the instanton fugacity of $\SU(2)$. For the original frame, $Q_g$ is rather related to one of the Coulomb branch moduli of the pure $\SO(2N+4)$ gauge theory and $Q_B$ is related to the instanton fugacity of $\SO(2N+4)$. 

It is possible to determine the precise relations between the K\"ahler parameters $Q_B$, $Q_{-2}, Q_{-1}, Q_g, \{Q_a\}$ and the Coulomb branch moduli and the instanton fugacity of the pure $\SO(2N+4)$ gauge theory. Let $C_f$ be the curve whose K\"ahler parameter is $Q_f$ for $f=-2, -1, g, 1, \cdots, N-1$. The $N+2$ curves $C_f, f=-2, -1, g, 1, \cdots, N-1$ form the $F_{\mathfrak{so}(2N+4)}$ fiber whose shape is the Dynkin diagram of $\mathfrak{so}(2N+4)$. Therefore, they are associated to the simple roots of the Lie algebra $\mathfrak{so}(2N+4)$ and we can parameterize 
\bea
Q_{i} &=& e^{-(a_{N-i} - a_{N-i+1})}, \; i=1, \cdots, N-1, \quad Q_g = e^{-(a_N - a_{N+1})}, \nn\\
Q_{-1} &=& e^{-(a_{N+1} - a_{N+2})}, \quad Q_{-2} = e^{-(a_{N+1} + a_{N+2})}, \label{CoulombSO2Np4}
\eea
where $a_i, i=1, \cdots, N+2$ are the Coulomb branch moduli of the pure $\SO(2N+4)$ gauge theory. 

One the other hand, the instanton fugacity $u_{\SO(2N+4)}$ is related to the size of the base $C_B$ and hence it is equal to $Q_B$ up to a factor consisting of $Q_f, f=-2, -1, g, 1, \cdots, N-1$,
\be
u_{\SO(2N+4)} = Q_B h(Q_{-2}, Q_{-1}, Q_g, \{Q_a\}),
\label{instantonSO2Np4pre}
\ee
where $h$ is a certain monomial of arguments.
In order to fix the factor $h$, let us see the intersection numbers between the curves $C_i, i=-2, -1, g, 1, \cdots, N-1, B$ and the surface $S_f$ which has the $C_f$ fibration over $C_B$ where $f=-2, -1, g, 1, \cdots, N-1$. Due to the Dynkin diagram structure of the fiber $F_{\mathfrak{so}(2N+4)}$, the intersection matrix between $C_f$ and $S_{f'}$ for $f, f'=-2, -1, g, 1, \cdots, N-1$ forms the negative of the Cartan matrix of the $\mathfrak{so}(2N+4)$ Lie algebra. Furthermore, $C_B$ intersects only with $S_g$ with the intersection number $-2$. The intersection numbers are summarized as in table \ref{tb:intersectionSO2Np4}. 
\begin{table}[t]
\begin{center}
\begin{tabular}{c|ccccccc}
& $S_{N-1}$ & $S_{N-2}$ & $\cdots$ & $S_1$ & $S_g$ & $S_{-1}$ & $S_{-2}$\\
\hline
$C_{N-1}$& -2 & 1 &  $\cdots$ & 0 & 0 & 0 & 0\\
$C_{N-2}$ & 1 & -2 & $\cdots$ & 0 & 0 & 0 & 0 \\
$\vdots$ &&&&&&&\\
$C_1$ & 0 & 0 & $\cdots$ & -2 & 1 & 0 & 0\\
$C_g$& 0 & 0 & $\cdots$ & 1 & -2 & 1 & 1 \\
$C_{-1}$ & 0 & 0 & $\cdots$ & 0 & 1 & -2 & 0\\
$C_{-2}$ & 0 & 0 & $\cdots$ & 0 & 1 & 0 & -2\\
\hline
$C_B$ & 0 & 0 & $\cdots$ & 0 & -2 & 0 & 0
\end{tabular}
\caption{The intersection numbers between the surfaces $S_f, f=-2, -1, g, 1, \cdots, N-1$ and the curves $C_i, i=-2, -1, g, 1, \cdots, N-1, B$. }
\label{tb:intersectionSO2Np4}
\end{center}
\end{table}
In other words, the intersection numbers imply the Coulomb branch moduli dependence for the K$\ddot{\text{a}}$hler parameter. Since the instanton fugacity does not depend on the Coulomb branch moduli, the factor $h(Q_{-2}, Q_{-1}, Q_g, \{Q_a\})$ in \eqref{instantonSO2Np4pre} should be chosen so that $u_{\mathfrak{so}(2N+4)}$ does not depend on the Coulomb branch moduli or equivalently the corresponding curve has the zero intersection number with any surface $S_f, f=-2, -1, g, 1, \cdots, N-1$. This uniquely fixes the factor $h(Q_{-2}, Q_{-1}, Q_g, \{Q_a\})$ and the instanton fugacity is given by
\be
u_{\mathfrak{so}(2N+4)} = Q_BQ_g^{-2N}Q_{-1}^{-N}Q_{-2}^{-N}\prod_{a=1}^{N-1}Q_a^{-2N+2a}. \label{instantonSO2Np4}
\ee

Therefore, we conjecture that the partition function \eqref{NekSO2Np4} yields the Nekrasov partition function of the pure $\SO(2N+4)$ gauge theory after inserting the gauge theory parameters given by the relations \eqref{CoulombSO2Np4} and \eqref{instantonSO2Np4}\footnote{In this paper, we ignore the perturbative partition function from vector multiplets in the Cartan subalgebra of a gauge group $G$. The contribution cannot be captured from the topological vertex calculation but it can be easily recovered by the general formula
\be
Z_{\text{Cartan}} = \mathcal{H}(1)^{\text{rank}(G)},
\ee
where $\text{rank}(G)$ is the rank of the gauge group $G$. }.

\subsubsection{Example: 5d pure $\SO(8)$ gauge theory}

Let us explicitly compute the partition function \eqref{NekSO2Np4} obtained from the $\SU(2)$ trivalent gauging for an example. We work on the simplest case when $N=2$, namely the 5d pure $\SO(8)$ gauge theory. Inserting $N=2$ to \eqref{NekSO2Np4} yields
\bea
Z_{\SO(8)}(Q_B, Q_g, Q_1, Q_{-1}, Q_{-2}) &=& \sum_{\lambda, \mu} Q_g^{|\lambda| + |\mu|}Z_{\lambda, \mu}^{\SU(2)\;\text{vector}}(Q_B)Z_{\lambda, \mu}^{\hat{D}_2(\SU(2))}(Q_{-1})\nn\\
&&Z_{\lambda, \mu}^{\hat{D}_2(\SU(2))}(Q_{-2})Z_{\lambda, \mu}^{\hat{D}_2(\SU(2))}(Q_B, \{Q_1\}), 
\label{eq:SO8vertex2}
\eea
and we argued that this gives rise to the Nekrasov partition function of the 5d pure $\SO(8)$ gauge theory. The Coulomb branch moduli $a_i, i=1, 2, 3, 4$ of the $\SO(8)$ gauge theory are given by \eqref{CoulombSO2Np4} with $N=2$, namely
\bea
Q_{1} = e^{-(a_{1} - a_{2})}, \quad Q_g = e^{-(a_2 - a_{3})}, \quad Q_{-1} = e^{-(a_{3} - a_{4})}, \quad Q_{-2} = e^{-(a_{3} + a_{4})}, \label{CoulombSO8}
\eea
and from \eqref{instantonSO2Np4} the instanton fugacity $u_{\SO(8)}$ is 
\be
u_{\SO(8)} = \frac{Q_B}{Q_{-2}^2Q_{-1}^2Q_g^4Q_1^2}. \label{instantonSO8}
\ee

\paragraph{Perturbative part}
Since the instanton fugacity is written by \eqref{instantonSO8}, the perturbative part is obtained at the order $\mathcal{O}(Q_B^0)$. Its explicit form is given by
\bea
Z_{\SO(8)}^{\text{Pert}}(Q_g, Q_1, Q_{-1}, Q_{-2}) &=& \mathcal{H}(Q_{-2})^2 \mathcal{H}(Q_{-1})^2\mathcal{H}(Q_1)^2\nn\\
&&\sum_{\lambda, \mu}Q_g^{|\lambda|+|\mu|}\frac{\prod_{i=-2,-1,1}s_{\lambda}(q^{-\rho}, Q_{i}q^{-\rho})s_{\mu^t}(q^{-\rho}, Q_{i}q^{-\rho})}{s_{\lambda}(q^{-\rho})s_{\mu^t}(q^{-\rho})}.\nn\\ \label{SO8pert1}
\eea
Indeed we have checked that
\bea
&&\sum_{\lambda, \mu}Q_g^{|\lambda|+|\mu|}\frac{\prod_{i=-2,-1,1}s_{\lambda}(q^{-\rho}, Q_iq^{-\rho})s_{\mu^t}(q^{-\rho}, Q_iq^{-\rho})}{s_{\lambda}(q^{-\rho})s_{\mu^t}(q^{-\rho})}\nn\\
&=& \mathcal{H}(Q_g)^2\mathcal{H}(Q_{-2}Q_g)^2\mathcal{H}(Q_{-1}Q_g)^2\mathcal{H}(Q_1Q_g)^2\mathcal{H}(Q_{-2}Q_{-1}Q_g)^2\mathcal{H}(Q_{-2}Q_1Q_g)^2\mathcal{H}(Q_{-1}Q_1Q_g)^2\nn\\
&&\mathcal{H}(Q_{-2}Q_{-1}Q_1Q_g)^2\mathcal{H}(Q_{-2}Q_{-1}Q_1Q_g^2)^2, \label{SO8pert2}
\eea
until the order of $Q_g^8$. Combining \eqref{SO8pert1} with \eqref{SO8pert2} yields
\bea
Z_{\SO(8)}^{\text{Pert}} &=&\mathcal{H}(Q_{-2})^2 \mathcal{H}(Q_{-1})^2\mathcal{H}(Q_1)^2\mathcal{H}(Q_g)^2\mathcal{H}(Q_{-2}Q_g)^2\mathcal{H}(Q_{-1}Q_g)^2\mathcal{H}(Q_1Q_g)^2\nn\\\
&&\mathcal{H}(Q_{-2}Q_{-1}Q_g)^2\mathcal{H}(Q_{-2}Q_1Q_g)^2\mathcal{H}(Q_{-1}Q_1Q_g)^2\mathcal{H}(Q_{-2}Q_{-1}Q_1Q_g)^2\mathcal{H}(Q_{-2}Q_{-1}Q_1Q_g^2)^2,\nn\\
\eea
which precisely reproduces the perturbative partition function of the pure $\SO(8)$ gauge theory except for the Cartan part which cannot be captured from the topological vertex.

\paragraph{Instanton part}
Next we turn to the instanton partition function of the pure $\SO(8)$ gauge theory. The instanton part is obtained by normalize the full partition function by the perturbative partition function,
\be
Z_{\SO(8)}^{\text{Inst}}(Q_B, Q_g, Q_1, Q_{-1}, Q_{-2}) = \frac{Z_{\SO(8)}(Q_B, Q_g, Q_1, Q_{-1}, Q_{-2})}{Z_{\SO(8)}^{\text{Pert}}(Q_g, Q_1, Q_{-1}, Q_{-2})}. \label{instpartSO8}
\ee
We checked that \eqref{instpartSO8} agrees with the result obtained from the localization \eqref{localizationSO} until the order of $Q_g^5$ for the one-instanton part and also the two-instanton part.
Note that the $q$ dependence of the unrefined one-instanton partition function is just $-\frac{q}{(1-q)^2}$ coming from the center of mass mode.
This behavior alone is nontrivial from \eqref{eq:SO8vertex2}.

\subsection{Adding flavors}
\label{sec:SO2Np4wflvrs}

As described in section \ref{sec:5dSO2Np4}, the trivalent gauging also provides us with a web-like diagram for the 5d theory which is dual to the $\SO(2N+4)$ gauge theory with $M_1+M_2$ vector multiplets in the vector representation. The figure is depicted in figure \ref{fig:webforSO2Np4wflvrs} and the quiver--like description of the dual theory is given in \eqref{SO2Np4wflvrs}. In this section, we assume $M_1 \leq N$ and $M_2 \leq N$ and excludes the case $M_1 + M_2 = 2N-1$. This is a technical assumption which eases the computation of the partition function, but it is straightforward to apply the trivalent gluing method to the case of $M_1 + M_2 = 2N-1$. 

\begin{figure}[t]
\centering
\includegraphics[width=5cm]{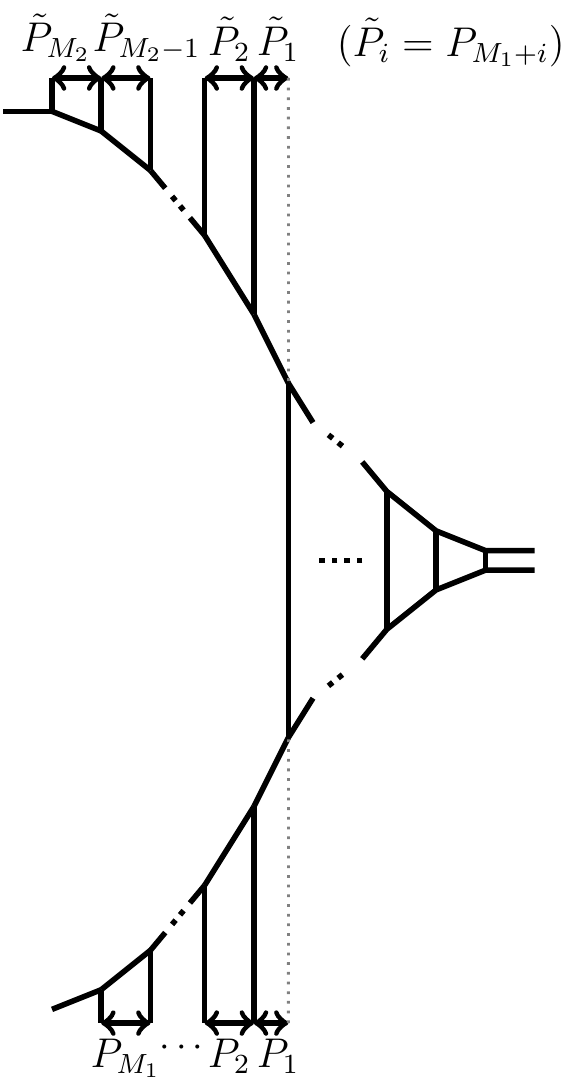}
\caption{The assignment of the newly introduced K$\ddot{\text{a}}$hler parameters  $\{P_c \}=\{P_1, \cdots, P_{M_1+M_2}\}$ to the web diagram of the $\hat{D}_N^{M_1,M_2}(\SU(2))$ matter theory. The Young diagrams $\lambda_i, (i=1, \cdots, N+M_1-1), \mu_i, (i=1, \cdots, N+M_2-1)$ are assgined by generalizing the Young diagram assignment in figure \ref{fig:DNSU2top} and the K\"ahler parameters $\{Q_a\} = \{Q_1, \cdots, Q_{N-1}\}$ are assigned in the same way as in figure \ref{fig:DNSU2top}. }
\label{fig:DNMSU2}
\end{figure}

In order to compute the Nekrasov partition function of the $\SO(2N+4)$ gauge theory with $M_1 + M_2$ flavors, we first calculate the partition function of the $\hat{D}_N^{M_1,M_2}(SU(2))$ matter. We then assign Young diagrams $\{\nu_a\} = \{\nu_1, \cdots, \nu_{N}\}$, $\{\lambda_b\} = \{\lambda_1, \cdots, \lambda_{2N-1}\}$,  $\{\mu_b\} = \{\mu_1, \cdots, \mu_{2N-1}\}$ to the internal lines in figure \ref{fig:DNMSU2}  by generalizing the Young diagram assignment in figure \ref{fig:DNSU2top}. The assignment of the K\"ahler parameters $\{Q_a\} = \{Q_1, \cdots, Q_{N-1}\}$ is the same as the assignment in figure \ref{fig:DNSU2top}. We further introduce new labels $\{P_c \}=\{P_1, \cdots, P_{M_1+M_2}\}$ to the lines in the web as in 
figure \ref{fig:DNMSU2}. 
Then the application of the topological vertex to the web in figure \ref{fig:DNMSU2} yields
\bea
\hat{Z}^{\hat{D}_N^{M_1,M_2}(SU(2))}_{\lambda, \mu}(Q_B, \{Q_a\}, \{P_c\}) &=& \sum_{\{\nu_a\}}\sum_{\{\lambda_b\}}\sum_{\{\mu_b\}}\prod_{a=1}^N\Big[(-Q_BQ_1^2Q_2^4\cdots Q_{a-1}^{2(a-1)})^{|\nu_a|}f_{\nu_a}^{-2a+1}(q)\Big]\nn\\
&&\left[\prod_{a=1}^{N-1}C_{\lambda_{a-1}^t\lambda_a\nu_a}(q)(-Q_a)^{|\lambda_a|}f_{\lambda_a^t}(q)\right]C_{\lambda_{N-1}^t\lambda_N\nu_N}(q)\nn\\
&&\prod_{a=1}^{M_1}C_{\lambda_{N+a}\lambda_{N+a-1}^t\emptyset}(q)(-P_a)^{|\lambda_{N+a-1}|}\prod_{a=2}^{M_1}f_{\lambda_{N+a-1}^t}(q)^{-1}\nn\\
&&\left[\prod_{a=1}^{N-1}C_{\mu_a\mu_{a-1}^t\nu_a^t}(q)(-Q_a)^{|\mu_a|}f^{-1}_{\mu_a^t}(q)\right]C_{\mu_N\mu_{N-1}^t\nu_N^t}(q)\nn\\
&&\prod_{a=1}^{M_2}C_{\mu_{N+a-1}^t\mu_{N+a}\emptyset}(q)(-P_{N+a})^{|\mu_{N+a-1}|}f_{\mu_{N+a}^t}(q),
\eea
where $\lambda_0 = \lambda, \lambda_{N+M_1} = \emptyset, \mu_0 = \mu, \mu_{N+M_2} = \emptyset$. A tedious but straightforward calculation gives rise to 
\bea
\hat{Z}^{\hat{D}_N^{M_1,M_2}(SU(2))}_{\lambda, \mu}(Q_B, \{Q_a\}, \{P_c\}) &=& q^{-\frac{1}{2}||\mu||^2 + \frac{1}{2}||\mu^t||^2}\nn\\
&&\sum_{\{\nu_a\}}\prod_{a=1}^N\Big[(Q_BQ_1^2\cdots Q_{a-1}^{2(a-1)})^{|\nu_a|}q^{a||\nu_a^t||^2 - (a-1)||\nu_a||^2}\tilde{Z}_{\nu_a}(q)\tilde{Z}_{\nu_a^t}(q)\Big]\nn\\
&&\prod_{1 \leq a \leq b \leq N-1}\mathcal{H}(Q_a Q_{a+1} \cdots Q_b)^{2}\nn\\
&&\prod_{1 \leq a \leq b \leq N-1}\mathcal{I}^+_{\nu_a, \nu_{b+1}}(Q_aQ_{a+1} \cdots Q_b)^{2}\prod_{1 \leq b < a \leq N}\mathcal{I}_{\nu_a, \nu_b}^-(Q_bQ_{b+1} \cdots Q_{a-1})^{2}\nn\\
&&\prod_{2 \leq a \leq b \leq M_1}\mathcal{H}(P_{a}P_{a+1} \cdots P_b)\prod_{2 \leq a \leq b \leq M_2}\mathcal{H}(P_{M_1+a}P_{M_1+a+1} \cdots P_{M_1+b})\nn\\
&&\prod_{1 \leq a \leq N,\; 1\leq b \leq M_1}\mathcal{H}(Q_aQ_{a+1}\cdots Q_{N-1}P_1P_{2} \cdots P_b)^{-1}\nn\\
&&\prod_{1 \leq a \leq N,\; 1\leq b \leq M_1}\mathcal{I}_{\nu_a, \emptyset}^{+}(Q_aQ_{a+1}\cdots Q_{N-1}P_1P_{2} \cdots P_b)^{-1}\nn\\
&&\prod_{1 \leq a \leq N,\; 1\leq b \leq M_2}\mathcal{H}(Q_aQ_{a+1}\cdots Q_{M-1}P_{M_1+1}P_{M_1+2} \cdots P_{M_1+b})^{-1}\nn\\
&&\prod_{1 \leq a \leq N,\; 1\leq b \leq M_2}\mathcal{I}_{\nu_a, \emptyset}^+(Q_aQ_{a+1}\cdots Q_{N-1}P_{M_1+1}P_{M_1+2} \cdots P_{M_1+b})^{-1}\nn\\
&&\sum_{\alpha}s_{\lambda/\alpha}\left(q^{-\rho - \nu_1}, Q_1q^{-\rho - \nu_2}, Q_1Q_2 q^{-\rho - \nu_3}, \cdots, \mathcal{Q}q^{-\rho - \nu_N}\right)\nn\\
&&s_{\alpha^t}\left(-\mathcal{Q}P_1q^{-\rho}, -\mathcal{Q}P_1P_2q^{-\rho}, \cdots, -\mathcal{Q}\mathcal{P}_1q^{-\rho}\right),\nn\\
&&\sum_{\beta}s_{\mu^t/\beta}\left(q^{-\rho - \nu_1}, Q_1q^{-\rho - \nu_2}, Q_1Q_2 q^{-\rho - \nu_3}, \cdots, \mathcal{Q}q^{-\rho - \nu_N}\right)\nn\\
&&s_{\beta^t}\left(-\mathcal{Q}P_{M_1+1}q^{-\rho}, -\mathcal{Q}P_{M_1+1}P_{M_1+2}q^{-\rho}, \cdots, -\mathcal{Q}\mathcal{P}_2q^{-\rho}\right),\label{DNMSU2pre}
\eea
where we defined
\bea
\mathcal{P}_1 = \prod_{a=1}^{M_1}P_a, \quad \mathcal{P}_2 = \prod_{a=1}^{M_2}P_{M_1+a}.
\eea

The partition function \eqref{DNMSU2pre} contains an extra factor associated to the parallel external legs in figure \ref{fig:DNMSU2}. In order to recover the partition function of a 5d theory realized on a 5-brane web, one needs to divide the topological vertex result by the extra factor \cite{Bergman:2013ala, Hayashi:2013qwa, Bao:2013pwa, Bergman:2013aca}. The contribution of the extra factor is 
\bea
Z_{\text{extra}, 1}(Q_B, \{Q_a\}, \{P_c\}) &=&\prod_{2 \leq a \leq b \leq M_1}\mathcal{H}(P_{a}P_{a+1} \cdots P_b)\prod_{2 \leq a \leq b \leq M_2}\mathcal{H}(P_{M_1+a}P_{M_1+a+1} \cdots P_{M_1+b})\nn\\
&&\mathcal{H}\left(Q_B\prod_{a=1}^NQ_{a-1}^{2(a-1)}(P_aP_{N+a})^{N-a+1}\right)\delta_{M_1,N}\delta_{M_2,N}. \label{extra1}
\eea 
Note that when $M_1 = M_2 = N$, another two parallel external legs appear and the extra factor from the parallel external legs is the last factor in \eqref{extra1}. Furthermore, we also divide \eqref{DNMSU2pre} by the half of the vector multiplet contribution of $\SU(2)$ given by \eqref{half}. Hence the partition function of the $\hat{D}_N^{M_1,M_2}(\SU(2))$ matter is 
\be
Z^{\hat{D}_{N}^{M_1,M_2}(\SU(2))}_{\lambda, \mu}(Q_B, \{Q_a\}, \{P_c\}) = \frac{\hat{Z}^{M_1,M_2}_{\lambda, \mu}(Q_B, \{Q_a\}, \{P_c\})}{Z_{\lambda, \mu}^{\text{Half}}(Q_B)Z_{\text{extra}, 1}(Q_B, \{Q_a\}, \{P_c\}) }. \label{DNMSU2}
\ee

Having identified the partition function of the $\hat{D}_N^{M_1,M_2}(\SU(2))$ matter, the Nekrasov partition function of the $\SO(2N+4)$ gauge theory with $M_1+M_2$ hypermultiplets in the vector representation is obtained by the trivalent gauging of $\hat{D}_N^{M_1, M_2}(\SU(2))$ matter with the two $\hat{D}_2(\SU(2))$ matter. Namely the partition function is given by 
\bea
\tilde{Z}_{\SO(2N+4)}^{M_1, M_2}(Q_B, Q_g, \{Q_a\}, \{P_c\}, Q_{-1}, Q_{-2})  &=& \sum_{\lambda, \mu}Q_g^{|\lambda|+|\mu|}Z_{\lambda,\mu}^{\SU(2)\;\text{vector}}(Q_B)\nn\\
&&Z_{\lambda, \mu}^{\hat{D}_2(\SU(2))}(Q_{-2})Z_{\lambda, \mu}^{\hat{D}_2(\SU(2))}(Q_{-1})\nn\\
&&Z^{\hat{D}_{N}^{M_1,M_2}(\SU(2))}_{\lambda, \mu}(Q_B, \{Q_a\}, \{P_c\}).\nn\\
\eea

However, this is not the final partition function. We argue that the trivalent gluing yields another extra factor. The presence of the another extra factor may be understood from the 5-brane web with an $O5$--plane in figure \ref{fig:SO2Np4wflvrs}. The extra factor is associated to the parallel external legs. In figure \ref{fig:SO2Np4wflvrs}, the diagram has $M_1$ flavor D5-branes on the left and $M_2$ flavor D5-branes on the right. In the absence of the $O5$-plane, the extra factor contribution may be given by a half of the perturbative vector multiplet contribution of $\SU(M_1)$ from the left and that of $\SU(M_2)$ from the right. Here $\SU(M_1)$ and $\SU(M_2)$ are the flavor symmetries associated to the $M_1$ flavor D5-branes on the left and $M_2$ D5-branes on the right respectively. In the presence of the $O5$-plane, the $M_1$ flavor D5-branes imply an $\Sp(M_1)$ flavor symmetry and the $M_2$ flavor D5-branes imply an $\Sp(M_2)$ flavor symmetry since the flavor branes are on top of an $O5^+$-plane. Therefore, the extra factor from the parallel external branes on the left may be a half of the perturbative contribution of $\Sp(M_1)$ vector multiplets and the extra factor from the parallel external branes on the right may be a half of the perturbative partition function of $Sp(M_2)$ vector multiplets. The combined partition function is written as 
\bea
Z_{\text{extra, pert}}&=&\prod_{2 \leq a \leq b \leq M_1}\mathcal{H}(P_{a}P_{a+1} \cdots P_b)\prod_{1\leq a \leq b \leq M_1}\mathcal{H}(\mathcal{R}\prod_{k=1}^aP_k\prod_{j=1}^bP_j)\nn\\
&&\prod_{2 \leq a \leq b \leq M_2}\mathcal{H}(P_{M_1+a}P_{M_1+a+1} \cdots P_{M_1+b})\prod_{1\leq a \leq b \leq M_2}\mathcal{H}(\mathcal{R}\prod_{k=1}^aP_{M_1+k}\prod_{j=1}^bP_{M_1+j}),\nn\\\label{extrapert}
\eea
where
\be
\mathcal{R}=Q_{-2}Q_{-1}Q_g^2\left[\prod_{a=1}^{N-1}Q_a^2\right]
\ee
Compared with the extra factor \eqref{extra1} of the $\hat{D}_N^{M_1,M_2}(\SU(2))$ matter with \eqref{extrapert}, we can deduce the additional extra factor
\bea
Z_{\text{extra}, 2}(Q_g, \{Q_a\}, \{P_c\}, Q_{-1}, Q_{-2}) &=& \prod_{1\leq a \leq b \leq M_1}\mathcal{H}(\mathcal{R}\prod_{k=1}^aP_k\prod_{j=1}^bP_j)\nn\\
&&\prod_{1\leq a \leq b \leq M_2}\mathcal{H}(\mathcal{R}\prod_{k=1}^aP_{M_1+k}\prod_{j=1}^bP_{M_1+j}).
\eea

Hence we claim that the final expression for the Nekrasov partition function of the $\SO(2N+4)$ gauge theory with $M_1 + M_2$ hypermultiplets in the vector representation is 
\be
Z_{\SO(2N+4)}^{M_1, M_2}(Q_B, Q_g, \{Q_a\}, \{P_c\}, Q_{-1}, Q_{-2})  = \frac{\tilde{Z}_{\SO(2N+4)}^{M_1, M_2}(Q_B, Q_g, \{Q_a\}, \{P_c\}, Q_{-1}, Q_{-2})  }{Z_{\text{extra}, 2}(Q_g, \{Q_a\}, \{P_c\}, Q_{-1}, Q_{-2}) }.  \label{NekSO2Np4wflvrs}
\ee

Lastly, we relate the K\"ahler parameters with the Coulomb branch moduli, mass parameters and the instanton fugacity of the 5d $SO(2N+4)$ gauge theory. The Coulomb branch moduli of the $\SO(2N+4)$ gauge theory are again related to the K\"ahler parameters $\{Q_a\}, Q_{-2}, Q_{-1}$ by \eqref{CoulombSO2Np4}. The mass parameters $m_i, i=1, \cdots, M_1+M_2$ are related to the height of the flavor branes of the 5-brane web in figure \ref{fig:SO2Np4wflvrs}. This leads to the parameterization
\be
P_i = e^{-(m_i - m_{i-1})}, \quad P_1 = e^{-(m_1 - a_1)}, \quad P_{M_1+1} = e^{-(m_{M_1+1} - a_1)}, \label{massSO2Np4wflvrs}
\ee
for $i=2, \cdots, M_1, M_1+2, \cdots, M_1+M_2$. The determination of the instanton fugacity is more involved. First we note that from the localization result of \eqref{localizationSO} one needs to redefine the instanton fugacity when one decouples one flavor as
\be
u_{\SO(2N+4), N_f} e^m = u_{\SO(2N+4), N_f-1}, \label{instanton.decouple}
\ee
where $m$ is the mass parameter 
which we send to infinity. In other words, the combination on the lefthand side of \eqref{instanton.decouple} remains finite in the limit $m \rightarrow \infty$. Therefore, decoupling $M_1+M_2$ flavors imply
\be
u_{\SO(2N+4), M_1+M_2}  = u_{\SO(2N+4), 0}e^{-\sum_{i=1}^{M_1+M_2}m_i}. \label{instantonredefinition}
\ee
Eq.~\eqref{instantonredefinition} suggests that the instanton fugacity may be written as
\be
u_{\SO(2N+4), M_1+M_2}=Q_B\left[\prod_{a=1}^{M_1}P_{a}^{M_1-a+1}\right]\left[\prod_{a=1}^{M_2}P_{M_1+a}^{M_2-a+1}\right]h(Q_{-2}, Q_{-1}, Q_g, Q_1, \cdots, Q_{N-1}).
\ee
The remaining task is to determine $h(Q_{-2}, Q_{-1}, Q_g, Q_1, \cdots, Q_{N-1})$ so that the instanton fugacity does not have the Coulomb branch moduli dependence. For that we denote the curve whose K\"ahler parameter is $Q_f$ by $C_f$ for $f=-2, -1, g, 1, \cdots, N-1$. We further introduce $C', C''$ whose K$\ddot{\text{a}}$hler parameters are $P_{1}^{M_1}, P_{M_1+1}^{M_2}$ respectively. Note that only $P_{1}, P_{M_1+1}$ in the set $\{P_c\}$ have the Coulomb branch moduli dependence. Let $S_i$ be the surface which is $C_f$ fibration over $C_B$ for $f=-2, -1, g, 1, \cdots, N-1$. Then the intersection numbers are summarized in table \ref{tb:intersectionSO2Np4wflvrs}.
\begin{table}[t]
\begin{center}
\begin{tabular}{c|ccccccc}
& $S_{N-1}$ & $S_{N-2}$ & $\cdots$ & $S_1$ & $S_g$ & $S_{-1}$ & $S_{-2}$\\
\hline
$C_{N-1}$& -2 & 1 &  $\cdots$ & 0 & 0 & 0 & 0\\
$C_{N-2}$ & 1 & -2 & $\cdots$ & 0 & 0 & 0 & 0 \\
$\vdots$ &&&&&&&\\
$C_1$ & 0 & 0 & $\cdots$ & -2 & 1 & 0 & 0\\
$C_g$& 0 & 0 & $\cdots$ & 1 & -2 & 1 & 1 \\
$C_{-1}$ & 0 & 0 & $\cdots$ & 0 & 1 & -2 & 0\\
$C_{-2}$ & 0 & 0 & $\cdots$ & 0 & 1 & 0 & -2\\
\hline
$C_B$ & 0 & 0 & $\cdots$ & 0 & -2 & 0 & 0\\
\hline
$C'$& $M_1$ & 0 & $\cdots$ & 0 & 0 & 0 & 0\\
$C''$ & $M_2$ & 0 & $\cdots$ & 0 & 0 & 0 & 0\\ 
\end{tabular}
\caption{The intersection numbers between the surfaces $S_f, f=-2, -1, g, 1, \cdots, N-1$ and the curves $C', C'', C_i, i=-2, -1, g, 1, \cdots, N-1, B$. }
\label{tb:intersectionSO2Np4wflvrs}
\end{center}
\end{table}
By making use of table \ref{tb:intersectionSO2Np4wflvrs}, we can fix the remaining factor $h(Q_{-2}, Q_{-1}, Q_g, Q_1, \cdots, Q_{N-1})$ and the instanton fugacity 
is given by
\bea
u_{\SO(2N+4), M_1+M_2} &=& Q_B\left[\prod_{a=1}^{M_1}P_{a}^{M_1-a+1}\right]\left[\prod_{a=1}^{M_2}P_{M_1+a}^{M_2-a+1}\right]Q_g^{-2N+M_1+M_2}\nn\\
&&Q_{-1}^{-N+\frac{M_1+M_2}{2}}Q_{-2}^{-N+\frac{M_1+M_2}{2}}\prod_{a=1}^{N-1}Q_a^{-2N+M_1+M_2+2a}, \label{instantonSO2Np4wflvrs}
\eea
in the case of the $M_1+M_2$ flavors.

Therefore, we claim that the partition function \eqref{NekSO2Np4wflvrs} with the relations \eqref{CoulombSO2Np4}, \eqref{massSO2Np4wflvrs} and \eqref{instantonSO2Np4wflvrs} yields the Nekrasov partition function of the $\SO(2N+4)$ gauge theory with $M_1+M_2$ hypermultiplets in the vector representation.

\subsubsection{Example: 5d $\SO(8)$ gauge theory with four flavors}

Let us then see an explicit simple example with $N=2$ and $M_1=M_2=2$. Then the 5d theory in \eqref{SO2Np4wflvrs} with $N=2$ and $M_1=M_2=2$ is dual to the 5d $\SO(8)$ gauge theory with four flavors. The full partition function is given by 
\bea
Z_{\SO(8)}^{2, 2}(Q_B, Q_g, Q_1, \{P_c\}, Q_{-1}, Q_{-2})  &=&\sum_{\lambda, \mu}Q_g^{|\lambda|+|\mu|}Z_{\lambda,\mu}^{\SU(2)\;\text{vector}}(Q_B)Z_{\lambda, \mu}^{\hat{D}_2(\SU(2))}(Q_{-2})\nn\\
&&Z_{\lambda, \mu}^{\hat{D}_2(\SU(2))}(Q_{-1})Z^{\hat{D}_{2}^{2, 2}(\SU(2))}_{\lambda, \mu}(Q_B, Q_1, \{P_c\})\nn\\
&&Z_{\text{extra}, 2}(Q_g, \{Q_a\}, \{P_c\}, Q_{-1}, Q_{-2})^{-1},  \label{NekSO8wflvrs}
\eea
where 
\bea
Z_{\text{extra}, 2}(Q_g, \{Q_a\}, \{P_c\}, Q_{-1}, Q_{-2}) &=& \mathcal{H}(P_1^2Q_1^2Q_g^2Q_{-2}Q_{-1}) \mathcal{H}(P_2P_1^2Q_1^2Q_g^2Q_{-2}Q_{-1})\nn\\
&&\mathcal{H}(P_2^2P_1^2Q_1^2Q_g^2Q_{-2}Q_{-1}) \mathcal{H}(P_3^2Q_1^2Q_g^2Q_{-2}Q_{-1})\nn\\
&&\mathcal{H}(P_4P_3^2Q_1^2Q_g^2Q_{-2}Q_{-1}) \mathcal{H}(P_4^2P_3^2Q_1^2Q_g^2Q_{-2}Q_{-1}).\nn\\
\eea
The Coulomb branch moduli $a_i, i=1, 2, 3, 4$ of the $\SO(8)$ gauge theory are the same as \eqref{CoulombSO8} and four mass parameters $m_i, i=1, 2, 3, 4$ are given by
\be
P_1=e^{-(m_1-a_1)}, \quad P_2 = e^{-(m_2 - m_1)}, \quad P_3 = e^{-(m_3 - a_1)} \quad P_4 = e^{-(m_4-m_3)}.\label{massSO8wflvrs}
\ee
The instanton fugacity $u_{\SO(8), 4}$ is 
\be
u_{\SO(8), 4} = Q_BP_1^2P_2P_3^2P_4Q_1. \label{instSO8wflvrs}
\ee

The comparison with the Nekrasov partition function of the $\SO(8)$ gauge theory with four flavors can be achieved by using the partition function \eqref{NekSO8wflvrs} with the maps \eqref{CoulombSO8}, \eqref{massSO8wflvrs} and \eqref{instSO8wflvrs}. Indeed we checked that our proposal agrees with the localization result \eqref{localizationSO} until the order of $Q_g^6$ for the perturbative part and the one-instanton part, and we also checked the agreement until the order of $Q_g^2$ for the two-instanton part.

\bigskip

\section{5d gauge theory with $E$-type gauge group}
\label{sec:5dE}
In section \ref{sec:partSO2Np4}, we computed the Nekrasov partition function of the 5d $\SO(2N+4)$ gauge theory with or without hypermultiplets in the vector representation by making use of the topological vertex and the gluing rule for the trivalent gauging. In fact, the technique can be applied to the calculation of the Nekrasov partition functions of the 5d pure $\E_6, \E_7$ and $\E_8$ gauge theories by using their dual descriptions \eqref{E6}, \eqref{E7} and \eqref{E8}. In this section we will obtain the Nekrasov partition functions of the 5d pure $\E_6, \E_7$ and $\E_8$ gauge theories and perform non-trivial checks with the general one-instanton result \eqref{oneinstsimplylaced}. 

\subsection{5d pure $\E_6$ gauge theory}
\label{sec:E6}
\begin{figure}[t]
\centering
\includegraphics{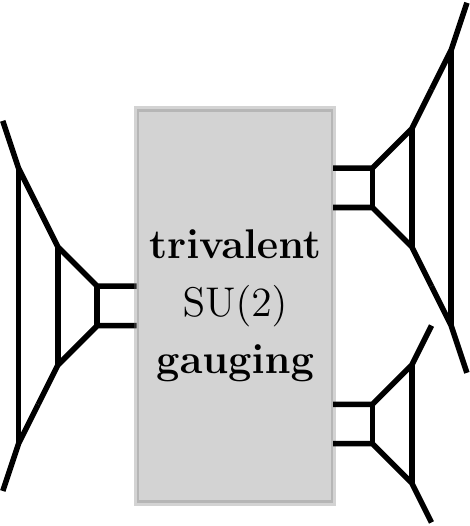}
\caption{A web-like description of the 5d theory which is dual to the pure $\E_6$ gauge theory.}
\label{fig:E6}
\end{figure}

The dual description of the 5d pure $\E_6$ gauge theory is given by \eqref{E6}. Namely the theory is realized by the $\SU(2)$ gauging of the diagonal part of the $\SU(2)$ flavor symmetries of the $\hat{D}_2(\SU(2))$theory and two $\hat{D}_3(\SU(2))$ theories. A web-like description of the dual theory is given in figure \ref{fig:E6}.

\begin{figure}[t]
\centering
\includegraphics[width=7cm]{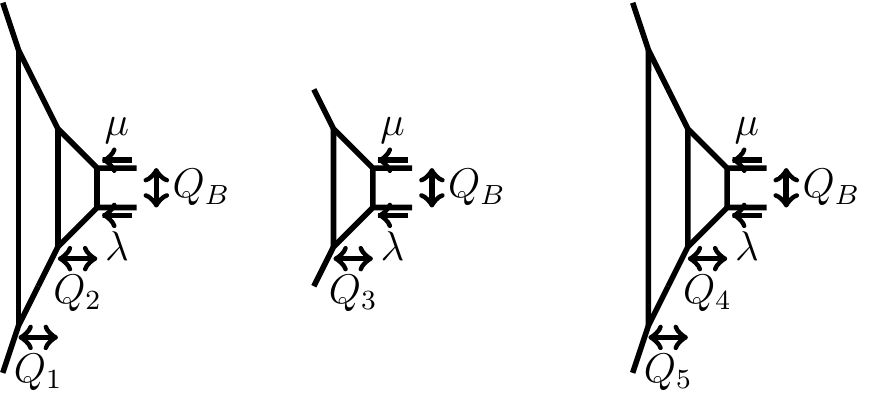}
\caption{The assignment of the K\"ahler parameters to the web in figure \ref{fig:E6}. We write the three webs separately for simplicity.}
\label{fig:E6Kahler}
\end{figure}

The partition function of the theory \eqref{E6} can be computed by using exactly the same technique obtained in section \ref{sec:partSO2Np4}. The assignment of the K\"ahler parameters to the web is summarized in figure \ref{fig:E6Kahler}.
$Q_1, Q_2, Q_3, Q_4, Q_5$ are related to the size of the spheres in the $F_{\mathfrak{e}_6}$ fiber other than $C_g$ and hence they correspond to the five Coulomb branch moduli of the pure $\E_6$ theory. In the dual frame, $Q_1, Q_2$ correspond to the Coulomb branch moduli of the $\hat{D}_3(\SU(2))$ theory coming from the leftmost web in figure \ref{fig:E6Kahler}, $Q_3$ corresponds to the Coulomb branch modulus of the $\hat{D}_2(\SU(2))$ theory from the middle web in figure \ref{fig:E6Kahler} and $Q_4, Q_5$ correspond to the Coulomb branch moduli of the $\hat{D}_3(\SU(2))$ theory coming from the rightmost web in figure \ref{fig:E6Kahler}. The other Coulomb branch moduli of the pure $\E_6$ gauge theory comes from trivalent gluing parameter $Q_g$ 
in the dual picture. $Q_B$ is the K\"ahler parameter for the size of the base $C_B$, related to the instanton fugacity of the pure $\E_6$ gauge theory. 

Since the web diagram in figure \ref{fig:E6} preserves the structure of the Dynkin diagram of $\mathfrak{e}_6$, $Q_1, Q_2, Q_3, Q_4, Q_5$ and $Q_g$ are related to the simple roots of $\mathfrak{e}_6$. Hence, we can read off the explicit Coulomb branch moduli dependence for the K\"ahler parameters as
\bea
Q_1 &=& e^{-\frac{1}{2}\left(a_1 - a_2 -a_3 - a_4 - a_5 + \sqrt{3}a_6 \right)},\quad Q_2 = e^{-(a_4 +  a_5)}, \quad Q_3 = e^{-(a_2 - a_3)},\nn\\
Q_g &=& e^{-(a_3 - a_4)}, \quad Q_4 = e^{-(a_4 - a_5)}, \quad Q_5 = e^{-\frac{1}{2}(a_1 - a_2 - a_3- a_4 + a_5 - \sqrt{3}a_6)}, \label{CoulombE6}
\eea
where $a_i, i=1, \cdots, 6$ are the Coulomb branch moduli of the pure $\E_6$ gauge theory. The instanton fugacity $u_{E_6}$ is equal to $Q_B$ up to a factor made from $Q_i, i=1, \cdots, 5$ and $Q_g$. We can determine the factor from the geometric data as done in section \ref{sec:pureSO2Np4}. Let $C_i$ be the curve whose K\"ahler parameter is given by $Q_i$ for $i=1, \cdots, 5$. We also define $S_i$ as the surfaces which has the $C_i$ fibration over $C_B$ for $i=1, \cdots, 5, g$. The intersection matrix between $S_i$ and $C_j$ inside $\tilde{X}_3$ is the negative of the Cartan matrix of $\mathfrak{e}_6$ as in table \ref{tb:intersectionE6}. On the other hand the curve $C_B$ has $-2$ intersection number only with $S_g$ but does not intersect with the other divisors. The intersection numbers are then summarized in table \ref{tb:intersectionE6}. 
\begin{table}[t]
\begin{center}
\begin{tabular}{c|ccccccc}
& $S_1$ & $S_2$ & $S_g$ & $S_4$ & $S_5$ & $S_3$\\
\hline
$C_1$ & -2 & 1 & 0 & 0 & 0 & 0\\
$C_2$ & 1 & -2 & 1 & 0 & 0 & 0 \\
$C_g$ & 0 & 1 & -2 & 1 & 0 & 1 \\
$C_4$ & 0 & 0 & 1& -2 & 1 & 0 \\
$C_5$ & 0 & 0 & 0 & 1 & -2 & 0\\
$C_3$ & 0 & 0 & 1 & 0 & 0 & -2\\
\hline
$C_B$ & 0 & 0 &  -2 & 0 & 0 & 0 & 
\end{tabular}
\caption{The matrix of the intersection numbers between the divisors $S_i, i=1, \cdots, 5, g$ and the curves $C_i, i=1, \cdots, 5, g, B$. }
\label{tb:intersectionE6}
\end{center}
\end{table}
Then the instanton fugacity $u_{\E_6}$ can be obtained by multiplying $Q_B$ by a combination of $Q_1, \cdots Q_5$ and $Q_g$ so that the intersection numbers with any surfaces $S_i, i=1, \cdots, 5, g$ vanish. This uniquely fixes the instanton fugacity as
\be
u_{\E_6} = \frac{Q_B}{Q_1^4 Q_2^8Q_g^{12} Q_4^8 Q_5^4Q_3^6}. \label{instantonE6}
\ee

The partition function of the dual theory \eqref{E6} is then given by the trivalent $SU(2)$ gauge of the partition functions of the two $\hat{D}_3(SU(3))$ theories and the $\hat{D}_2(SU(2))$ theory, 
\bea
Z_{E_6}(Q_g, Q_B, Q_{1,2,3,4,5}) &=& \sum_{\lambda, \mu}Q_g^{|\lambda| + |\mu|}Z_{\lambda, \mu}^{\SU(2)\;\text{vector}}(Q_B)Z^{\hat{D}_3(\SU(2))}_{\lambda, \mu}(Q_B, \{Q_2, Q_1\})\nn\\
&&Z^{\hat{D}_2(\SU(2))}_{\lambda, \mu}(Q_B, \{Q_3\})Z^{\hat{D}_3(\SU(2))}_{\lambda, \mu}(Q_B, \{Q_4, Q_5\}), \label{partE6}
\eea
where the K\"ahler parameters $Q_g, Q_1, \cdots, Q_5$ and $Q_B$ are related to the Coulomb branch moduli by \eqref{CoulombE6} and the instanton fugacity of the pure $E_6$ gauge theory by \eqref{instantonE6}. We checked that the partition function \eqref{partE6} perfectly agrees with the perturbative part \eqref{pert.vector} until the order of $Q_g^6$. We also checked that it matches with the known result of the $E_6$ instanton \eqref{oneinstsimplylaced} until the order of $Q_g^2$ for the one-instanton part. 

The fact that $Z_{E_6}$ is a positive power series of $Q_{1,2,3,4,5,g}$ combined with \eqref{instantonE6} alone gives non-trivial information. The one-instanton partitions function is $u_{E_6}$ times \eqref{oneinstsimplylaced}, and that should be a positive power series of $Q_{1,2,3,4,5,g}$. Therefore, for $G=E_6$, the numerator of \eqref{oneinstsimplylaced} should be proportional to $Q_1^4Q_2^8Q_g^{12}Q_4^8Q_5^4Q_3^6$, which is alone non-trivial from the form of \eqref{oneinstsimplylaced}.

\subsection{5d pure $\E_7$ gauge theory}
\begin{figure}[t]
\centering
\includegraphics[width=5cm]{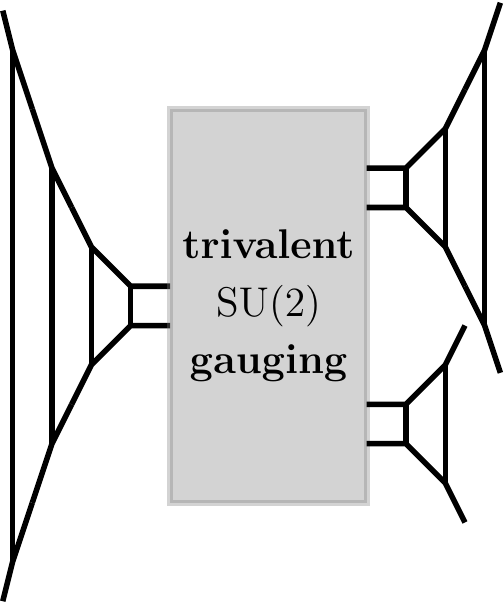}
\caption{A web-like of the 5d theory which is dual to the pure $\E_7$ gauge theory.}
\label{fig:E7}
\end{figure}

We then move onto the calculation of the partition function of the pure $\E_7$ gauge theory. The dual theory is given in \eqref{E7} and its web-like description is depicted in figure \ref{fig:E7}.
\begin{figure}[t]
\centering
\includegraphics[width=7cm]{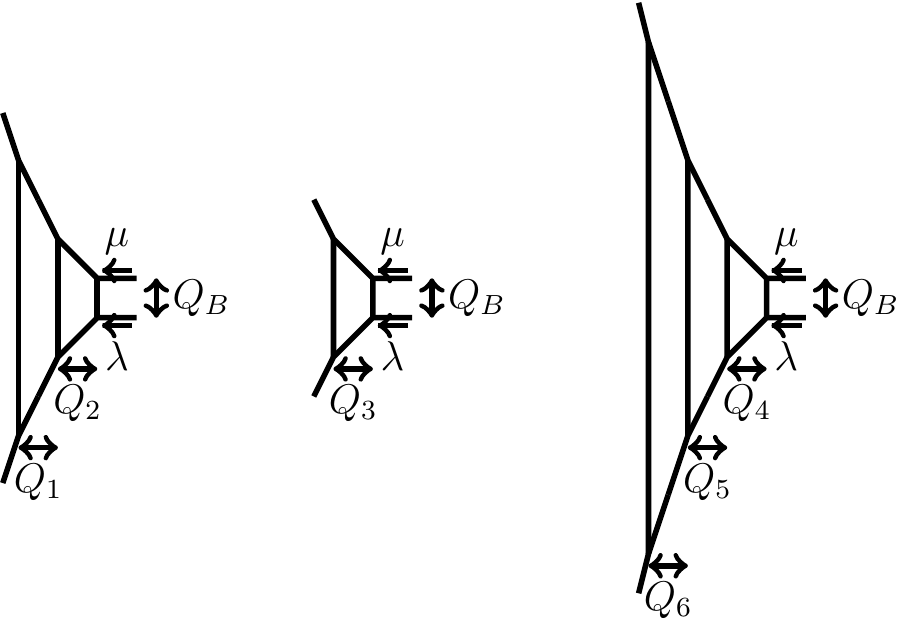}
\caption{The assignment of the K\"ahler parameters to the web in figure \ref{fig:E7}. We write the three webs separately for simplicity.}
\label{fig:E7Kahler}
\end{figure}
We then assign K\"ahler parameters to the web as in figure \ref{fig:E7Kahler}. Like the case in section \ref{sec:E6}, $Q_1, \cdots, Q_6$ and $Q_g$ correspond to the Coulomb branch moduli of the pure $\E_7$ gauge theory. Here $Q_g$ is again the K\"ahler parameter for the trivalent gluing. Since $Q_1, \cdots, Q_6$ and $Q_g$ are associated to the simple roots of $\mathfrak{e}_7$, the relations between $Q_1, \cdots, Q_6, Q_g$ and the Coulomb branch moduli $a_i, i=1, \cdots, 7$ of the pure $\E_7$ gauge theory are given by
\bea
Q_1 &=& e^{-(a_2 - a_3)}, \quad Q_2 = e^{-(a_3 - a_4)}, \quad Q_3 = e^{-(a_5 + a_6)}, \quad Q_g = e^{-(a_4 - a_5)},\nn\\
Q_4 &=& e^{-(a_5 - a_6)}, \quad Q_5 = e^{-\frac{1}{2}(a_1 - a_2 - a_3 - a_4 - a_5 + a_6 - \sqrt{2}a_7)}, \quad Q_6 = e^{-\sqrt{2}a_7}.\label{CoulombE7}
\eea

On the other hand, the instanton fugacity $u_{E_7}$ of the pure $\E_7$ gauge theory can be read off from the intersection numbers between complex surfaces and curves inside $\tilde{X}_3$. Let $C_i, i=1, \cdots, 6$ be the curves whose size is given by the K$\ddot{a}$hler parameters $Q_1, \cdots Q_6$ respectively. We also denote $S_i$ by the surface which has the $C_i$ fibration over $C_B$ for $i=1, \cdots, 6, g$. Then the intersection matrix between $S_i$ and $C_{i'}$ for $i, i'=1, \cdots, 6, g$ and also $C_B$ inside $\tilde{X}_3$ is summarized in table \ref{tb:intersectionE7}. 
\begin{table}[t]
\begin{center}
\begin{tabular}{c|cccccccc}
& $S_1$ & $S_2$ & $S_g$ & $S_4$ & $S_5$ & $S_6$ & $S_3$\\
\hline
$C_1$ & -2 & 1 & 0 & 0 & 0 & 0 & 0\\
$C_2$ & 1 & -2 & 1 & 0 & 0 & 0 & 0\\
$C_g$ & 0 & 1 & -2 & 1 & 0 & 0 & 1 \\
$C_4$ & 0 & 0 & 1& -2 & 1 & 0 & 0\\
$C_5$ & 0 & 0 & 0 & 1 & -2 & 1 & 0\\
$C_6$ & 0 & 0 & 0 & 0 & 1 & -2 & 0\\
$C_3$ & 0 & 0 & 1 & 0 & 0 & 0 & -2\\
\hline
$C_B$ & 0 & 0 & -2 & 0 & 0 & 0 & 0 
\end{tabular}
\caption{The matrix of the intersection numbers between the divisors $S_i, i=1, \cdots, 6, g$ and the curves $C_i, i=1, \cdots, 6, g, B$. }
\label{tb:intersectionE7}
\end{center}
\end{table}
The instanton fugacity $u_{\E_7}$ of the pure $\E_7$ gauge theory is equal to $Q_B$ up to a factor made from $Q_i, i=1, \cdots, 6, g$. The factor can be determined by requring that the intersection numbers with all the surfaces$S_i, i=1, \cdots, 6, g$ vanish. The condition leaves the unique choice 
\be
u_{\E_7} = \frac{Q_B}{Q_1^8Q_2^{12}Q_g^{24}Q_4^{18}Q_5^{12}Q_6^6Q_3^{16}}. \label{instantonE7}
\ee

The partition function of the pure $\E_7$ gauge theory can be calculated from the trivalent $SU(2)$ gauging of the $\hat{D}_3(\SU(2))$ matter, the $\hat{D}_2(\SU(2))$ matter and the $\hat{D}_4(\SU(2))$ matter. Hence its expression becomes 
\bea
Z_{\E_7}(Q_g, Q_B, Q_{1,2,3,4,5, 6}) &=& \sum_{\lambda, \mu}Q_g^{|\lambda| + |\mu|}Z_{\lambda, \mu}^{\SU(2)\;\text{vector}}(Q_B)Z^{\hat{D}_3(\SU(2))}_{\lambda, \mu}(Q_B, \{Q_2, Q_1\})\nn\\
&&Z^{\hat{D}_2(\SU(2))}_{\lambda, \mu}(Q_B, \{Q_3\})Z^{\hat{D}_4(\SU(2))}_{\lambda, \mu}(Q_B, \{Q_4, Q_5, Q_6\}), \label{partE7}
\eea
where the relations between the K\"ahler parameters and the Coulomb branch moduli and the instanton fugacity are given by \eqref{CoulombE7} and \eqref{instantonE7} respectively. Then We found \eqref{partE7} agrees with the perturbative partition function \eqref{pert.vector} of the pure $\E_7$ gauge theory until the order of $Q_g^6$. We also checked that the partition function \eqref{partE7} agrees with the known result of \eqref{oneinstsimplylaced} until the order of $Q_g^3$ for the one-instanton part. 
Again, \eqref{instantonE7} indicates that the numerator of \eqref{oneinstsimplylaced} is proportional to $Q_1^8Q_2^{12}Q_g^{24}Q_4^{18}Q_5^{12}Q_6^6Q_3^{16}$ for $G=E_7$.

\subsection{5d pure $\E_8$ gauge theory} 

\begin{figure}[t]
\centering
\includegraphics[width=5cm]{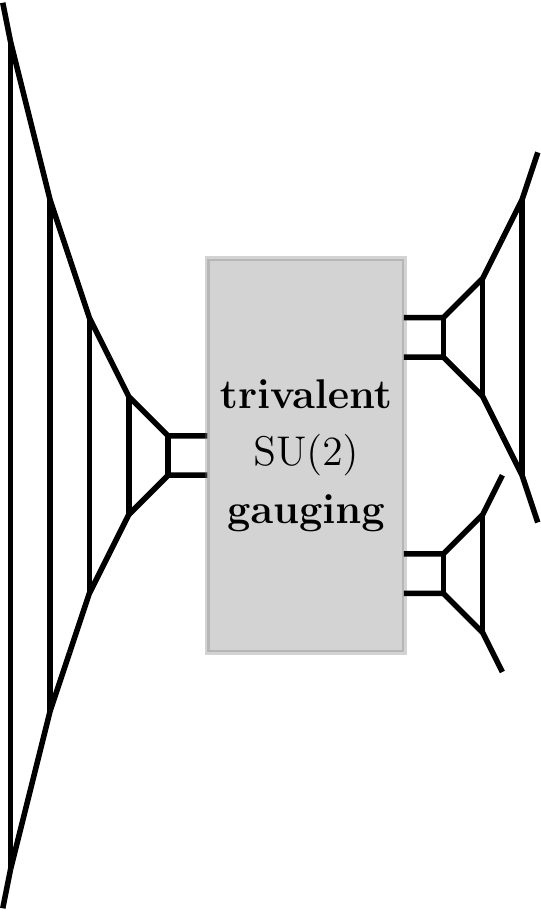}
\caption{A web-like of the 5d theory which is dual to the pure $\E_8$ gauge theory.}
\label{fig:E8}
\end{figure}

\begin{figure}[t]
\centering
\includegraphics[width=7cm]{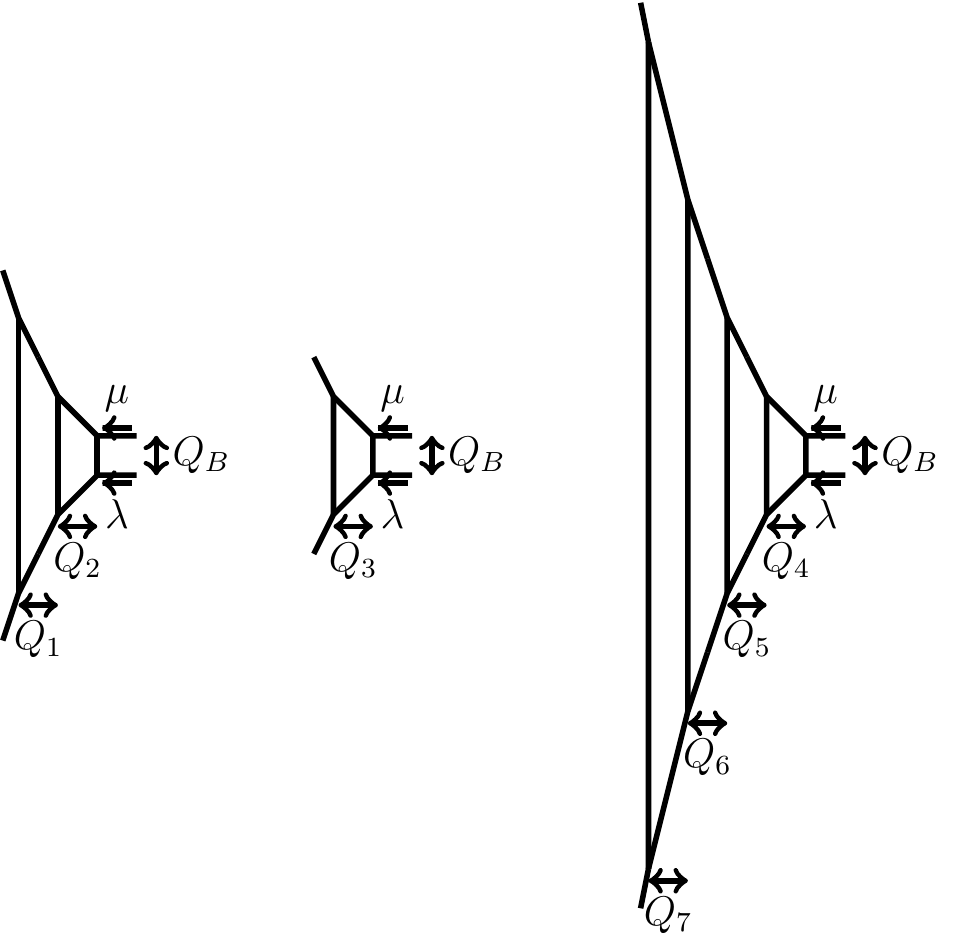}
\caption{The assignment of the K\"ahler parameters to the web in figure \ref{fig:E8}. We write the three webs separately for simplicity.}
\label{fig:E8Kahler}
\end{figure}

Finally we consider the 5d pure $E_8$ gauge theory. The dual theory is given by \eqref{E8} and its web-like is drawn in figure \ref{fig:E8}, 
The assignment of the K\"ahler parameters to the web is summarized in figure \ref{fig:E8Kahler}. 
$Q_1, \cdots, Q_7$ and $Q_g$ correspond to the simple roots of the $\mathfrak{e}_8$ Lie algebra and are related to the Coulomb branch moduli of the pure $\E_8$ gauge theory,
\bea
Q_1 &=& e^{-\frac{1}{2}(a_1 - a_2 - a_3 - a_4 - a_5 - a_6 - a_7 + a_8)}, \quad Q_2 = e^{-(a_7 - a_8)}, \quad Q_3 = e^{-(a_7 + a_8)}, \nn\\
Q_g &=& e^{-(a_6 - a_7)},\quad Q_4 = e^{-(a_5 - a_6)}, \quad Q_5 = e^{-(a_4 - a_5)}, \quad Q_6 = e^{-(a_3 - a_4)}. \quad Q_7 = e^{-(a_2 - a_3)}, \nn\\ \label{CoulombE8}
\eea
where $a_i, i=1, \cdots, 8$ are the Coulomb branch moduli of the pure $\E_8$ gauge theory. The instanton fugacity $u_{E_8}$ is again equal to $Q_B$ up to a factor consisting of $Q_1, \cdots, Q_7$ and $Q_g$. Let $C_i, i=1, \cdots, 7$ be the curves whose size is given by the K\"ahler parameters $Q_1, \cdots, Q_7$ respectively. Let $S_i$ be the surface which has the $C_i$ fiber over $C_B$ for $i=1, \cdots, 7, g$. Then the intersection matrix is given by the numbers summarized in table \ref{tb:intersectionE8}. 
\begin{table}[t]
\begin{center}
\begin{tabular}{c|ccccccccc}
& $S_1$ & $S_2$ & $S_g$ & $S_4$ & $S_5$ &$S_6$ & $S_7$ &  $S_3$\\
\hline
$C_1$ & -2 & 1 & 0 & 0 & 0 & 0 & 0 & 0\\
$C_2$ & 1 & -2 & 1 & 0 & 0 & 0 & 0 & 0\\
$C_g$ & 0 & 1 & -2 & 1 & 0 & 0 & 0 & 1 \\
$C_4$ & 0 & 0 & 1& -2 & 1 & 0 & 0 & 0\\
$C_5$ & 0 & 0 & 0 & 1 & -2 & 1 & 0 & 0\\
$C_6$ & 0 & 0 & 0 & 0 & 1 & -2 & 1 & 0\\
$C_7$ & 0 & 0 & 0 & 0 & 0 & 1 & -2 & 0\\
$C_3$ & 0 & 0 & 1 & 0 & 0 & 0 & 0 & -2\\
\hline
$C_B$ & 0 & 0 & -2 & 0 & 0 & 0 & 0  & 0
\end{tabular}
\caption{The matrix of the intersection numbers between the divisors $S_i, i=1, \cdots, 7, g$ and the curves $C_i, i=1, \cdots, 7, g, B$. }
\label{tb:intersectionE8}
\end{center}
\end{table}
The instanton fugacity $u_{\E_8}$ of the pure $\E_8$ gauge theory is then
\be
u_{\E_8} = \frac{Q_B}{Q_1^{20}Q_2^{40}Q_g^{60}Q_4^{48}Q_5^{36}Q_6^{24}Q_7^{12}Q_3^{30}}. \label{instantonE8}
\ee

From the dual description \eqref{E6}, the Nekrasov partition function of the pure $\E_8$ gauge theory is given by the trivalent $\SU(2)$ gauging of the $\hat{D}_3(\SU(2))$ matter, the $\hat{D}_2(\SU(2))$  matter and the $\hat{D}_5(\SU(2))$ matter. Therefore, the partition function can be written by
\bea
Z_{E_8}(Q_g, Q_B, Q_{1,2,3,4,5,6, 7}) &=& \sum_{\lambda, \mu}Q_g^{|\lambda| + |\mu|}Z_{\lambda, \mu}^{\SU(2)\;\text{vector}}(Q_B)Z^{\hat{D}_3(\SU(2))}_{\lambda, \mu}(Q_B, \{Q_2, Q_1\})\nn\\
&&Z^{\hat{D}_2(\SU(2))}_{\lambda, \mu}(Q_B, \{Q_3\})Z^{\hat{D}_4(\SU(2))}_{\lambda, \mu}(Q_B, \{Q_4, Q_5, Q_6, Q_7\}). \nn\\\label{partE8}
\eea
The relations between the K\"ahler parameters $Q_1, \cdots, Q_7, Q_g, Q_B$ and the Coulomb branch moduli and the instanton fugacity of the pure $\E_8$ gauge theory are \eqref{CoulombE8} and \eqref{instantonE8}. Then We found \eqref{partE8} agrees with the perturbative partition function \eqref{pert.vector} of the pure $\E_8$ gauge theory until the order of $Q_g^6$.

In principle, we can also compare the partition function \eqref{partE8} with the parameterization \eqref{CoulombE8} and \eqref{instantonE8} with the general result of the one--instanton \eqref{oneinstsimplylaced}.
However, because of the same reasoning explained of $E_{6,7}$ case, the numerator of \eqref{oneinstsimplylaced} is non-trivially proportional to the denominator of \eqref{instantonE8}, which is a very high power.
This means that one needs to expand the expression \eqref{oneinstsimplylaced} to the order $60$th in $Q_g$ for the comparison. This is computationally difficult and hence we performed the comparison by inserting some specific values to $Q_1, Q_2, Q_3, Q_4, Q_5, Q_6, Q_7$ and found the agreement until the order of $Q_g^1$ for the one--instanton part. When we take $Q_2$ generic, then we checked the agreement until the order of $Q_g^0$.

\bigskip

\section{A 5d description of non-Higgsable clusters}
\label{sec:6dminimal}
So far we have considered 5d theories which are dual to the 5d gauge theories with a gauge group of type $G=\D_{N+2}, \E_6, \E_7, \E_8$ by utilizing the trivalent $\SU(2)$ gauging.
In this section, we further make use of the trivalent gauging and construct 5d theories given by a circle compactification of certain 6d SCFTs called non-Higgsable clusters \cite{Morrison:2012np, Heckman:2013pva}.
Non-Higgsable cluster theories with one tensor multiplet on a tensorial Coulomb branch are called 6d minimal SCFTs. We will mainly focus on these examples and also comment on another non-Higgsable cluster in the last subsection. 

6d minimal SCFTs can be obtained from an F-theory compactification on a Calabi-Yau threefold $X_3$ which has an elliptic fibration over the Hirzebruch surface $\mathbb{F}_n$ with $n=1, \cdots, 8, 12$ \cite{Morrison:1996na, Morrison:1996pp}.
We may take the field theory limit by sending the size of the fiber $\mathbb{P}^1_F$ in $\mathbb{F}_n$ to infinite.
Then the Calabi-Yau threefold $X_3$ becomes non-compact and the non-compact direction is given by a line bundle $\mathcal{O}(-n)$ over the base $\mathbb{P}^1_B$.
The low energy effective field theory is a 6d minimal SCFT with one tensor multiplet on a tensorial Coulomb branch, and we denote it by $\mathcal{O}(-n)$ model.
When $n \geq 3$, the 6d theories preserve eight supercharges and have no flavor symmetry, and they are in a class of non-Higgsable clusters. 
The 6d minimal SCFTs are important building blocks to construct more general 6d SCFTs \cite{Heckman:2013pva, DelZotto:2014hpa, Heckman:2015bfa}.

In fact, the geometry of some of the $\mathcal{O}(-n)$ models has an orbifold limit \cite{Witten:1996qb} given by $(T^2\times\mathbb{C}^2)/\Gamma$ where the orbifold action is
\be
g = (\omega^2; \omega^{-1}, \omega^{-1}) \label{6dorbifold}
\ee
with $\omega^{n} = 1$.
Here $n$ should be restricted to $n=2, 3, 4, 6, 8, 12$ so that the orbifold action consistently acts on the torus.
In \eqref{6dorbifold}, the first component acts on the complex coordinate of $T^2$ and the other two components act on the two complex coordinates of $\mathbb{C}^2$.

The case of $n=2$ is special since it corresponds to 6d $\mathcal{N}= (2, 0)$ SCFT of $A_1$ type. The self-dual strings of the theory are called M-strings \cite{Haghighat:2013gba, Haghighat:2013tka}. On the other hand the other cases of $n=3, 4, 6, 8, 12$ yield 6d $\mathcal{N}=(1, 0)$ SCFTs and hence we will focus on these cases.

A 5d description from a circle compactification of the 6d minimal SCFTs for $n=4, 6, 8, 12$ has been obtained in \cite{DelZotto:2015rca}.
In fact, the 5d theory also exhibits the structure of the $\SU(2)$ gauging of three or four non-Lagrangian theories.
The fact that the similar structure appears to the cases of the dual theories of the pure gauge theories of $\D_{N+2}, \E_6, \E_7, \E_8$-type is not coincidence.
Let $r_{S^1}$ be the radius of the $S^1$ for the circle compactification and consider a limit where $r_{S^1} \rightarrow 0$.
Then the limit makes the Kaluza--Klein modes associated to the circle compactification decouple and leads to a 5d $\mathcal{N}=1$ theory with a 5d UV completion. In fact, the $\mathcal{O}(-n)$ models with $n=4, 6, 8, 12$ reduce to the 5d pure $\SO(8), \E_6, \E_7, \E_8$ gauge theories in the limit respectively. Hence, the 5d description of the 6d minimal models in the cases of $n=4. 6, 8, 12$ should be closely related to the dual description of the 5d pure gauge theories of $\D_{N+2}, \E_6, \E_7, \E_8$-type after taking the limit. Since we have developed the technique of the trivalent gluing for computing the partition functions of the pure gauge theories with a gauge group $G=\SO(2N+4), \E_6, \E_7, \E_8$ in section \ref{sec:gluing}, it is also possible to apply the method to the 5d description of the 6d minimal SCFTs. 

Furthermore, we will also propose a 5d description from a circle compactification of the $\mathcal{O}(-3)$ model and compute the partition function of the 5d theory. The comparison with the elliptic genus of the $\mathcal{O}(-3)$ model obtained in \cite{Kim:2016foj} gives non-trivial support for the 5d description as well as the trivalent gluing rule. 

One of other non-Higssable clusters can be also realized by an orbifold construction \cite{DelZotto:2015rca} and we will propose its 5d description.

\subsection{$\mathcal{O}(-4)$ model}

We first consider a 5d description of the $\mathcal{O}(-4)$ model on a circle.
The 6d theory is the $\SO(8)$ gauge theory with no flavor symmetry accompanied by one tensor multiplet on its tensorial Coulomb branch.
The orbifold geometry for the $\mathcal{O}(-4)$ model is $X_3 = (T^2\times\mathbb{C}^2)/\Gamma$ where the orbifold action $\Gamma$ is
\be
g = (\omega^2; \omega^{-1}, \omega^{-1}), \quad \text{with}\;\; \omega^4 = 1. \label{6dorbifoldOm4}
\ee
F-theory compactification on $X_3 \times S^1$ is dual to an M-theory compactification on $X_3$ \cite{Vafa:1996xn}. Hence, a 5d description of the $\mathcal{O}(-4)$ model is given by a low energy effective field theory from M-theory on $X_3 = (T^2\times\mathbb{C}^2)/\Gamma$ with the orbifold action \eqref{6dorbifoldOm4}.

Let us review the 5d construction in \cite{DelZotto:2015rca}. First, since $g^2 = (1; -1, -1)$, the orbifold yields an $A_1$ singularity over the torus, leading to an $\SU(2)$ gauge symmetry. Along the torus direction, the orbifold reduces to $\mathbb{Z}_2$. Hence the torus becomes a sphere $C_g$ with four fixed points. 
Around each fixed point, the orbifold becomes $\mathbb{C}^3/\Gamma$ where the orbifold action is the same as \eqref{6dorbifoldOm4}. This is exactly the same orbifold geometry considered in section \ref{sec:5dSO2Np4} and yields the 5d SCFT, $\hat{D}_2(\SU(2))$. Therefore, the 5d description of the $\mathcal{O}(-4)$ model is given by the $SU(2)$ gauging of four $\hat{D}_2(\SU(2))$ theories,
\begin{align}
  {\overset{\overset{\overset{\overset{\text{\large$\hat{D}_2(\SU(2))$}}{\textstyle\vert}}{\text{\large $\hat{D}_2(\SU(2)) -$SU(2)$-\hat{D}_2(\SU(2))$}}}{\textstyle\vert}}{\text{\large$\hat{D}_2(\SU(2))$}}}\label{Om4}
\end{align}
Note that when one sends a Coulomb branch modulus of one of the $\hat{D}_2(\SU(2))$ theories to infinite, the 5d description reduces to \eqref{SO2Np4} with $N=2$, namely it is a dual description of the pure $\SO(8)$ gauge theory.
This is consistent with the fact that the $r_{S^1} \rightarrow 0$ limit of the $\mathcal{O}(-4)$ models yields the 5d pure $\SO(8)$ gauge theory.
In terms of the geometry, \eqref{Om4} comes from a surface which has a fiber consisting of a collection of spheres whose shape is the affine Dynkin diagram of $\mathfrak{so}(8)$.
The limit reduces the affine Dynkin diagram of $\mathfrak{so}(8)$ to the Dynkin diagram of $\mathfrak{so}(8)$ and the 5d theory reduces to \eqref{SO2Np4} with $N=2$ from \eqref{Om4}. 

Let us also see the number of 5d gauge theory parameters can be reproduced from a circle compactification of the $\mathcal{O}(-4)$ model. The $\mathcal{O}(-4)$ model has four vector multiplets in the Cartan subalgebra of $\SO(8)$ and one tensor multiplet. After the circle compactification, both become 5d vector multiplets in the Cartan subalgebra Hence the number of the Coulomb branch moduli in the 5d theory should be five. Indeed we have five Coulomb branch moduli from the 5d theory \eqref{Om4}. One comes from the $\SU(2)$ gauging and four come from the four rank one $\hat{D}_2(\SU(2))$ theories. Since the 6d theory has no flavor symmetry, the 5d theory should has only one mass parameter originating from the radius of the compactification circle. From the 5d description \eqref{Om4}, we have only one mass parameter associated to the gauge coupling of the middle $\SU(2)$ gauging. 

We then compute the partition function of the 5d theory \eqref{Om4}. The gluing procedure is essentially the same as the one for the trivalent gluing even when we gauge four copies of the $\hat{D}_2(\SU(2))$ matter. The partition function is given by
\bea
Z_{\mathcal{O}(-4)}(Q_g, Q_B, Q_{1,2,3,4}) &=& \sum_{\lambda, \mu}Q_g^{|\lambda| + |\mu|}Z_{\lambda, \mu}^{\SU(2)\;\text{vector}}(Q_B)\nn\\
&&Z^{\hat{D}_2(\SU(2))}_{\lambda, \mu}(Q_B, \{Q_1\})Z^{\hat{D}_2(\SU(2))}_{\lambda, \mu}(Q_B, \{Q_2\})\nn\\
&&Z^{\hat{D}_2(\SU(2))}_{\lambda, \mu}(Q_B, \{Q_3\})Z^{\hat{D}_2(\SU(2))}_{\lambda, \mu}(Q_B, \{Q_4\}). \label{partOm4}
\eea
$Q_1, \cdots, Q_4$ and $Q_g$ correspond to the five Coulomb branch moduli and $Q_B$ is related to the instanton fugacity of the $SU(2)$ gauging. 

Since Eq.~\eqref{partOm4} is the Nekrasov partition function of the 5d description \eqref{Om4} of the 6d $\mathcal{O}(-4)$ model on a circle, it should also agree with the elliptic genus of the self-dual strings of the $\mathcal{O}(-4)$ model. The full elliptic genus is in general given by the sum of the elliptic genus of $k$ strings $Z_k$,
\be
Z_{\text{elliptic}}(Q, Q_{\tau}) = Z_0(Q, Q_{\tau})\left(1+\sum_{k=1}^{\infty}Z_k(Q, Q_{\tau})Q_s^k\right)
\ee
where $Q_s$ is the string fugacity counting the self--dual strings and $Q_{\tau}$ is $Q_{\tau} = e^{2\pi i \tau}$ with the complex structure modulus $\tau$ of a torus for the elliptic genus computation which is given by localization of a two-dimensional theory on a torus.
$Q$ represents the other fugacities.
The elliptic genus of the $\mathcal{O}(-4)$ model has been computed in \cite{Haghighat:2014vxa} and the $1$ string contribution is given by\footnote{We put an overall minus sign compared to \cite{Haghighat:2014vxa}. This sign is needed so that in the 5d limit the partition function reduces to the 5d $\SO(8)$ Nekrasov Partition function written in \eqref{localizationSO}. }
\bea
&&Z_1^{\mathcal{O}(-4)}(Q_{m_1}, Q_{m_2}, Q_{m_3}, Q_{m_4}, Q_{\tau}) \nn\\
&=& -\frac{1}{2}\frac{\eta^2}{\theta(q)\theta(t^{-1})}\sum_{i=1}^4\left[\frac{\theta\left(\frac{q}{t}Q_{m_i}^2\right)\theta\left(\frac{q^2}{t^2}Q_{m_i}^2\right)}{\eta^2}\prod_{j\neq i}\prod_{s=\pm 1}\frac{\eta^2}{\theta(Q_{m_i}Q_{m_j}^s)\theta\left(\frac{q}{t}Q_{m_i}Q_{m_j}^s\right)} + (Q_{m_i} \rightarrow Q_{m_i}^{-1})\right], \nn\\ \label{elliOm4}
\eea
where $t = q$ for the unrefined case. $Q_{m_i}, i=1, 2, 3, 4$ are the fugacities for the $\SO(8)$ symmetry. $\eta$ is the Dedekind eta function and $\theta(Q)$ is an elliptic theta function, which are defined by
\bea
\eta &=& Q_{\tau}^{\frac{1}{24}}\prod_{n=1}^{\infty}\left(1-Q_{\tau}^n\right),\\
\theta(Q) &=& \theta_1\left(Q_{\tau}, Q\right) = -iQ_{\tau}^{\frac{1}{8}}Q^{\frac{1}{2}}\prod_{n=1}^{\infty}\left(1-Q_{\tau}^n\right)\left(1-Q_{\tau}^nQ\right)\left(1-Q^{n-1}Q^{-1}\right).
\eea

In order to compare \eqref{elliOm4} with \eqref{partOm4}, one needs to find a map between the parameters.
The map has been also worked out in \cite{Haghighat:2014vxa}, and we reproduce a part of it since we can apply it to other cases.
The K$\ddot{\text{a}}$hler parameters $Q_1, Q_2, Q_3, Q_4, Q_g$ are related to the size of spheres which form the affine Dynkin diagram of $\mathfrak{so}(8)$. Shrinking spheres which form the shape of the $\mathfrak{so}(8)$ Dynkin diagram leads to the $\SO(8)$ gauge symmetry in the 6d theory. Therefore, the K$\ddot{\text{a}}$hler parameters for the shrunken spheres are related to $Q_{m_i}, i=1, \cdots, 4$ and we for example choose $Q_1, Q_2, Q_3, Q_g$ for giving the 6d $SO(8)$ gauge symmetry. Note that $Q_1, Q_2, Q_3, Q_g$ correspond to the simple roots of $\mathfrak{so}(8)$ whereas $Q_{m_i}, i=1, \cdots ,4$ take values at the Cartan subalgebra of $\mathfrak{so}(8)$. Therefore their relations are 
\bea
Q_1 = Q_{m_1}Q_{m_2}^{-1}, \quad Q_g = Q_{m_2}Q_{m_3}^{-1}, \quad Q_2 = Q_{m_3}Q_{m_4}^{-1}, \quad Q_3= Q_{m_3}Q_{m_4}. \label{CoulombOm4}
\eea
Furthermore, $Q_{\tau}$ can be written by $Q_{\tau} = \prod_{i=1,2,3,4,g}Q_i^{c_i}$ where $c_i$ is the comark associated to a simple root of $\mathfrak{so}(8)$ and $c_4=1$ for the extended node. Therefore, we obtain
\be
Q_{\tau} = Q_1Q_2Q_g^2Q_3Q_4. \label{tauOm4}
\ee
The final parameter which we need to identify is the string fugacity $Q_s$ which counts the self--dual strings in the elliptic genus calculation. Since the self-dual strings arise from D3-branes wrapping the base $\mathbb{P}^1_B$, it should be related to $Q_B$ by
\be
Q_s = Q_B h(Q_1, Q_2, Q_3, Q_4, Q_g).
\ee
One can restrict the explicit form of $h(Q_1, Q_2, Q_3, Q_4, Q_g)$ by requiring that the string fugacity has no Coulomb branch moduli dependence.
As in section \eqref{sec:pureSO2Np4}, the curve $\mathbb{P}^1_B$ has nonzero intersection number $-2$ with $S_g$ which has the $C_g$ fibration over $\mathbb{P}^1_B$.
Then, we can deduce that  
\be
Q_s = Q_B\frac{Q_4^a}{(Q_1Q_2Q_3Q_g^2)^b}
\ee
with $a+b=2$.
The precise value of $a, b$ cannot be determined from the conditions so far but we may determine it by the explicit comparison between \eqref{partOm4} and \eqref{elliOm4}.
In fact, it should be easy to determine $a, b$ since we can just observe the overall rescaling difference between \eqref{partOm4} and \eqref{elliOm4} at the order $\mathcal{O}(Q_B^1)$. We here simply quote the result of \cite{Haghighat:2014vxa}, 
\be
Q_s = Q_B\frac{Q_4}{Q_1Q_2Q_3Q_g^2}. \label{stringOm4}
\ee

With the relations \eqref{CoulombOm4}, \eqref{tauOm4} and \eqref{stringOm4}, one can perform the explicit comparison of \eqref{partOm4} with \eqref{elliOm4}.
Since \eqref{elliOm4} is the one--string contribution, we can use the one-instanton result of \eqref{partOm4}.
Furthermore, the partition function \eqref{partOm4} is expanded by $Q_g$ and hence we need to expand \eqref{elliOm4} by $Q_g$ for the comparison.
Although it would be in principle possible to perform the comparison by the double expansion in $Q_B$ and $Q_g$, we need an exact expression for $Q_{\tau}$. In order to use the truncated form of the elliptic theta function, we further expand the both equations by $Q_4$ which appears only in $Q_{\tau}$, not in $Q_{m_i}, i=1,2,3,4$. Then we have found the complete agreement between the two results until the order of $Q_g^2Q_4^3$ for the one-string part. Hence, the gluing rule indeed works for the case when the 5d theory has a 6d UV completion.

\subsection{$\mathcal{O}(-n)$ model with $n=6,  8, 12$}

The analysis for the $\mathcal{O}(-n)$ models with $n=6, 8, 12$ is parallel to the case of the $\mathcal{O}(-4)$ model.
The elliptic genus of one--string for the $\mathcal{O}(-6)$ model has been computed in \cite{Gadde:2015xta, Putrov:2015jpa, DelZotto:2016pvm}.
Furthermore, Ref.~\cite{DelZotto:2016pvm} also computed the elliptic genus of one--string for the $\mathcal{O}(-n)$ model with $n=8, 12$. The calculation in terms of the BPS invariants from the Calabi--Yau geometry by using the mirror symmetry has been done in \cite{Haghighat:2014vxa}. We here present another approach to compute the elliptic genus of the $\mathcal{O}(-n)$ models with $n=6, 8, 12$ from the Nekrasov partition functions of their 5d descriptions. 
Although comparisons between the result here and in the literature is possible in principle, it is technically difficult because unrefined limit $t\to q$ we take is not compatible with the results in the literature, at least in a naive way.
Thus, unfortunately, we do not provide comparison of the results for $\mathcal{O}(-n)$ theory with $n=6,8,12$.

\paragraph{$\mathcal{O}(-6)$ model} When $n=6$, the F-theory geometry is given by $(T^2 \times \mathbb{C}^2)/\Gamma$ where the orbifold $\Gamma$ is given by 
\be
g = (\omega^2; \omega^{-1}, \omega^{-1}), \quad \text{with}\;\; \omega^6 = 1. \label{6dorbifoldOm6}
\ee
Then a 5d description of the $\mathcal{O}(-6)$ model is obtained by considering M-theory on the same background geometry. Since $g^3=(1; -1, -1)$, we again have an $A_1$ singularity over the torus, leading to an $\SU(2)$ gauge symmetry. In the torus direction the orbifold action is $\mathbb{Z}_3$ and the torus becomes a sphere with three $\mathbb{Z}_3$ fixed points. Around each fixed point, the geometry becomes $\mathbb{C}^3/\Gamma$ with $\Gamma$ given by \eqref{6dorbifoldOm6}. Hence each fixed point gives rises to the 5d $\hat{D}_3(\SU(2))$ theory, and the three $\hat{D}_3(\SU(2))$ theories are coupled by the $\SU(2)$ gauging. In summary, the 5d description of the $\mathcal{O}(-6)$ model is given by the trivalent $\SU(2)$ gauging of the three $\hat{D}_3(\SU(2))$ theories,
\begin{align}
\hat{D}_3(\SU(2)) - {\overset{\overset{\text{\large$\hat{D}_3(\SU(2))$}}{\textstyle\vert}}{\SU(2)}} - \hat{D}_3(\SU(2)) \label{Om6}
\end{align}
Note that sending one Coulomb branch modulus of one of the $\hat{D}_3(\SU(2))$ theories to infinity reproduces the 5d description of the pure $\E_6$ theory given by \eqref{E6}. 

We can again see the number of 5 gauge theory parameters can be reproduced from a circle compactification of the 6d $\mathcal{O}(-6)$ model. The 6d $\mathcal{O}(-6)$ model has six vector multiplets in the Cartan subalgebra and one tensor multiplet. Hence we should have $6+1$ vector multiplets in the Cartan subalgebra in 5d after a circle compactification. Indeed the $\SU(2)$ gauging provides one Coulomb branch moduli and each of the three $\hat{D}_3(\SU(2))$ gives two Coulomb branch moduli, leading to the seven-dimensional Coulomb branch moduli space. Since the 6d theory does not have any flavor symmetry, we expect one mass parameter in 5d. This agrees with the one instanton fugacity from the trivalent $\SU(2)$ gauging. 

The computation of the partition function is straightforward by using the trivalent gauging as well as the partition function of the $\hat{D}_3(\SU(2))$ theories. The proposed partition function is then
\bea
Z_{\mathcal{O}(-6)}(Q_g, Q_B, Q_{1,2,3,4,5, 6}) &=& \sum_{\lambda, \mu}Q_g^{|\lambda| + |\mu|}Z_{\lambda, \mu}^{\SU(2)\;\text{vector}}(Q_B)Z^{\hat{D}_3(\SU(2))}_{\lambda, \mu}(Q_B, \{Q_2, Q_1\})\nn\\
&&Z^{\hat{D}_3(\SU(2))}_{\lambda, \mu}(Q_B, \{Q_3, Q_4\})Z^{\hat{D}_3(\SU(2))}_{\lambda, \mu}(Q_B, \{Q_5, Q_6\}), \nn\\\label{partOm6}
\eea

For the relation to the elliptic genus of the $\mathcal{O}(-6)$ model, $Q_{\tau}$ can be written by $Q_{\tau} = \prod_{i=1,2,3,4,5,6,g}Q_i^{c_i}$ where $c_i$ is the comark associated to a simple root of $\mathfrak{e}_6$ and $c_i = 1$ for the extended node. Hence we get
\be
Q_{\tau} = Q_1Q_2^2Q_3^2Q_4Q_5^2Q_6Q_g^3.
\ee 
The string fugacity $Q_s$ is again proportional to $Q_B$. 

\paragraph{$\mathcal{O}(-8)$ model} When $n=8$, the F-theory geometry is an orbifold $(T^2 \times \mathbb{C}^2)/\Gamma$ where the orbifold $\Gamma$ is given by 
\be
g = (\omega^2; \omega^{-1}, \omega^{-1}), \quad \text{with}\;\; \omega^8 = 1. \label{6dorbifoldOm8}
\ee
A 5d description of the $\mathcal{O}(-8)$ model is realized by M-theory on the same background geometry. Since $g^4=(1; -1, -1)$, there is an $A_1$ singularity over the torus. On the torus, we have a $\mathbb{Z}_4$ orbifold which induces a sphere with one $\mathbb{Z}_2$ fixed point and two $\mathbb{Z}_4$ fixed points. The $\mathbb{Z}_2$ fixed point gives rise to the 5d $\hat{D}_2(\SU(2))$ theory and each of the two $\mathbb{Z}_4$ fixed points yields the 5d $\hat{D}_4(\SU(2))$ theory. Therefore, the 5d description of the $\mathcal{O}(-8)$ model is 
\begin{align}
\hat{D}_4(\SU(2)) - {\overset{\overset{\text{\large$\hat{D}_2(\SU(2))$}}{\textstyle\vert}}{\SU(2)}} - \hat{D}_4(\SU(2)) \label{Om8}
\end{align}
Again sending one Coulomb branch modulus in one of the $\hat{D}_4(\SU(2))$ theories yields the 5d theory in \eqref{E7}, which is dual to the 5d pure $\E_7$ gauge theory.

The partition function of the 5d theory is then given by the trivalent $\SU(2)$ gauging of the partition functions of one $\hat{D}_2(\SU(2))$ theory and two $\hat{D}_4(\SU(2))$ theories, and it is given by 
\bea
Z_{\mathcal{O}(-8)}(Q_g, Q_B, Q_{1,2,3,4,5,6,7}) &=& \sum_{\lambda, \mu}Q_g^{|\lambda| + |\mu|}Z_{\lambda, \mu}^{\SU(2)\;\text{vector}}(Q_B)Z^{\hat{D}_4(\SU(2))}_{\lambda, \mu}(Q_B, \{Q_3, Q_2, Q_1\})\nn\\
&&Z^{\hat{D}_2(\SU(2))}_{\lambda, \mu}(Q_B, \{Q_4\})Z^{\hat{D}_4(\SU(2))}_{\lambda, \mu}(Q_B, \{Q_5, Q_6, Q_7\}), \nn\\\label{partOm8}
\eea
The relation to the complex structure modulus of the torus for the elliptic genus calculation is 
\be
Q_{\tau} = Q_1Q_2^2Q_3^3Q_4^2Q_5^3Q_6^2Q_7Q_g^4.
\ee 

\paragraph{$\mathcal{O}(-12)$ model} Finally we turn to the case of $n=-12$. The F-theory geometry is is given by $(T^2 \times \mathbb{C}^2)/\Gamma$ where the orbifold $\Gamma$ is given by 
\be
g = (\omega^2; \omega^{-1}, \omega^{-1}), \quad \text{with}\;\; \omega^{12} = 1. \label{6dorbifoldOm12}
\ee
For a 5d description of the $\mathcal{O}(-12)$ model, we consider M-theory on the same orbifold background. Since $g^6 = (1; -1, -1)$, we have an $A_1$ singularity on the torus. On the torus, the orbifold action is $\mathbb{Z}_6$ which gives rise to a sphere with a $\mathbb{Z}_2$ fixed point, a $\mathbb{Z}_3$ fixed point and also a $\mathbb{Z}_6$ fixed point. Each fixed point is associated to the $\hat{D}_2(\SU(2))$ theory, the $\hat{D}_3(\SU(2))$ theory, and the $\hat{D}_6(\SU(2))$ theory respectively. Therefore, the 5d description of the $\mathcal{O}(-12)$ model is
\begin{align}
\hat{D}_6(\SU(2)) - {\overset{\overset{\text{\large$\hat{D}_2(\SU(2))$}}{\textstyle\vert}}{\SU(2)}} - \hat{D}_3(\SU(2)) \label{Om12}
\end{align}
Again when we send one Coulomb branch modulus of the $\hat{D}_6(\SU(2))$ theory, we recover the 5d theory \eqref{E8} which is dual to the 5d pure $E_8$ gauge theory.

The partition function of the 5d theory \eqref{Om12} is then calculated by the trivalent $\SU(2)$ gauging of the partition functions of the $\hat{D}_2(\SU(2))$ theory, the $\hat{D}_3(\SU(2))$ theory and the $\hat{D}_6(\SU(2))$ theory, and we propose that
\bea
Z_{\mathcal{O}(-12)}(Q_g, Q_B, Q_{1,2,3,4,5,6,7,8}) &=& \sum_{\lambda, \mu}Q_g^{|\lambda| + |\mu|}Z_{\lambda, \mu}^{\SU(2)\;\text{vector}}(Q_B)Z^{\hat{D}_3(\SU(2))}_{\lambda, \mu}(Q_B, \{Q_2, Q_1\})\nn\\
&&Z^{\hat{D}_2(\SU(2))}_{\lambda, \mu}(Q_B, \{Q_3\})Z^{\hat{D}_6(\SU(2))}_{\lambda, \mu}(Q_B, \{Q_4, Q_5, Q_6, Q_7, Q_8\}), \nn\\\label{partOm12}
\eea
The relation to the complex structure modulus of the torus for the elliptic genus calculation is 
\be
Q_{\tau} = Q_1^2Q_2^4Q_3^3Q_4^5Q_5^4Q_6^3Q_7^2Q_8Q_g^6.
\ee

\subsection{$\mathcal{O}(-3)$ model}
\label{sec:Om3}

So far we have considered a 5d description of the 6d $\mathcal{O}(-n)$ models with $n=4, 6, 8, 12$ and all of them are described by the $\SU(2)$ gauging of three or four non-Lagrangian theories of type $\hat{D}_p(\SU(2))$. We here consider a 5d description of the $\mathcal{O}(-3)$ model, which has a slight difference from the other cases.

The F-theory geometry of the 6d $\mathcal{O}(-3)$ model is given by an orbifold  $(T^2 \times \mathbb{C}^2)/\Gamma$ where the orbifold $\Gamma$ is given by 
\be
g = (\omega^2; \omega^{-1}, \omega^{-1}), \quad \text{with}\;\; \omega^3 = 1. \label{6dorbifoldOm3}
\ee
Again a 5d description of the $\mathcal{O}(-3)$ theory is obtained from M-theory on the same orbifold. One difference from the other cases is that we do not have an $A_1$ singularity or other singularity which exists over the torus. However the $\mathbb{Z}_3$ action acts on the torus and it becomes a sphere $C_g$ with three $\mathbb{Z}_3$ fixed points. 
Around each fixed point the geometry becomes $\mathbb{C}^3/\mathbb{Z}_3$ and the resolved geometry is a local $\mathbb{P}^2$ Calabi--Yau threefold. 
The 5d SCFT obtained from the fixed point by shrinking the $\mathbb{P}^2$ is called $E_0$ theory \cite{Morrison:1996xf}. The $E_0$ theory has one Coulomb branch modulus but no other parameter. Therefore, a 5d description of the $\mathcal{O}(-3)$ consists of three $E_0$ theories coupled to each other. Note that since the $E_0$ theory does not have a flavor symmetry we cannot couple them by gauging flavor symmetries. 

\begin{figure}[t]
\centering
\includegraphics{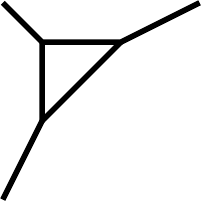}
\caption{A 5-brane web for the $E_0$ theory.}
\label{fig:localP2}
\end{figure}

\begin{figure}[t]
\centering
\includegraphics{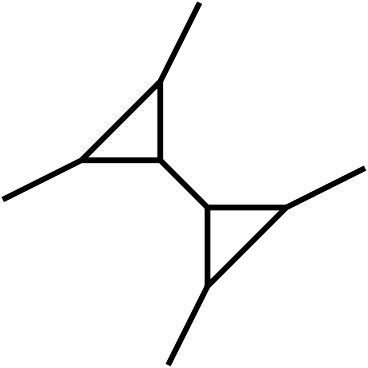}
\caption{A 5-brane web for the connecting two $E_0$ theories.}
\label{fig:twolocalP2}
\end{figure}

\begin{figure}[t]
\centering
\includegraphics{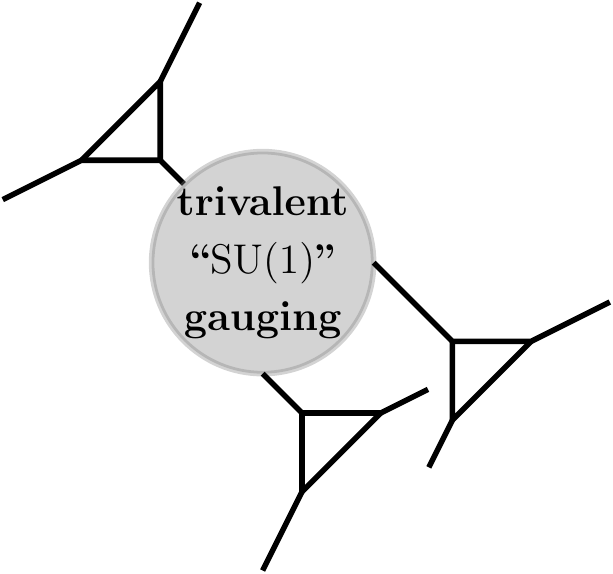}
\caption{A web--like description for the connecting three $E_0$ matter which is a 5d description of the $\mathcal{O}(-3)$ model.}
\label{fig:threelocalP2}
\end{figure}

From the geometric picture the three $E_0$ theories are coupled by the presence of the sphere $C_g$. Before considering coupling the three $E_0$ matter, let us think of coupling two $E_0$ matter. The 5-brane web picture for the $E_0$ theory is given in figure \ref{fig:localP2}. 
Then we can connect the two $E_0$ theories as in figure \ref{fig:twolocalP2}.
Note that a resolved conifold appears along the gluing and hence this geometry corresponds to a local Calabi--Yau threefold whose compact base is $\mathbb{P}^1$ on which we have two fixed points described by $\mathbb{C}^3/\mathbb{Z}_3$ in a singular limit.
Similarly, for coupling three $E_0$ matter along the $C_g \simeq \mathbb{P}^1$, we glue the three copies of the 5-brane web corresponding to a local $\mathbb{P}^2$ manifold by a single line as in figure \ref{fig:threelocalP2}.
We call the gluing trivalent ``$\SU(1)$'' gluing. 
Schematically, we may write
\begin{align}
E_0 - {\overset{\overset{\text{\large$E_0$}}{\textstyle\vert}}{\SU(1)}} - E_0
\label{Om3}
\end{align}
We propose that the ``$\SU(1)$'' gauging of three 5d $E_0$ matter is a 5d description of the 6d $\mathcal{O}(-3)$ model on a circle. 

One can check the number of gauge theory parameters in 5d agrees with the expectation from 6d.
In 6d, we have two vector multiplets in the Cartan subalgebra of $\mathfrak{su}(3)$ and one tensor multiplets.
Hence, the 5d description should have three vector multiplets and indeed the three copies of the $E_0$ theories provide three 5d vector multiplets in the Cartan subalgebra.
Since the 6d theory has no global symmetry, the 5d theory should have only one mass parameter from the radius of the compactification circle. This corresponds to the instanton fugacity of the trivalent ``$\SU(1)$'' gauging or the gluing parameter. 

Another consistency check is that the 6d $\mathcal{O}(-3)$ model in the limit $r_{S^1} \rightarrow 0$ should give rise to a pure $\SU(3)$ gauge theory. When one decouples one $E_0$ theory we arrive at a 5d theory whose web diagram is given in figure \ref{fig:twolocalP2}. A flop transition with respect to the gluing 5-brane indeed reproduces a 5-brane web diagram for a pure $\SU(3)$ gauge theory.

\begin{figure}[t]
\centering
\includegraphics{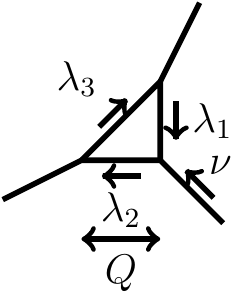}
\caption{The assignment of Young diagrams and K$\ddot{\text{a}}$hler parameter for computing the $E_0$ matter.}
\label{fig:localP2matter}
\end{figure}

\begin{figure}[t]
\centering
\includegraphics{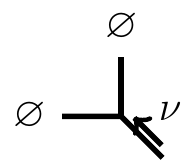}
\caption{A web diagram for the ``half'' of the conifold.}
\label{fig:halfconifold}
\end{figure}

We then present a prescription for the trivalent ``$\SU(1)$'' gauging of the three $E_0$ matter. The essential point is the same as the trivalent $\SU(2)$ gluing done in section \ref{sec:gluing}. Namely, in order to get the partition function of the $E_0$ matter for the ``$\SU(1)$'' gauging, we divide the partition function for the $E_0$ theory with non-trivial Young diagram on one external leg by a ``half'' of the partition function of the resolved conifold. More concretely, we compute the partition function of the $E_0$ theory with non-trivial Young diagram on one external leg corresponding to a web in figure \ref{fig:localP2matter}. 
The application of the topological vertex to the web in figure \ref{fig:localP2matter} yields 
\bea
\hat{Z}_{\nu}^{E_0}(Q) &=&\sum_{\lambda_1, \lambda_2, \lambda_3}(-Q)^{|\lambda_1| + |\lambda_2| + |\lambda_3|}f_{\lambda_1}(q)^{-2}f_{\lambda_2^t}(q)^2f_{\lambda_3^t}(q)^2C_{\lambda_3\lambda_1^t\nu^t}(q)C_{\lambda_2\lambda_3^t\emptyset}(q)C_{\lambda_1\lambda_2^t\emptyset}(q).\nn\\
\eea
The explicit calculation gives 
\bea
\hat{Z}_{\nu}^{E_0}(Q)&=&q^{\frac{1}{2}||\nu^t||^2}\tilde{Z}_{\nu^t}(q)\nn\\
&&\sum_{\lambda_1, \lambda_2, \lambda_3, \eta, \eta', \eta''}(-Q)^{|\lambda_1| + |\lambda_2| + |\lambda_3|}q^{-\frac{3}{2}\sum_{i=1}^3(||\lambda_i||^2 - ||\lambda_i^t||^2)}\nn\\
&&s_{\lambda_1^t/\eta}(q^{-\rho-\nu})s_{\lambda_1^t/\eta'}(q^{-\rho})s_{\lambda_2^t/\eta'}(q^{-\rho})s_{\lambda_2^t/\eta''}(q^{-\rho})s_{\lambda_3^t/\eta}(q^{-\rho-\nu^t})s_{\lambda_3^t/\eta''}(q^{-\rho}). \nn\\\label{localP2matterpre}
\eea
We then divide \eqref{localP2matterpre} by a ``half'' of the partition function of the resolved conifold represnted by a web in figure \ref{fig:halfconifold}.
Its partition function is simply given by
\be
Z^{\text{Half}\;\SU(1)}_{\nu} =  C_{\emptyset\emptyset\nu^t} = q^{\frac{1}{2}||\nu^t||^2}\tilde{Z}_{\nu^t}(q).
\ee
Then we claim that the partition function for the $E_0$ matter is 
\be
Z_{\nu}^{E_0}(Q) = \frac{\hat{Z}_{\nu}^{E_0}(Q)}{Z^{\text{Half}\;\SU(1)}}. \label{localP2matter}
\ee

\begin{figure}[t]
\centering
\includegraphics{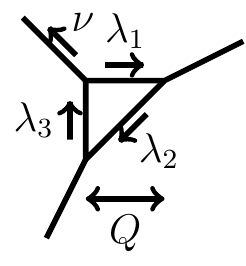}
\caption{Another web diagram for the $E_0$ matter with an opposite direction for the arrow associated to $\nu$ compared to the one in figure \ref{fig:localP2matter}}
\label{fig:localP2matter2}
\end{figure}
The partition function after the $\pi$ rotation compared to the one in figure \ref{fig:localP2matter} but with an opposite direction of the arrow for $\nu$ as in figure \ref{fig:localP2matter2} in fact gives the same answer as \eqref{localP2matter} after the division by the partition function of another ``half'' of the partition function of the resolved conifold. Hence, this means that the partition function \eqref{localP2matter} can be used for gluing from the left and also from the right. Therefore, one can use \eqref{localP2matter} for each of the contribution of the three $E_0$ matter. 

Finally, we couple the three $E_0$ matter system by the ``$\SU(1)$'' gauging corresponding the resolved conifold. The ``$\SU(1)$'' gauging contribution is 
\be
Z_{\nu}^{\SU(1)} = (-1)^{|\nu|}q^{\frac{1}{2}||\nu||^2+\frac{1}{2}||\nu^t||^2}\tilde{Z}_{\nu}(q)\tilde{Z}_{\nu^t}(q).
\ee
Therefore, the partition function of the 5d theory \eqref{Om3} is 
\bea
Z_{\mathcal{O}(-3)}(Q_g, Q_1, Q_2, Q_3) &=& \sum_{\nu}Q_g^{|\nu|}Z_{\nu}^{\SU(1)}Z^{E_0}_{\nu}(Q_1)Z^{E_0}_{\nu}(Q_2)Z^{E_0}_{\nu}(Q_3). \label{partOm3}
\eea

Since Eq.~\eqref{partOm3} is the partition function of the 5d theory for a circle compactification of the 6d $\mathcal{O}(-3)$ model, it should agree with the elliptic genus of the 6d $\mathcal{O}(-3)$ model. The elliptic genus of the 6d $\mathcal{O}(-3)$ model has been calculated in \cite{Kim:2016foj} and one string contribution is 
\bea
Z_{1}^{\mathcal{O}(-3)} (Q_{\tau}, Q_{m_1, m_2, m_3}) = \frac{\eta^2}{\theta(q)\theta(q^{-1})}\sum_{i=1}^3\frac{\eta^4\theta(Q_{m_i}^2)\theta(Q_{m_i})}{\prod_{j\neq i}\theta(Q_{m_i}Q_{m_j}^{-1})\theta(Q_{m_i}^{-1}Q_{m_j})\theta(Q_{m_j})}. \label{elliOm3}
\eea
Here $Q_{m_i}, i=1, 2, 3$ are the fugacities for the $\SU(3)$ symmetry and satisfy $Q_{m_1}Q_{m_2}Q_{m_3}=1$.

For the comparison of \eqref{elliOm3} with \eqref{partOm3}, we need to perform a flop transition with respect to a curve for the gluing. As mentioned before, the limit $Q_3 \rightarrow 0$ in \eqref{partOm3} does not directly yield the web diagram of the pure $\SU(3)$ gauge theory but we needed to perform a flop transition associated to $Q_g$. Therefore, in order to compare \eqref{partOm3} with \eqref{elliOm3}, we need to compute the partition function after the flop transition. 

\begin{figure}[t]
\centering
\includegraphics[width=9cm]{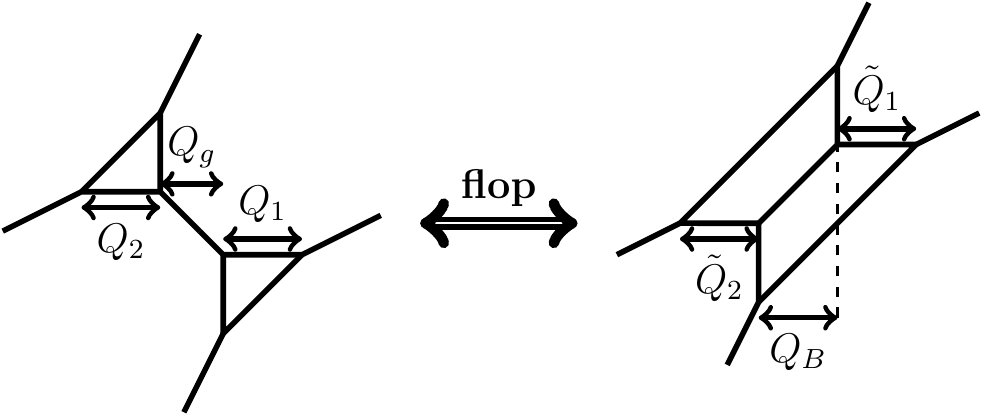}
\caption{A flop transition for the gluing cuve in figur \ref{fig:twolocalP2}. }
\label{fig:flop}
\end{figure}

Before going to the flop transition for web in figure \ref{fig:threelocalP2}, let us consider the flop transition for the web in figure \ref{fig:twolocalP2}. The partition function from the web in figure \ref{fig:twolocalP2} is 
\bea
Z_{\SU(3)}(Q_g, Q_1, Q_2) &=& \sum_{\nu}Q_g^{|\nu|}Z_{\nu}^{\SU(1)}Z^{E_0}_{\nu}(Q_1)Z^{E_0}_{\nu}(Q_2). \label{partSU3}
\eea
The flop transition is given in figure \ref{fig:flop}.
From the web in figure \ref{fig:flop}, the flop transition relates the K$\ddot{\text{a}}$hler parameters in the two webs by
\be
Q_g = Q_B^{-1}, \quad Q_1Q_g =\tilde{Q}_1, \quad Q_2Q_g = \tilde{Q}_2, \label{flop1}
\ee
where $\tilde{Q}_1, \tilde{Q}_2$ correspond to the Coulomb branch moduli of the pure $\SU(3)$ gauge theory and $Q_B$ is related to the instanton fugacity of the pure $\SU(3)$ gauge theory. Inserting \eqref{flop1} into \eqref{partSU3} does not work since it seems to include the negative power of $Q_B$ which should not appear in the pure $\SU(3)$ partition function. A trick is that when we divide it by partition function of the resolved conifold, namely
\be
\tilde{Z}_{\SU(3)}(Q_B, \tilde{Q}_1, \tilde{Q}_2) =  \frac{Z_{\SU(3)}(Q_B^{-1}, \tilde{Q}_1Q_B, \tilde{Q}_2Q_B) }{\mathcal{H}(Q_B^{-1})^{-1}} \label{partSU3pre}
\ee
then the terms with negative power of $Q_B$ disappears \cite{Iqbal:2004ne, Konishi:2006ev, Gukov:2007tf, Taki:2008hb, Awata:2009sz, Iqbal:2012mt}. Since we divide \eqref{partSU3} by $\mathcal{H}(Q_B^{-1})^{-1}$, we need to multiply \eqref{partSU3pre} by $\mathcal{H}(Q_B^{-1})^{-1}$. Here we can use the flop invariance of the partition function of the resolved conifold \cite{Iqbal:2004ne}
\be
\mathcal{H}(Q_B^{-1})^{-1} =\mathcal{H}(Q_B)^{-1}, \label{flopinvariance}
\ee
up to some factor which we neglect. Therefore, the partition function after the flop transition as in figure \ref{fig:flop} is given by
\be
Z_{\SU(3)}^{\text{flop}}(Q_B, \tilde{Q}_1, \tilde{Q}_2) =  \frac{\tilde{Z}_{\SU(3)}(Q_B^{-1}, \tilde{Q}_1Q_B, \tilde{Q}_2Q_B) }{\mathcal{H}(Q_B^{-1})^{-1}}\mathcal{H}(Q_B)^{-1}.  \label{partSU3flop}
\ee
Eq.~\eqref{partSU3flop} should agree with the $\SU(3)$ Nekrasov partition function given by
\bea
Z_{\SU(3)}^{\text{Nek}} = Z_{\SU(3)}^{\text{Pert}}\left(1 + u_{\SU(3)}^kZ_{\SU(3), k}^{\text{Inst}}\right),
\eea
where
\bea
Z_{\SU(3), k}^{\text{Inst}} &=& \sum_{\sum_i|Y_i| = k}\prod_{i, j=1}^3\prod_{s \in Y_i}\frac{1}{2\sinh\frac{E_{ij}(s)}{2}2\sinh\frac{E_{ij}(s)-(\epsilon_1 + \epsilon_2)}{2}}, \label{SU3Nekrasov}\\
E_{ij}(s)&=& a_i - a_j -\epsilon_1\ell_i(s) + \epsilon_2(a_j(s) + 1).
\eea
Here $Y_i$'s are Young diagrams and $a_1, a_2, a_3$ with $a_1 + a_2 + a_3 = 0$ are the Coulomb branch moduli related to $Q_1, Q_2, Q_3$ by
\be
e^{-a_1} = \tilde{Q}_1^{\frac{2}{3}}\tilde{Q}_2^{\frac{1}{3}}, \quad e^{-a_2} = \tilde{Q}_1^{-\frac{1}{3}}\tilde{Q}_2^{\frac{1}{3}}.
\ee 
$Z_{\SU(3)}^{\text{Pert}}$ is the perturbative part of the $\SU(3)$ partition function and 
\be
Z_{\SU(3)}^{\text{Pert}} = \mathcal{H}(\tilde{Q}_1)^2\mathcal{H}(\tilde{Q}_2)^2\mathcal{H}(\tilde{Q}_1\tilde{Q}_2)^2.
\ee
We checked that Eq.~\eqref{partSU3flop} agrees with the Nekrasov partition function \eqref{SU3Nekrasov}  of the pure $\SU(3)$ gauge theory by identifying the instanton fugacity $u_{\SU(3)}$ as
\be
u_{\SU(3)} = - \frac{Q_B}{\tilde{Q}_1\tilde{Q}_2}.
\ee
until the order of $Q_B^3\tilde{Q}_1^2\tilde{Q}_2^2$ and $Q_B^2\tilde{Q}_1^3\tilde{Q}_2^2$. 

We are now ready to apply the flop transition to the partition function \eqref{partOm3}. We assume that the same prescription for the flop transition apply for the trivalent ``$\SU(1)$'' gauging. We conjecture that the partition function of the 5d theory \eqref{partOm3} after the flop transition is given by
\bea
Z_{\mathcal{O}(-3)}^{\text{flop}}(Q_B, \tilde{Q}_1, \tilde{Q}_2, \tilde{Q}_3) =  \frac{\tilde{Z}_{\mathcal{O}(-3)}(Q_B^{-1}, \tilde{Q}_1Q_B, \tilde{Q}_2Q_B, \tilde{Q}_3Q_B) }{\mathcal{H}(Q_B^{-1})^{-1}}\mathcal{H}(Q_B)^{-1}.  \label{partOm3flop}
\eea
The partition function \eqref{partOm3flop} can be directly compared with the elliptic genus \eqref{elliOm3}. The K$\ddot{\text{a}}$hler parameters $\tilde{Q}_1, \tilde{Q}_2, \tilde{Q}_3$ form the affine Dynkin diagram of $\mathfrak{su}(3)$ and we can for example choose $\tilde{Q}_1, \tilde{Q}_2$ for the simple roots of the $\mathfrak{su}(3)$ corresponding to the 6d $\SU(3)$ symmetry. Then a map between $Q_{m_1}, Q_{m_2}, Q_{m_3}$ and $\tilde{Q}_1, \tilde{Q}_2$ is 
\be
\tilde{Q}_1 = Q_{m_1}Q_{m_2}^{-1}, \quad \tilde{Q}_2 = Q_{m_2}Q_{m_3}^{-1},
\ee
which can be written by
\be
Q_{m_1} = \tilde{Q}_1^{\frac{2}{3}}\tilde{Q}_2^{\frac{1}{3}}, \quad Q_{m_2} = \tilde{Q}_1^{-\frac{1}{3}}\tilde{Q}_2^{\frac{1}{3}}, \label{CoulombOm3}
\ee
with $Q_{m_3} = Q_{m_1}^{-1}Q_{m_2}^{-1}$. From the comarks of the affine Dynkin diagram of $\mathfrak{su}(3)$, the complex structure modulus of the torus is  
\be
Q_{\tau} = \tilde{Q}_1\tilde{Q}_2\tilde{Q}_3. \label{tauOm3}
\ee
By using the maps \eqref{CoulombOm3} and \eqref{tauOm3}, we checked that \eqref{partOm3flop} agrees with $Q_B\frac{Q_{\tau}^{\frac{1}{2}}}{\tilde{Q}_1\tilde{Q}_2}Z_{1}^{\mathcal{O}(-3)} (Q_{\tau}, \{Q_{m_i}\})$ until the order of $\tilde{Q}_1^2\tilde{Q}_2^2\tilde{Q}_3^2$ for the one-string part. This also implies that the string fugacity is given by
\be
Q_s = Q_B\frac{\tilde{Q}_3^{\frac{1}{2}}}{\tilde{Q}_1^{\frac{1}{2}}\tilde{Q}_2^{\frac{1}{2}}}. 
\ee

\subsection{Another non-Higgsable cluster}
\label{sec:6dNHC}

So far we have focused on the $\mathcal{O}(-n)$ models which contain only one tensor multiplet or equivalently one $\mathbb{P}^1$ base.
In particular when $n=3, 4, 6, 8, 12$ the $\mathcal{O}(-n)$ model has an orbifold description of $(T^2 \times \mathbb{C}^2)/\Gamma$ with the orbifold action given by \eqref{6dorbifold}, leading to its 5d description after a circle compactification. There are still another non-Higgsable cluster theories which contain multiple tensor multiplets or more than one base curves \cite{Morrison:2012np, Heckman:2013pva}. The 6d theories again have no flavor symmetry. The F-theory geometry has a compact base which is an elliptic fibration over a collection of spheres given in table \ref{tb:NHC}.
\begin{table}[t]
\begin{center}
\begin{tabular}{|c|c|c|c|} 
\hline
base & $3, 2$ & $3, 2, 2$ & $2, 3, 2$\\
\hline
gauge & $\mathfrak{g}_2 \times \mathfrak{su}(2)$ & $\mathfrak{g}_2 \times \mathfrak{sp}(1)$ & $\mathfrak{su}(2) \times \mathfrak{so}(7) \times \mathfrak{su}(2)$\\
\hline 
matter & $\frac{1}{2}({\bf 7} + {\bf 1}, {\bf 2})$ & $\frac{1}{2}({\bf 7} + {\bf 1}, {\bf 2})$ & $\frac{1}{2}({\bf 2}, {\bf 8}, {\bf 1}) + \frac{1}{2}({\bf 1}, {\bf 8}, {\bf 2})$\\
\hline
\end{tabular}
\caption{Non-Higgsable clusters with multiple base curves. The first row of the table represents the negative of the self--intersection numbers of the base spheres.}
\label{tb:NHC}
\end{center}
\end{table}
They are also important ingredients for constructing 6d SCFTs. 

Among the three non-Higgsable clusters, the last entry in table \ref{tb:NHC} has an orbifold description \cite{DelZotto:2015rca}. The F-theory geometry is $(T^2 \times \mathbb{C}^2)/\Gamma$ with the orbifold action
\be
g = (\omega^{-6}; \omega, \omega^5), \label{orbifoldNHC}
\ee
with $\omega^8 = 1$. On the torus the orbifold action is $\mathbb{Z}_4$. The torus then becomes a sphere with one $\mathbb{Z}_2$ fixed point and two $\mathbb{Z}_4$ fixed points. 

\begin{figure}[t]
\centering
\includegraphics[width=5cm]{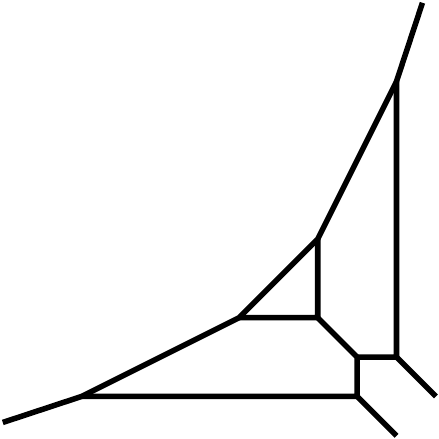}
\caption{A 5-brane web diagram corresponding to the orbifold $\mathbb{C}^3/\Gamma$ with the orbifold action \eqref{orbifoldNHC}.}
\label{fig:NHC}
\end{figure}

Then we consider a 5d description of this 6d theory. We can simply consider M-theory on the same orbifold geometry. Since $g^4 = (1; -1, -1)$, the orbifold action induces an $A_1$ singularity and the 5d theory has an $\SU(2)$ gauge symmetry. Around the $\mathbb{Z}_2$ fixed point, the geometry becomes $\mathbb{C}^3/\Gamma'$ with the action
\be
g' = g^2 = (\omega^{-12},\omega^2, \omega^{10}) = (\omega^{-4}, \omega^2, \omega^2) = (\omega^{\prime -2}, \omega^{\prime} , \omega^{\prime}),
\ee
with $\omega^{\prime 4} = 1$. This is the same geometry as \eqref{5dorbifold} with $p=2$. Namely, the 5d theory is the $\hat{D}_2(\SU(2))$ theory at the $\mathbb{Z}_2$ fixed point. Around the $\mathbb{Z}_4$ fixed point, the geometry is an orbifold $\mathbb{C}^3/\Gamma$ with the orbifold action \eqref{orbifoldNHC}. It is possible to write a 5-brane web corresponding to the orbifold geometry and it is depicted in figure \ref{fig:NHC}.  
The 5d theory has an $\SU(2)$ flavor symmetry with three Coulomb branch moduli. We denote the 5d theory by $\hat{D}_{\Gamma}(\SU(2))$, Therefore, the 5d theory for the non-Higgsable cluster is
\begin{align}
\hat{D}_{\Gamma}(\SU(2)) - {\overset{\overset{\text{\large$\hat{D}_2(\SU(2))$}}{\textstyle\vert}}{\SU(2)}} - \hat{D}_{\Gamma}(\SU(2)) \label{NHC}
\end{align}
The 5d theory is again given by the $\SU(2)$ trivalent gauging. 

Let us see whether the numbers of 5d gauge theory parameters agrees with the expectation from 6d. 
The number of vector multiplets in the Cartan subalgebra in 6d is $1 + 3 + 1 = 5$.
The number of tensor multiplets is $3$.
After a circle compactification they should become $5 + 3 = 8$ vector multiplets in the Cartan subalgebra and the 5d theory should have an eight--dimensional Coulomb branch moduli space.
In 5d, $\hat{D}_2(\SU(2))$ theory has one Coulomb branch modulus and two $\hat{D}_{\Gamma}(\SU(2))$ theories have $2\times 3 = 6$ Coulomb branch moduli.
By adding one Coulomb branch modulus from the trivalent $\SU(2)$ gauging, the 5d theory has an eight-dimensional Coulomb branch moduli space which agrees with the expectation.
Since the 6d theory has no flavor symmetry, the 5d theory should have only one mass parameter, Indeed the 5d theory \eqref{NHC} has one mass parameter coming from the instanton fugacity of the $\SU(2)$ trivalent gauging.

\bigskip

\section{Refinement}
\label{sec:refined}
So far we have considered the unrefined partition function where the two $\Omega$--deformation parameters $\epsilon_1, \epsilon_2$ are set to $\epsilon_1 = - \epsilon_2$. In this section, we extend the rule for the trivalent $\SU(2)$ gluing to the refined topological vertex formalism. Instead of performing the calculation in full generality, we will focus on a specific example of the pure $\SO(8)$ gauge theory and describe how the trivalent $\SU(2)$ gauging can be generalized to the refined case. The application to other cases will be carried out in a similar manner in principle.

\subsection{Refined partition function of $\hat{D}_2(\SU(2))$ matter from flop transition}

In order to perform the computation for the trivalent $\SU(2)$ gauging for the refined case, we first need to determine the refined partition function of the $\hat{D}_N(\SU(2))$ matter corresponding to the web in figure \ref{fig:DNSU2top}. Similarly to the topological vertex formalism, we assign the refined topological vertex which is labeled by three Young diagrams corresponding to three legs at each vertex of a 5-brane web. However the role of the three legs is not symmetric and we assign $t$, $q$ and a preferred direction for each leg. Furthermore, when one glues a leg with $t (\text{or}\; q)$ with another leg, then the another leg should be labeled by $q (\text{or}\; t)$. 

Let us first think about the case when we choose the vertical directions in figure \ref{fig:DNSU2top} for the preferred direction, then the gluing leg in the horizontal direction should be labeled by $t$ or $q$. In order to have the consistent gluing for the refined topological vertex, one needs to label $t$ or $q$ in a different way for the horizontal legs in the web for the other $\hat{D}_{N}(\SU(2))$ matter. When we glue two $\hat{D}_{N}(\SU(2))$ matter system then this gluing rule causes no problem. However when we consider the trivalent gluing with three $\hat{D}_{N}(\SU(2))$ matter system, then it is difficult to 
glue three webs consistently with the gluing rule for the refined topological vertex.

This problem can be avoided when we choose the horizontal direction in figure \ref{fig:DNSU2top} for the preferred direction.
This is also conceptually plausible. The equation \eqref{eq:vectorvertex} which we relied on can be generalized to the refined case only when the preferred direction is taken to be horizontal.
However another problem arises since some vertex does not have a leg in the preferred direction and we cannot apply the refined topological vertex to such a vertex. 

\begin{figure}[t]
\centering
\includegraphics[width=5cm]{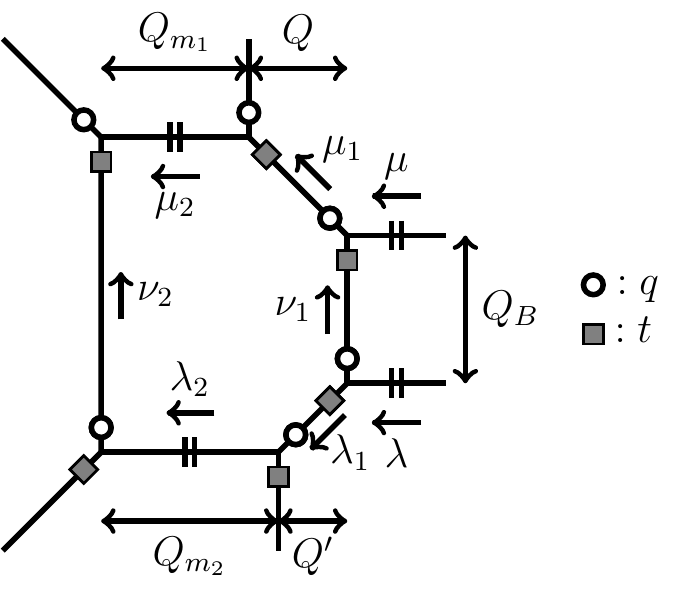}
\caption{A 5-brane web for a theory which is S--dual to the $\SU(2)$ gauge theory with two flavors. The K$\ddot{\text{a}}$hler parameters satisfy $Q_{m_1}Q = Q_{m_2}Q'$.}
\label{fig:SU2w2flvrs}
\end{figure}

\begin{figure}[t]
\centering
\includegraphics[width=7cm]{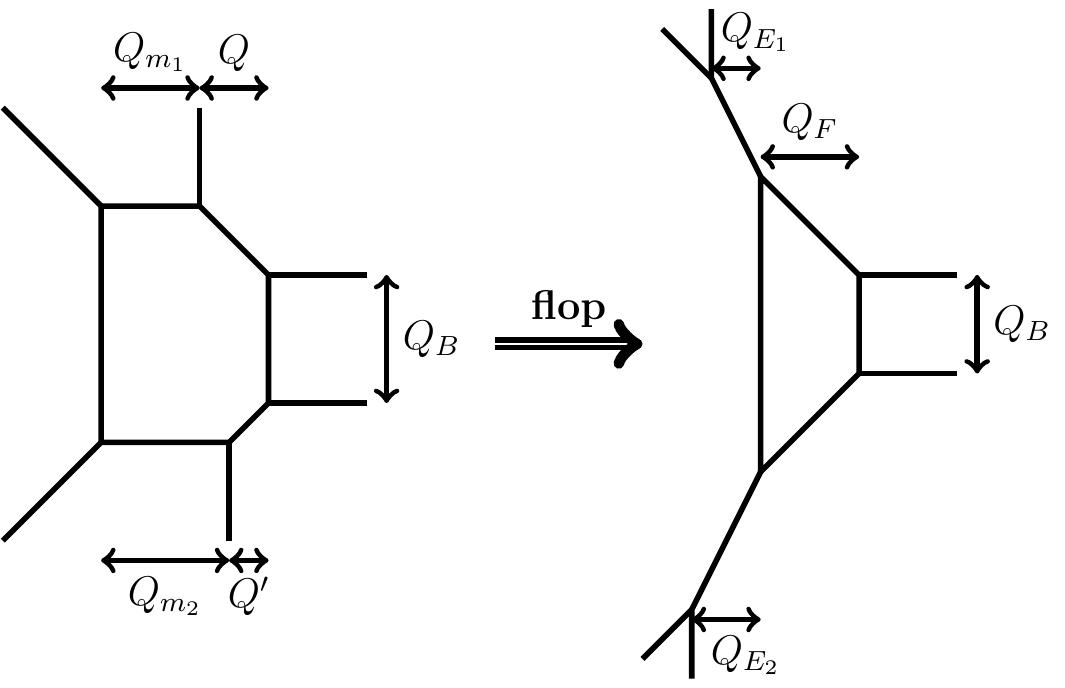}
\caption{Two flop transitions from the web in figure \ref{fig:SU2w2flvrs}.}
\label{fig:flop2}
\end{figure}

In fact, there is a way to solve the second problem by using a flop transition. To see that we focus on the case of the $\hat{D}_2(\SU(2))$ matter which we will use for the computation of the refined partition function of the pure $\SO(8)$ gauge theory. Although we cannot apply the refined topological vertex to the web for the $\hat{D}_2(\SU(2))$ matter with the horizontal direction chosen for the preferred direction, we can first apply the refined topological vertex to a different but a related to web in figure \ref{fig:SU2w2flvrs}. 
From the web in figure \ref{fig:SU2w2flvrs}, we can perform a flop transition with respect to the curves whose K$\ddot{\text{a}}$hler parameters are $Q_{m_1}$ and $Q_{m_2}$ as in figure \ref{fig:flop2}. Then we obtain a web on the right in figure \ref{fig:flop2}.
From the right web in figure \ref{fig:flop2}, one can send $Q_{m_1}, Q_{m_2} \rightarrow 0$, giving rise to a web in figure \ref{fig:SU2w2flvrs}. From the comparison between the webs in figure \ref{fig:flop2}, the K$\ddot{\text{a}}$hler parameters are related by
\be
Q_{m_1} = Q_{E_1}^{-1}, \quad  Q= Q_FQ_{E_1}, \quad Q_{m_2} = Q_{E_2}^{-1}, \quad Q' = Q_FQ_{E_2}. \label{flop2}
\ee
The same trick has been used to obtain the refined partition function for the $\hat{D}_{\mathbb{P}^2}$ theory \cite{Iqbal:2012mt}. 

We then first compute the refined partition function for the web in figure \ref{fig:SU2w2flvrs}. The application of the refined topological vertex to the web in figure \ref{fig:SU2w2flvrs} yields
\bea
\tilde{\hat{Z}}_{\lambda, \mu}^L(Q_B, Q, Q', Q_{m_1}, Q_{m_2}) &=& \sum_{\lambda_1, \lambda_2, \mu_1, \mu_2, \nu_1, \nu_2}C_{\nu_2^t\emptyset\mu_2^t}(t, q)C_{\emptyset\nu_2\lambda_2^t}(t, q)(-Q_BQQ')^{|\nu_2|}\tilde{f}_{\nu_2^t}(t, q)\nn\\
&&C_{\emptyset\mu_1^t\mu_2}(q, t)(-Q)^{|\mu_1|}C_{\nu_1^t\mu_1\mu^t}(t, q)(-Q_B)^{|\nu_1|}\tilde{f}_{\nu_1^t}(t, q)\nn\\
&&C_{\lambda_1\nu_1\lambda^t}(t, q)(-Q')^{|\lambda_1|}C_{\lambda_1^t\emptyset\lambda_2}(q, t)(-Q_{m_1})^{|\mu_2|}(-Q_{m_2})^{|\lambda_2|},\nn\\
\eea
where $QQ_{m_1} = Q'Q_{m_2}$. After a calculation, we get 
\bea
\tilde{\hat{Z}}_{\lambda, \mu}^L(Q_B, Q, Q', Q_{m_1}, Q_{m_2})&=& q^{\frac{1}{2}||\mu^t||^2 + ||\lambda^t||^2}\tilde{Z}_{\lambda^t}(t, q)\tilde{Z}_{\mu^t}(t, q)\nn\\
\sum_{\lambda_1, \mu_1, \nu_1, \nu_2, \eta, \eta'}&&Q_B^{|\nu_1| + |\nu_2|}(-1)^{|\mu_1| + |\lambda_1|}Q^{|\mu_1|+|\nu_2|}Q^{\prime |\lambda_1| + |\nu_2|}\nn\\
&&s_{\mu_1/\eta}(t^{-\mu}q^{-\rho})s_{\nu_1/\eta}(t^{-\rho}q^{-\mu^t})s_{\nu_1/\eta'}(t^{-\lambda}q^{-\rho})s_{\lambda_1^t/\eta'}(t^{-\rho}q^{-\lambda^t})\nn\\
&&\left(\frac{q}{t}\right)^{\frac{1}{2}(|\eta| + |\eta'| - |\nu_2| - |\nu_1|)}Z^{\text{RC}}_{\mu_1^t\nu_2}(Q_{m_1}; t, q)Z^{\text{RC}}_{\lambda_1\nu_2}(Q_{m_2}; q, t),\nn\\ \label{SU2w2flvrs}
\eea
where 
\be
Z^{\text{RC}}_{\mu_1\mu_2}(Q; t, q)=\sum_{\nu}(-Q)^{|\nu|}q^{\frac{1}{2}||\nu||^2}t^{\frac{1}{2}||\nu^t||^2}\tilde{Z}_{\nu}(t, q)\tilde{Z}_{\nu^t}(q, t)s_{\mu_1}(t^{-\rho}q^{-\nu})s_{\mu_2}(t^{-\rho}q^{-\nu}).
\ee

In order to apply the flop transition in figure \ref{fig:flop2}, 
we use a similar trick which we used in section \ref{sec:Om3}. The insertion of \eqref{flop2} into \eqref{SU2w2flvrs} gives
\bea
\tilde{\hat{Z}}_{\lambda, \mu}^L(Q_B, Q_FQ_{E_1}, Q_FQ_{E_2}, Q_{E_1}^{-1}, Q_{E_2}^{-1})&=& q^{\frac{1}{2}||\mu^t||^2 + ||\lambda^t||^2}\tilde{Z}_{\lambda^t}(t, q)\tilde{Z}_{\mu^t}(t, q)\nn\\
\sum_{\lambda_1, \mu_1, \nu_1, \nu_2, \eta, \eta'}&&Q_B^{|\nu_1| + |\nu_2|}(-1)^{|\mu_1| + |\lambda_1|}Q_F^{|\mu_1|+|\lambda_1| + 2|\nu_2|}\nn\\
&&s_{\mu_1/\eta}(t^{-\mu}q^{-\rho})s_{\nu_1/\eta}(t^{-\rho}q^{-\mu^t})s_{\nu_1/\eta'}(t^{-\lambda}q^{-\rho})s_{\lambda_1^t/\eta'}(t^{-\rho}q^{-\lambda^t})\nn\\
&&\left(\frac{q}{t}\right)^{\frac{1}{2}(|\eta| + |\eta'| - |\nu_2| - |\nu_1|)}\nn\\
&&Z^{\text{RC}}_{\mu_1^t\nu_2}(Q_{E_1}^{-1}; t, q)Z^{\text{RC}}_{\lambda_1\nu_2}(Q_{E_2}^{-1}; q, t)Q_{E_1}^{|\mu_1| + |\nu_2|}Q_{E_2}^{|\lambda_1| + |\nu_2|}.\nn\\ \label{SU2w2flvrs1}
\eea
Then we consider the quantity 
\be
G_{\mu\nu}(Q; t, q) = \frac{Z_{\mu\nu}^{\text{RC}}(Q; t, q)}{Z_{\text{conifold}}(Q)}, \label{Gmunu}
\ee
where
\be
Z_{\text{conifold}}(Q) = \prod_{i, j=1}\left(1 - Qq^{i-\frac{1}{2}}t^{j-\frac{1}{2}}\right) \label{refinedconifold}
\ee
In fact, $G_{\mu\nu}(Q)$ is a polynomial of degree $|\mu| +|\nu|$ in $Q$ \cite{Gukov:2007tf, Awata:2009sz}. Therefore, the following limit is well-defined
\be
\tilde{G}_{\mu\nu}(t, q)= \lim_{Q\rightarrow 0} G_{\mu\nu}(Q^{-1}; t, q)Q^{|\mu|+|\nu|}. 
\ee
By using the flop invariance for the partition function of the resolved conifold \eqref{refinedconifold}, the limit for $Z_{\mu\nu}^{\text{RC}}(Q; t, q)$ can be taken as
\bea
Z_{\mu\nu}^{\text{flop}}(t, q) &=& \lim_{Q \rightarrow 0} Z_{\text{conifold}}(Q) \times \frac{Z_{\mu\nu}^{\text{RC}}(Q^{-1}; t, q)}{Z_{\text{conifold}}(Q^{-1})}Q^{|\mu|+|\nu|}\nn\\
&=& \left[\frac{Z_{\mu\nu}^{\text{RC}}(Q^{-1}; t, q)}{Z_{\text{conifold}}(Q^{-1})}Q^{|\mu|+|\nu|}\right]_{Q^0},
\eea
where $[F(Q)]_{Q^0}$ implies that we take the zeroth order of $Q$ from $F(Q)$. Therefore, applying the limit $Q_{E_1}, Q_{E_2} \rightarrow 0$ to \eqref{SU2w2flvrs1}, we obtain
\bea
\hat{Z}_{\lambda, \mu}^L(Q_B, Q_F) &=& q^{\frac{1}{2}(||\mu^t||^2 + ||\lambda^t||^2)}\tilde{Z}_{\lambda^t}(t, q)\tilde{Z}_{\mu^t}(t, q)\nn\\
\sum_{\lambda_1, \mu_1, \nu_1, \nu_2, \eta, \eta'}&&Q_B^{|\nu_1| + |\nu_2|}(-1)^{|\mu_1| + |\lambda_1|}Q_F^{|\mu_1|+|\lambda_1| + 2|\nu_2|}\nn\\
&&s_{\mu_1/\eta}(t^{-\mu}q^{-\rho})s_{\nu_1/\eta}(t^{-\rho}q^{-\mu^t})s_{\nu_1/\eta'}(t^{-\lambda}q^{-\rho})s_{\lambda_1^t/\eta'}(t^{-\rho}q^{-\lambda^t})\nn\\
&&\left(\frac{q}{t}\right)^{\frac{1}{2}(|\eta| + |\eta'| - |\nu_2| - |\nu_1|)}Z_{\mu_1^t\nu_2}^{\text{flop}}(t, q)Z_{\lambda_1\nu_2}^{\text{flop}}(q, t). \label{refinedD2SU2pre}
\eea

For the refined partition function of the $\hat{D}_2(\SU(2))$ matter, one needs to divide the refined partition function by a ``half'' of the partition function of the $\SU(2)$ vector multiplets
\bea
Z^{\text{Half}, L}_{\lambda, \mu}(Q_B) &=& q^{\frac{1}{2}(||\lambda^t||^2 + ||\mu^t||^2)}\tilde{Z}_{\lambda^t}(t, q)\tilde{Z}_{\mu^t}(t, q)\nn\\
&&\sum_{\nu}Q_B^{|\nu|}\left(\frac{q}{t}\right)^{\frac{1}{2}(- |\nu|)}
s_{\nu^t}(q^{-\mu^t}t^{-\rho})s_{\nu^t}(q^{-\rho}t^{-\lambda}).
\eea
Therefore, the refined version of the $\hat{D}_2(\SU(2))$ matter contribution is given by
\be
Z^{\hat{D}_2(\SU(2)), \text{ref}}_{\lambda, \mu}(Q_B, Q_F) = \frac{\hat{Z}_{\lambda, \mu}^L(Q_B, Q_F)}{Z^{\text{Half}, L}_{\lambda, \mu}(Q_B) }. \label{refinedD2SU2}
\ee

\begin{figure}[t]
\centering
\includegraphics[width=5cm]{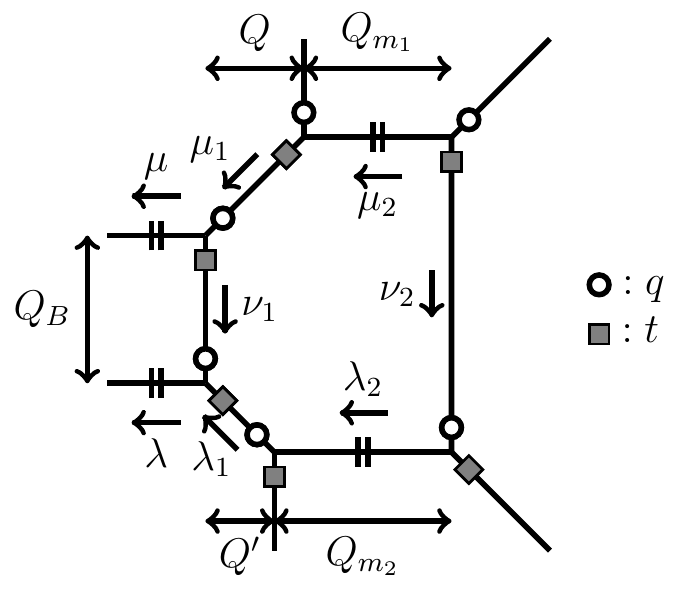}
\caption{Another 5-brane web for a theory which is S--dual to the $\SU(2)$ gauge theory with two flavors.}
\label{fig:SU2w2flvrs1}
\end{figure}

In order to treat \eqref{refinedD2SU2} for the $\hat{D}_2(\SU(2))$ matter, we check whether the web diagram which is given by the $\pi$ rotation compared to figure \ref{fig:SU2w2flvrs} but with the opposite direction for the arrows of $\lambda, \mu$ yields the same partition function. We then compute the partition function for the web in figure \ref{fig:SU2w2flvrs1} and apply the limit \eqref{flop2}.
By following the same steps, we obtain the partition function 
\bea
\hat{Z}_{\lambda, \mu}^R(Q_B, Q_F) &=& t^{\frac{1}{2}(||\mu||^2 + ||\lambda||^2)}\tilde{Z}_{\lambda}(q, t)\tilde{Z}_{\mu}(q, t)\nn\\
\sum_{\lambda_1, \mu_1, \nu_1, \nu_2, \eta, \eta'}&&Q_B^{|\nu_1| + |\nu_2|}(-1)^{|\mu_1| + |\lambda_1|}Q_F^{|\mu_1|+|\lambda_1| + 2|\nu_2|}\nn\\
&&s_{\mu_1/\eta}(t^{-\mu}q^{-\rho})s_{\nu_1/\eta}(t^{-\rho}q^{-\mu^t})s_{\nu_1/\eta'}(t^{-\lambda}q^{-\rho})s_{\lambda_1^t/\eta'}(t^{-\rho}q^{-\lambda^t})\nn\\
&&\left(\frac{t}{q}\right)^{\frac{1}{2}(|\eta| + |\eta'| - |\nu_2| - |\nu_1|)}Z_{\mu_1^t\nu_2}^{\text{flop}}(t, q)Z_{\lambda_1\nu_2}^{\text{flop}}(q, t). \label{refinedD2SU2preR}
\eea
Dividing \eqref{refinedD2SU2preR} by the another half of the partition function of the vector multiplet for the $\SU(2)$ theory gives
\be
Z_{\lambda, \mu}^{\prime \hat{D}_2(\SU(2)), \text{ref}}(Q_B, Q_F) = \frac{\hat{Z}_{\lambda, \mu}^R(Q_B, Q_F)}{Z^{\text{Half}, R}_{\lambda, \mu}(Q_B)},\label{refinedD2SU2R}
\ee
where 
\bea
Z^{\text{Half}, R}_{\lambda, \mu}(Q_B) &=& t^{\frac{1}{2}(||\lambda||^2 + ||\mu||^2)}\tilde{Z}_{\lambda}(q, t)\tilde{Z}_{\mu}(q, t)\nn\\
&&\sum_{\nu}Q_B^{|\nu|}\left(\frac{t}{q}\right)^{\frac{1}{2}(- |\nu|)}
s_{\nu^t}(q^{-\mu^t}t^{-\rho})s_{\nu^t}(q^{-\rho}t^{-\lambda}).
\eea
It is not clear whether \eqref{refinedD2SU2} is equal to \eqref{refinedD2SU2R} but we checked that they are indeed equal to each other until the order of $Q_B^2Q_F^4$ for the cases $(\mu, \lambda) = (\emptyset, \emptyset), (\mu, \lambda) = \left(\{2, 1\}, \{1, 1\}\right)$. Therefore, we can use \eqref{refinedD2SU2} for the refined partition function of the $\hat{D}_2(\SU(2))$ matter.

\subsection{Examples: 5d pure $SO(8)$ gauge theory and $\mathcal{O}(-4)$ model}

In the previous subsection, we computed the refined version of the partition function of the $\hat{D}_2(\SU(2))$ matter. In this section we apply the trivalent $\SU(2)$ gauging for the refined partition function and obtain the Nekrasov partition functions of the pure $\SO(8)$ gauge theory and the 5d theory from the $\mathcal{O}(-4)$ model on a circle . 

\paragraph{Pure $\SO(8)$ gauge theory} A 5d dual description of the pure $\SO(8)$ gauge theory is described by \eqref{SO2Np4} with $N=2$. The Nekrasov partition function of the pure $\SU(2)$ gauge theory is given by 
\be
Z_{\SU(2)}(Q_B, Q_g) = \sum_{\lambda, \mu}Q_g^{|\lambda| + |\mu|}Z_{\lambda, \mu}^{\SU(2)\;\text{vector,ref}}(Q_B),
\ee
where 
\be
Z_{\lambda, \mu}^{\SU(2)\;\text{vector,ref}}(Q_B) =  t^{\frac{1}{2}||\mu||^2-\frac{1}{2}||\lambda||^2}q^{-\frac{1}{2}||\mu^t||^2+\frac{1}{2}||\lambda^t||^2}Z^{\text{Half}, R}_{\lambda, \mu}(Q_B)Z^{\text{Half}, L}_{\lambda, \mu}(Q_B). 
\ee
Hence, we propose that the refined Nekrasov partition function of the pure $\SO(8)$ gauge theory is given by
\bea
Z_{\SO(8)}(Q_B, Q_g, Q_1, Q_2, Q_3) &=& \sum_{\lambda, \mu}Q_g^{|\lambda| + |\mu|}Z_{\lambda, \mu}^{\SU(2)\;\text{vector,ref}}(Q_B)Z_{\lambda, \mu}^{\hat{D}_2(\SU(2)),\text{ref}}(Q_B, \{Q_1\})\nn\\
&&Z_{\lambda, \mu}^{\hat{D}_2(\SU(2)), \text{ref}}(Q_B, \{Q_2\})Z_{\lambda, \mu}^{\hat{D}_2(\SU(2)), \text{ref}}(Q_B, \{Q_3\}), \label{refinedpartSO8}
\eea
where the K$\ddot{\text{a}}$hler parameters are related to the gauge theory parameters by \eqref{CoulombSO2Np4} and \eqref{instantonSO2Np4} with $N=2$. We checked that Eq.~\eqref{refinedpartSO8} agrees with the refined Nekrasov partition function of the pure $\SO(8)$ gauge theory until the order of $Q_1^3Q_2^3Q_3^3Q_g^3$ for the one-instanton part. 

\paragraph{$\mathcal{O}(-4)$ model} 

We can also make use of the refined $\hat{D}_2(\SU(2))$ matter contribution to compute the Nekrasov partition function of the 5d theory \eqref{Om4} which arises from a circle compactification of the $\mathcal{O}(-4)$ model. In this case, we gauge four refined partition functions of the $\hat{D}_2(\SU(2))$ matter and the full partition function is given by
\bea
Z_{\mathcal{O}(-4)}(Q_B, Q_g, Q_1, Q_2, Q_3, Q_4) &=& \sum_{\lambda, \mu}Q_g^{|\lambda| + |\mu|}Z_{\lambda, \mu}^{\SU(2)\;\text{vector,ref}}(Q_B)\nn\\
&&Z_{\lambda, \mu}^{\hat{D}_2(\SU(2)),\text{ref}}(Q_B, \{Q_1\})Z_{\lambda, \mu}^{\hat{D}_2(\SU(2)), \text{ref}}(Q_B, \{Q_2\})\nn\\
&&Z_{\lambda, \mu}^{\hat{D}_2(\SU(2)), \text{ref}}(Q_B, \{Q_3\})Z_{\lambda, \mu}^{\hat{D}_2(\SU(2)), \text{ref}}(Q_B, \{Q_4\}). \nn\\\label{refinedpartOm4}
\eea
We checked that \eqref{refinedpartOm4} agrees with the elliptic genus \eqref{elliOm4} of the $\mathcal{O}(-4)$ model until the order of $Q_1^2Q_2^2Q_3^2Q_4^2Q_g$ for the one-string part.



\bigskip

\section{Conclusion}
\label{sec:concl}

In this paper, we have proposed a novel method to compute the topological string partition functions/Nekrasov partition functions of 5d theories constructed by the trivalent gluing/gauging. A dual description of 5d pure gauge theories with a gauge group of $\D, \E$--type is given by the $\SU(2)$ trivalent gauging of three 5d $\hat{D}_N(\SU(2))$ matter theories. We have proposed a way to apply the topological vertex formalism to the trivalent gauging and successfully calculated their Nekrasov partition functions. We first computed the partition function of the 5d $\hat{D}_N(\SU(2))$ theory with non-trivial flavor instanton backgrounds, which can be used for a matter contribution for the $\SU(2)$ gauging. Then, combining the $\hat{D}_N(\SU(2))$ matter contributions with the partition function of the $\SU(2)$ vector multiplets yields the Nekrasov partition functions of the 5d pure gauge theories of $\D, \E$--type gauge groups. This method gives a new way to compute the Nekrasov partition functions and one advantage of this technique is that the higher-order instanton partition functions can be obtained systematically simply by summing over Young diagrams with more boxes. We also performed non--trivial checks with the known results of the $\SO(8)$ gauge theory with or without flavors and also the pure $\E_6, \E_7, \E_8$ gauge theories up to some order of the gluing parameters. 

Moreover, we will see in appendix \ref{sec:SOodd} that applying a Higgsing prescription to the Nekrasov partition function of a gauge theory with a $\D$--type gauge group and flavors may yield the Nekrasov partition function of a gauge theory with a $\mathrm{B}$--type gauge group. Therefore, with the Higgsing procedure as well as the trivalent gluing method, it is now possible to compute the Nekrasov partition functions of 5d pure gauge theories with a $\mathrm{ABCDE}$ gauge group from the topological vertex. 

Another application of the trivalent gluing method is that we can also compute the Nekrasov partition functions of 5d theories which have a 6d UV completion. In particular the 5d description of the $\mathcal{O}(-n)$ models with $n=4, 6, 8, 12$ is written by gauging four or three 5d $\hat{D}_N(\SU(2))$ matter theories. We applied the trivalent gauging method for the 5d theories and performed a non--trivial check for the case of $n=4$ by comparing the Nekrasov partition function with the elliptic genus of the one--string calculated in \cite{Haghighat:2014vxa}. We also proposed a 5d description of the $\mathcal{O}(-3)$ model and calculated its Nekrasov partition function. Remarkably, we found perfect agreement with the elliptic genus result of the one--string in \cite{Kim:2016foj} up to some orders. In every case, the computation for higher instantons can be achieved very systematically and the trivalent gauging method provides a powerful tool to compute their elliptic genera. We also determine a 5d description of another non-Higgsable cluster theory and the 5d theory can be again described by the $\SU(2)$ gauging of three 5d theories. 

Most of the computation in this paper have been done in the unrefined limit. We also argued that it is possible to extend the computation for the refined topological vertex when we choose the preferred direction to the gluing direction. Indeed we have checked that the trivalent gluing prescription works for the refined one--instanton partition function for the pure $\SO(8)$ gauge theory and also the refined one-string elliptic genus of the $\mathcal{O}(-4)$ model. We expect that the refined calculation can be generalized to other cases.

As for the comparison with the exceptional instantons of 5d theories, we restrict the check to the one--instanton order which can be computed from the general formula \eqref{oneinstsimplylaced}. 
The higher-instanton partition functions of the exceptional gauge groups have been calculated in \cite{Gaiotto:2012uq, Keller:2012da, Hanany:2012dm, Cremonesi:2014vla}.
However, a direct comparison of the results obtained in this paper with the results in \cite{Gaiotto:2012uq, Keller:2012da, Hanany:2012dm,Cremonesi:2014vla} may not be straightforward since the explicit expressions in the literature seems not to be compatible with the unrefined limit. It would be interesting to extend the computation for the Nekrasov partition function of the exceptional gauge groups to the refined one by using the technique in section \ref{sec:refined}. 
Similarly the unrefined limit also prevented us from comparing the results with computations from other methods in the literature about 6d $\mathcal{O}(-n)$ theories with $n=6,8,12$.
It would be interesting to extend the Nekrasov partition function computation for the 5d descriptions of the $\mathcal{O}(-6), \mathcal{O}(-8), \mathcal{O}(-12)$ models to the refined partition function computation and perform checks with the results in \cite{Gadde:2015xta, Putrov:2015jpa, DelZotto:2016pvm}.

We expect that our trivalent gauging method has vast applications. In this paper we only consider vector matter of the $\SO(2N+4)$ gauge group.
It will be interesting to generalize our method to include matter in different representations. Furthermore, our method is applicable to any $\SU(N)$ gluing of possibly non-Lagrangian matter. Finding more dualities among 5d/6d theories like what we argued in section \ref{sec:newweb} and computing Nekrasov partition functions would be fruitful.

\bigskip

\acknowledgments
We would like to thank Sung-Soo Kim, Kimyeong Lee, Mastato Taki, Futoshi Yagi, and Yuji Tachikawa for useful discussion and conversation. We would also like to thank the RIKEN workshop of the Progress in Mathematical Understanding of Supersymmetric Theories during a part of this work. H.H.\ would like to thank Korea Institute for Advanced Study for hospitality.
K.O.\ is partially supported by the Programs for Leading Graduate
Schools, MEXT, Japan, via the Advanced Leading Graduate Course for Photon
Science and by JSPS Research Fellowship for Young Scientists.
K.O.\ gratefully acknowledges support from the Institute for Advanced Study.

\bigskip

\appendix 


\section{5d $SO(2N+3)$ gauge theory}
\label{sec:SOodd}

In section \ref{sec:SO2Np4wflvrs}, we have computed the partition function of the $\SO(2N+4)$ gauge theory with $N_f = M_1 + M_2$ flavors by making use of the trivalent $\SU(2)$ gauging. On a Higgs branch of the $\SO(2N+4)$ gauge theory with $M_1+M_2$ flavors, it is possible to realize a 5d $\SO(2N+3)$ gauge theory with $N_f - 1$ flavors in the far infrared. A 5-brane web picture for the Higgsing has been presented in \cite{Zafrir:2015ftn}. Therefore, one can apply the Higgsing prescription for the Nekrasov partition function of the $\SO(2N+4)$ gauge theory with $N_f$ flavors to obtain the Nekrasov partition function of the $\SO(2N+3)$ gauge theory with $N_f - 1$ flavors.


From the Higgsing procedure of the 5-brane web with an $O5$-plane, the Higgsing from the $\SO(2N+4)$ gauge theory with $N_f$ flavors to the $\SO(2N+3)$ gauge theory with $N_f-1$ flavors may be achieved by setting one mass parameter and also one Coulomb branch modulus to be zero. We can for example choose
\be
a_{N+2} = m_{N_f} = 0. \label{tuning}
\ee
Here we denote the Coulomb branch moduli of $\SO(2N+4)$ by $a_i, i=1, \cdots, N+2$ and the mass parameters by $m_i, i=1, \cdots, N_f$. 

In fact, the tuning condition \eqref{tuning} can be directly applied to the Nekrasov partition function of the $\SO(2N+4)$ gauge theory with $N_f$ flavors. A similar Higgsing prescription has been used to compute the Nekrasov partition function of the rank one $E_7$ theory \cite{Hayashi:2013qwa} and also the rank one $E_8$ theory \cite{Hayashi:2014wfa}. In the refined case, the tuning is not as simple as \eqref{tuning} but the parameters are fixed to be $\left(\frac{q}{t}\right)^{\frac{1}{2}}$ or $\left(\frac{t}{q}\right)^{\frac{1}{2}}$. However, in the unrefined case, we can directly use the tuning condition of \eqref{tuning}. 

Let us see how the condition \eqref{tuning} works for the perturbative part. The perturbative partition function of the $\SO(2N+4)$ gauge theory with $N_f$ flavors can be written as\footnote{Note that the flop invariance of the partition function of the resolved conifold implies
\be
\mathcal{H}(Q) = \mathcal{H}(Q^{-1}). \label{floppert}
\ee 
We always make use of \eqref{floppert} to compare the perturbative partition functions from the topological vertex with the perturbative partition function from the localization result. Namely we check the equality between the two perturbative partition functions up to the flop transitions \eqref{floppert}.  
}
\be
Z^{\text{Pert}}_{\SO(2N+4), N_f} =\mathcal{H}(1)^{N+2}\left[\prod_{1 \leq i < j \leq N+2}\mathcal{H}\left(e^{-(a_i \pm a_j)}\right)^2\right]\left[\prod_{i=1}^{N+2}\prod_{f=1}^{N_f}\mathcal{H}\left(e^{-(m_f \pm a_i)}\right)^{-1}\right]. \label{pertSO2Np4wflvrsgeneral}
\ee
Inserting the condition \eqref{tuning} into \eqref{pertSO2Np4wflvrsgeneral} yields 
\bea
Z^{\text{Pert}}_{\SO(2N+4), N_f} |_{\text{Eq.~}\eqref{tuning}} &=& \mathcal{H}(1)^N\left[\prod_{i=1}^{N+1}\mathcal{H}\left(e^{-\alpha_i}\right)^2\right]\left[\prod_{1 \leq i < j \leq N+1}\mathcal{H}\left(e^{-(a_i \pm a_j)}\right)^2\right]\nn\\
&&\left[\prod_{i=1}^{N+1}\prod_{f=1}^{N_f-1}\mathcal{H}\left(e^{-(m_f \pm a_i)}\right)^{-1}\right]\left[\prod_{f=1}^{N_f-1}\mathcal{H}\left(e^{-m_f}\right)^{-1}\right]\nn\\
&=& \mathcal{H}(1)^{-1} Z_{\SO(2N+3), N_f-1}^{\text{Pert}},
\eea
up to flop transitions \eqref{floppert}. Therefore, the perturbative partition function of the $\SO(2N+3)$ gauge theory with $N_f-1$ flavors is reproduced except for the factor $ \mathcal{H}(1)^{-1}$ which can be understood as a singlet contribution o the Higgs vacuum. 

When one includes the instanton partition function, a natural expectation is that 
\be
Z_{\SO(2N+4), N_f}|_{\text{Eq.~}\eqref{tuning}} = \mathcal{H}(1)^{-1} Z_{\SO(2N+3). N_f-1}. \label{HiggstoSOodd}
\ee
We checked that \eqref{HiggstoSOodd} indeed holds for the one-instanton part of a simple case of $N=2, N_f = 2$ by using the localization result \eqref{localizationSO}. 

Assuming that \eqref{HiggstoSOodd} is correct, it is then possible to compute the Nekrasov partition function of the 5d $\SO(2N+3)$ gauge theory with flavors by combining \eqref{NekSO2Np4wflvrs} and \eqref{tuning} from the relation \eqref{HiggstoSOodd}. 



\bigskip

\section{Some formulae for computation}
\label{sec:formulae}
In this appendix, we collect formulae which we have used for the calculation of the (refined) topological vertex as well as the Nekrasov partition function in this paper. 

\subsection{Refined topological vertex}
\label{sec:topvertex}

The topological vertex is a powerful tool to compute the all genus topological string amplitude \cite{Iqbal:2002we, Aganagic:2003db} for a Calabi--Yau manifold $X_3$ of the form
\be
Z_{\text{top}} = \exp\left(\sum_{g=0}^{\infty}F_gg_{\text{top}}^{2g-2}\right),
\ee
where
\be
F_g = \sum_{C \in H_2(X_3, \mathbb{Z})}N_C^g Q_C.
\ee
$g_{\text{top}}$ is the topological string coupling constant, $N_g^C$ is the genus $g$ Gromov--Witten invariant for a curve $C$ and $Q_C = e^{-k_c}$ with the K$\ddot{\text{a}}$hler parameters $k_C$ for a curve $C$. The topological vertex is parameterized by the topological string coupling and it is possible to further generalize it to the refined topological vertex by introducing two parameters $q, t$ corresponding to the $\Omega$--deformation parameters by $q = e^{-\epsilon_1}, t = e^{\epsilon_2}$ \cite{Awata:2005fa, Iqbal:2007ii}. The unrefined limit is given by setting $q = t$. Although the original refined topological vertex has constructed for the application to toric Calabi--Yau threefolds, it can be also applied to certain non--toric Calabi--Yau threefolds by making use of a Higgsing or topology changing transition from a toric Calabi--Yau threefold \cite{Hayashi:2013qwa, Hayashi:2014wfa, Kim:2015jba, Hayashi:2015xla, Hayashi:2016abm, Hayashi:2016jak}. Here we summarize the rule for applying the refined topological vertex to a toric Calabi--Yau threefold or a dual 5-brane web. 

\begin{figure}[t]
\centering
\includegraphics[width=5cm]{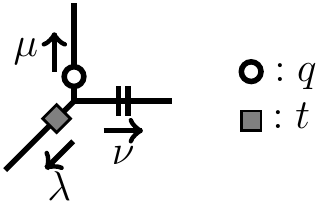}
\caption{A vertex whose three legs are labeled by a pair of $(t, \lambda)$, $(q, \mu)$ and $\nu$ with the preferred direction. The preferred direction is denoted by $||$.}
\label{fig:vertex}
\end{figure}

The refined topological vertex formalism provides us with a method to compute the all genus topological string amplitude on a background of a toric Calabi--Yau threefold by a way which is similar to the method using Feynman diagrams. We first decompose a toric diagram or  5-brane web into trivalent vertices with three legs. We assign a Young diagram to each leg with some orientation. When the leg is an external leg, then we assign a trivial Young diagram on it. We also need to choose a preferred direction in the diagram and one leg of the refined topological vertex should be in the preferred direction. We then assign $t, q$ for the other two legs of the vertex. The $t, q$ assignment should be compatible with the gluing rule which we will mention below. Let $\lambda, \mu, \nu$ be three Young diagrams. When the three legs of a vertex is labeled by a pair of $(t, \lambda)$, $(q, \mu)$ and $\nu$ with the preferred direction as in figure \ref{fig:vertex}, 
we assign to the vertex of a 5-brane web the refined topological vertex
\be
C_{\lambda\mu\nu}(t, q) = t^{-\frac{1}{2}||\mu^t||^2}q^{\frac{||\mu||^2 + ||\nu||^2}{2}}\tilde{Z}_{\nu}(t, q)\sum_{\eta}\left(\frac{q}{t}\right)^{\frac{|\eta| + |\lambda| - |\mu|}{2}}s_{\lambda^t/\eta}(t^{-\rho}q^{-\nu})s_{\mu/\eta}(t^{-\nu^t}q^{-\rho}), \label{reftopvertex}
\ee
where
\be
\tilde{Z}_{\nu}(t, q) = \prod_{s \in \nu}\left(1-q^{l_{\nu}(s)}t^{a_{\nu}(s) + 1}\right)^{-1}.
\ee
Here we also defined
\be
l_{\nu}(i, j) = \nu_i - j, \quad a_{\nu}(i, j) = \nu^t_j - i.
\ee
for $(i, j) \in \nu$.

\begin{figure}[t]
\centering
\includegraphics[width=7cm]{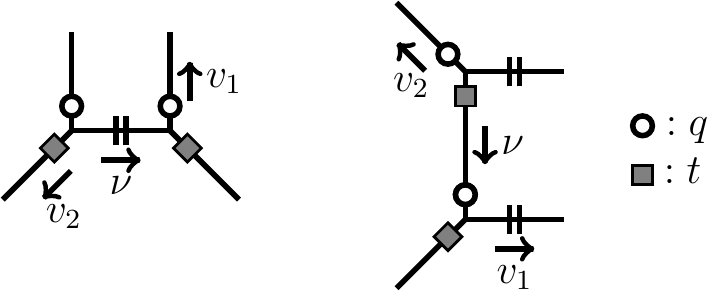}
\caption{The assignment of the vectors for determining the framing factor. }
\label{fig:framing}
\end{figure}

Then we need to glue the vertices for going back to the original 5-brane web. For each gluing of two legs, the assigned Young diagram on one leg should be transposed compared to the Young diagram on the other leg. Then the gluing is done by summing over a Young diagram $\nu$ associated to the two legs with a weight. When we glue along the preferred direction then the weight takes a form of 
\be
(-Q)^{|\nu|}f_{\nu}(t, q)^n, \label{weight1}
\ee
where the framing factor for the preferred direction is
\be
f_{\nu}(t, q) = (-1)^{|\nu|}t^{\frac{-||\nu^t||^2}{2}}q^{\frac{||\nu||^2}{2}}.
\label{eq:framing}
\ee
When we glue along the non-preferred direction then the weight has a form of 
\be
(-Q)^{|\nu|}\tilde{f}_{\nu}(t, q)^n, \label{weight2}
\ee
where the framing factor for the non-preferred direction is 
\be
\tilde{f}_{\nu}(t, q) = (-1)^{|\nu|}q^{-\frac{||\nu^t||^2}{2}}t^{\frac{||\nu||^2}{2}}\left(\frac{t}{q}\right)^{\frac{|\nu|}{2}}.
\ee
where $n$ is given by $n = \det(v_1, v_2)$ as in figure \ref{fig:framing}. 
$Q$ is given by $Q = e^{-k_C}$ where $k_C$ is the K$\ddot{\text{a}}$hler parameter for a curve associated to the glued internal line. When we glue along the non-preferred direction, we need to connect a leg on which $q$ is assigned with a leg on which $t$ is assigned.   

By assigning refined topological vertex \eqref{reftopvertex} for each vertex and also the weights \eqref{weight1} or \eqref{weight2}, the topological string partition function is given by summing all the assigned Young diagrams. The rules for the unrefined version can be obtained simply by setting $t = q$. 

\begin{figure}[t]
\centering
\includegraphics[width=5cm]{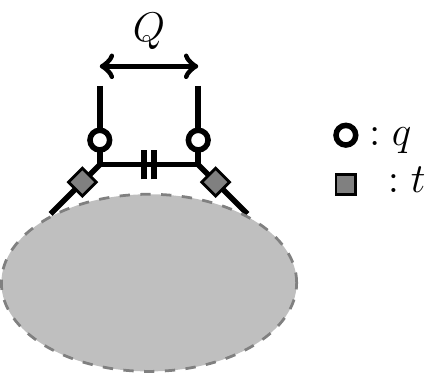}
\caption{A 5-brane web which contains extra factor from the topological vertex calculation.}
\label{fig:extra}
\end{figure}

An important point is that the topological string partition function for a certain local Calabi--Yau threefold $X_3$ is related to the Nekrasov partition function of a 5d theory with eight supercharges realized from M-theory compactification on $\tilde{X}_3$ or equivalently on  a 5-brane web dual to $\tilde{X}_3$ \cite{Iqbal:2003ix, Iqbal:2003zz, Eguchi:2003sj, Hollowood:2003cv, Taki:2007dh}. In fact, it turns out the topological string partition function calculated from the refined topological vertex contains contributions that are not present in the Nekrasov partition function and one needs to extract that factor \cite{Bergman:2013ala, Hayashi:2013qwa, Bao:2013pwa, Bergman:2013aca}. The factor is related to the contribution from strings between parallel external legs. Therefore the factor can be read off from a 5-brane web and for example the extra factor from a web in figure \ref{fig:extra}
 is given by 
\be
Z_{\text{extra}} = \prod_{i, j=1}^{\infty}\left(1 - Qq^{i}t^{j-1}\right)^{-1}.
\ee
We call such a factor extra factor. 

Therefore, the Nekrasov partition function of a 5d theory can be computed by the topological string partition function of the corresponding Calabi--Yau threefold by dividing it by the extra factor,
\be
Z_{\text{Nek}} = \frac{Z_{\text{top}}}{Z_{\text{extra}}}.
\ee
Note that the refined topological vertex computation does not include the perturbative contribution from vector multiplets in the Cartan subalgebra but it can be easily recovered since it has a general form
\be
Z_{\text{Cartan}} = \prod_{i, j=1}^{\infty}\left(1 - q^{i}t^{j-1}\right)^{-\frac{\text{rank}(G)}{2}}\left(1 - q^{i-1}t^{j}\right)^{-\frac{\text{rank}(G)}{2}}
\ee
for a gauge group $G$. 

\subsection{Nekrasov partition function}

In this section we summarize the result of the Nekrasov partition function for some 5d gauge theories with eight supercharges. 

For a gauge with a gauge group $G$, the perturbative partition function of the vector multiplets is given by
\be
Z_{\text{vec}}^{\text{Pert}} = Z_{\text{Cartan}}\prod_{i,j=1}^{\infty}\left[\prod_{\alpha\in\Delta_+}\left(1-e^{-\alpha\cdot a}q^it^{j-1}\right)\left(1-e^{-\alpha\cdot a}q^{i-1}t^j\right)\right]^{-1}, \label{pert.vector}
\ee
where $\Delta_+$ is a set of positive roots and $a=(a_1, \cdots, a_{\text{rank}G})$ represents the Coulomb branch moduli in the Cartan subalgebra. The perturbative partition function of hypermultiplets in the representation ${\bf r}$ is 
\be
Z_{\text{hyp}}^{\text{Pert}} =\prod_{i,j=1}^{\infty}\left[\prod_{ w \in{\bf r}}\left(1-e^{- (w \cdot a-m)}q^{i-\frac{1}{2}}t^{j-\frac{1}{2}}\right)\right], \label{pert.hyper}
\ee
where $w$ is a weight of the representation ${\bf r}$. Note that the comparison using the perturbative partition functions \eqref{pert.vector} and \eqref{pert.hyper} is done up to flop transitions. 

For the pure gauge theory with a gauge group $G$, the general result for the one--instanton part has been also known and it is given by \cite{Benvenuti:2010pq, Keller:2011ek, Zafrir:2015uaa,Billo:2015pjb, Billo:2015jyt} 
\bea
Z_{\text{1--inst}}^G = \frac{\left(\frac{q}{t}\right)^{\frac{h^{\vee}}{2}}}{(1-q)(1-t^{-1})}\sum_{\gamma \in \Delta_l} \frac{e^{\frac{(h^{\vee}-1)\gamma\cdot a}{2}}}{\left(1-\frac{q}{t}e^{\gamma\cdot a}\right)\left(e^{\frac{\gamma\cdot a}{2}} - e^{-\frac{\gamma\cdot a}{2}}\right)\prod_{\gamma^{\vee}\cdot\alpha=1}\left(e^{\frac{\alpha\cdot a}{2}} - e^{-\frac{\alpha\cdot a}{2}}\right)}.\nn\\ \label{oneinstgeneral}
\eea
$\alpha, \gamma$ are roots of the Lie algebra $\mathfrak{g}$, $h^{\vee}$ is the dual Coxeter number\footnote{The relevant numbers in this paper are $h^{\vee}_{E_6} = 12, h^{\vee}_{E_7} = 18, h^{\vee}_{E_8} = 30$.}, $\Delta_l$ is a set of long roots\footnote{In the case when $G$ is simply--laced, $\Delta_l$ is a set of all the roots.}. $\gamma^{\vee}$ is a coroot of $\gamma$. When $G$ is simply-laced and we take the unrefined case $q = t$, the expression after putting all the terms over a common denominator takes a form
\be
Z_{\text{1--inst}}^G = -\frac{q}{(1-q)^2}\frac{\sum_{\gamma \in \Delta_+}\left[(-1)^{1+n}e^{(\tilde{\gamma}_- + \gamma)\cdot a}(e^{(h^{\vee}-2)\gamma\cdot a}+1)\prod_{\beta \in \Delta_+}(e^{\beta\cdot a}-1)^{2-|\gamma^{\vee}\cdot\beta|}\right]}{\prod_{\alpha \in \Delta_+} (e^{\alpha\cdot a} - 1)^2}, \label{oneinstsimplylaced}
\ee
where $\Delta_+$ is again a set of positive roots and $\tilde{\gamma}_- =\sum_{\gamma^{\vee}\cdot\beta=-1, \beta\in\Delta_+}\beta$. $n$ stands for the number of positive roots $\beta$ which satisfy $\gamma^{\vee}\cdot\beta = -1$. 

Next we turn to the result of the instanton partition function from the localization technique \cite{Nekrasov:2002qd, Nekrasov:2003rj, Marino:2004cn, Nekrasov:2004vw, Shadchin:2005mx, Hwang:2014uwa}. The $k$--instanton partition function can be computed from the index of the one--dimensional ADHM quantum mechanics whose moduli space is given by the corresponding $k$--instanton moduli space. We here quote the result of the instanton partition function for the $\SO(N)$ gauge theory with hypermultiplets in the vector representation. 

The $k$--instanton partition function for the $\SO(N)$ gauge theory with $N_f$ hypermultiplets in the vector representation is given by a contour integral over the dual $\Sp(k)$ gauge group variables,
\be
Z_{k\text{--inst}} = \frac{1}{|W_k|}\oint\left[\prod_{I=1}^k\frac{d\phi_I}{2\pi i}\right]Z_{\text{vec}}Z_{\text{hyp}}, \label{localizationSO}
\ee
where 
\bea
Z_{\text{vec}}&=& \frac{\prod_{1\leq I < J \leq k}2\sinh\left(\frac{\pm\phi_I\pm\phi_J}{2}\right)2\sinh\left(\frac{\pm\phi_I\pm\phi_J + 2\epsilon_+}{2}\right)\prod_{I=1}^k2\sinh\left(\pm\phi_I\right)2\sinh\left(\pm\phi_I+\epsilon_+\right)}{\prod_{I=1}^k\prod_{i=1}^{n}2\sinh\left(\frac{\pm\phi_I\pm a_i + \epsilon_+}{2}\right)\prod_{1\leq I < J \leq k}2\sinh\left(\frac{\pm \phi_I \pm \phi_J + \epsilon_1}{2}\right)2\sinh\left(\frac{\pm \phi_I \pm \phi_J + \epsilon_2}{2}\right)}\nn\\
&\times&\frac{\left(2\sinh\epsilon_+\right)^k}{\left(2\sinh\frac{\epsilon_1}{2}2\sinh\frac{\epsilon_2}{2}\right)^k}
\eea
for $N=2n$ and 
\bea
Z_{\text{vec}}&=& \frac{\prod_{1\leq I < J \leq k}2\sinh\left(\frac{\pm\phi_I\pm\phi_J}{2}\right)2\sinh\left(\frac{\pm\phi_I\pm\phi_J + 2\epsilon_+}{2}\right)\prod_{I=1}^k2\sinh\left(\pm\phi_I\right)2\sinh\left(\pm\phi_I+\epsilon_+\right)}{\prod_{I=1}^k2\sinh\left(\frac{\pm\phi_I+\epsilon_+}{2}\right)\prod_{I=1}^k\prod_{i=1}^{n}2\sinh\left(\frac{\pm\phi_I\pm a_i + \epsilon_+}{2}\right)}\nn\\
&\times&\frac{\left(2\sinh\epsilon_+\right)^k}{\left(2\sinh\frac{\epsilon_1}{2}2\sinh\frac{\epsilon_2}{2}\right)^k\prod_{1\leq I < J \leq k}2\sinh\left(\frac{\pm \phi_I \pm \phi_J + \epsilon_1}{2}\right)2\sinh\left(\frac{\pm \phi_I \pm \phi_J + \epsilon_2}{2}\right)}
\eea
for $N=2n+1$. Here the notation $2\sinh(\pm x \pm y)$ means $2\sinh(\pm x \pm y) = 2\sinh(x+y)2\sinh(x-y)2\sinh(-x+y)2\sinh(-x-y)$. $a_i, i=1, \cdots, n$ are the Coulomb branch moduli of the $\SO(N)$ and we also defined $\epsilon_+ = \frac{\epsilon_1 + \epsilon_2}{2}$. $Z_{\text{hyp}}$ is the contribution from the hypermultiplets in the vector representation and it is given by
\be
Z_{\text{hyp}} = \prod_{I=1}^k\prod_{f=1}^{N_f}2\sinh\left(\frac{m_f - \phi_I}{2}\right)2\sinh\left(\frac{m_f + \phi_I}{2}\right). 
\ee
Finally $|W_k|$ is the order of the Weyl group of the $\Sp(k)$ which is the dual gauge group of $\SO(N)$. More concretely, $|W_k| = 2^kk!$. 

The contour integral \eqref{localizationSO} can be systematically evaluated by so-called the Jeffery-Kirwan residue rule \cite{Hwang:2014uwa}.

\subsection{Schur functions}
\label{sec:Schur}
Here we summarize the formulas on Schur functions which is needed to perform the topological vertex computations.
Schur polynomials $s_{\lambda}(x_1,\cdots,x_n)$ with finite variables can be defined by
\begin{align}
	s_{\lambda}(x_1,\cdots,x_n)&=\frac{\mathrm{det}A_{\lambda}}{\mathrm{det}A_{\phi}},\\
	(A_{\lambda})_{ij} &=
	\begin{cases}
		x_i^{\lambda_j+n-j} & j\le L\\
		x_i^{n-j} & j > L
	\end{cases},
	\label{eq:Schur1}
\end{align}
where $\lambda={\lambda_1,\cdots ,\lambda_{L}}$ is a integer partition.
Schur polynomials have a scaling property
\begin{equation}
	s_{\lambda}(a x_1,\cdots,a x_n)=a^{|\lambda|}s_{\lambda}(x_1,\cdots,x_n)
\end{equation}
with $|\lambda|=\sum \lambda_i$.
Schur functions are the infinite variables generalization of this polynomial.
In particular, we often use principal specialization of Schur function, defined by
\begin{equation}
	s_{\lambda}(q^{-\rho})=s_{\lambda}(q^{1/2},q^{3/2},q^{5/2},\cdots).
	\label{eq:Schurscale}
\end{equation}
Stanley's hook-length formula \cite{stanley1971theory} says
\begin{equation}
	s_{\lambda}(q^{-\rho})=q^{\frac{||\lambda||}{2}}\prod_{u\in\lambda}\frac{1}{1-q^{\mathrm{hook}(u)}}
	\label{eq:Stanley}
\end{equation}
where $u$ runs through boxes of the Young diagram $\lambda$, and $\mathrm{hook}(u)$ is $a(u)+\ell(u)+1$.
The important point is that the righthand side is finite product and thus this formula is exact with respect to $q$.
This formula is the reason why we can compute partition functions from topological vertices exactly with respect to the exponentiated $\epsilon$ parameters.

We also encounter Schur functions with arguments like
\begin{equation}
	s_{\lambda}(q^{-\rho-\nu})=s_\lambda(q^{1/2-\nu_1},q^{3/2-\nu_2}, \cdots q^{L'/2 -\nu_{L'}},q^{(L'+1)/2}\cdots).
	\label{eq:Schurrhonu}
\end{equation}
where $\nu=(\nu_1,\cdots,\nu_{L'})$ is another partition.
To compute this function explicitly, we make use of the formula
\begin{equation}
	s_\lambda(\mathbf{x},\mathbf{y})
	=\sum_{\mu,\nu\subset\lambda}c^{\lambda}_{\mu,\nu}s_{\mu}(\mathbf{x})s_\nu(\mathbf{y}),
	\label{eq:Schurcoprod}
\end{equation}
where $\mathbf{x},\mathbf{y}$ are sets of variables and $c^\lambda_{\mu,\nu}$ are Littlewood-Richardson coefficients.
Set $\mathbf{x}$ to be the first $L'$ variables of \eqref{eq:Schurrhonu} and $\mathbf{y}$ to be the remaining, and use \eqref{eq:Schur1} for the former and \eqref{eq:Stanley}, \eqref{eq:Schurscale} for the latter.
Using $\eqref{eq:Schurcoprod}$ repeatedly, we can also compute Schur functions like
\begin{equation}
	s_{\lambda}(q^{-\rho-\nu_1},q^{-\rho-\nu_2},\cdots,q^{-\rho-\nu_N}).
	\label{eq:Schurgen}
\end{equation}

We also encounter two variants of Schur functions, 
which are skew Schur functions
\begin{equation}
	s_{\lambda/\mu}(\mathbf{x})=\sum_{\nu\subset\lambda}c^{\lambda}_{\mu,\nu} s_\nu(\mathbf{x}),
\end{equation}
and super Schur functions
\begin{equation}
	s_{\lambda}(\mathbf{x}|\mathbf{y})
	=\sum_{\mu,\nu\subset\lambda}c^{\lambda}_{\mu,\nu}s_{\mu}(\mathbf{x})s_{\nu^t}(\mathbf{y}).
\end{equation}
The skew Schur function $s_{\lambda/\mu}$ is equal to the Schur function $s_\lambda$ when $\mu=\emptyset$, and 0 when $\mu$ is not included in $\lambda$.

A \texttt{Mathematica} implementation which automates computations of Schur functions like \eqref{eq:Schurgen} and those generalization to skew and super Schur functions is available online at \href{https://github.com/kantohm11/SchurFs}{https://github.com/kantohm11/SchurFs}.

In the main part of this paper, we used the following formulas \cite{macdonald1998symmetric}
\begin{align}
	\sum_{\eta} s_{\lambda/\eta}(\mathbf{x})s_{\eta}(\mathbf{y})&=s_{\lambda}(\mathbf{x},\mathbf{y})\\
	\sum_{\mu} s_{\mu/{\eta_1}}(\mathbf{x})s_{\mu/\eta_2}(\mathbf{y})&= 
	\prod_{i,j}(1-x_jy_j)^{-1}\sum_\tau s_{\eta_1/\tau}(\mathbf{y})s_{\eta_2/\tau}(\mathbf{x})\\
	\sum_{\mu} s_{\mu/{\eta_1}}(\mathbf{x})s_{\mu^t/\eta_2}(\mathbf{y})&= 
	\prod_{i,j}(1+x_jy_j)\sum_\tau s_{\eta_1^t/\tau^t}(\mathbf{y})s_{\eta_2^t/\tau}(\mathbf{x}).
\end{align}


\bigskip

\bibliographystyle{JHEP}
\bibliography{refs}
\end{document}